\documentclass[nofootinbib,
 amsmath,amssymb,
preprint,%
]{revtex4-2}
\usepackage{graphicx}
\usepackage{dcolumn}
\usepackage{bm}
\usepackage{xcolor}
\usepackage[breaklinks,colorlinks]{hyperref}
\hypersetup{
    colorlinks, 
    linktoc=all, 
    linkcolor=red,  
  citecolor=hyptxt,
  urlcolor=blue}
  \hypersetup{
  citebordercolor=,
  filebordercolor=green,
  linkbordercolor=blue
}
\definecolor{hyptxt}{rgb}{0.7, 0.4, 0.9}
\usepackage{appendix}
\usepackage{setspace} 
\usepackage{changes}

\definechangesauthor[name={Sandro}, color=red]{san}

\newcommand{\dd}{\mathrm{d}}

\newcommand{\ci}{\mathsf{i}}

\newcommand{\myfrac}[2]{%
    \setbox0\hbox{$#1$}        
    \dimen0=\wd0               
    \setbox1\hbox{$#2$}        
    \dimen1=\wd1               
    \ifdim\wd0<\wd1            
        \dfrac{#1\hfill}{#2}   
    \else
        \dfrac{#1}{#2\hfill}   
    \fi
}

\newcommand{\ddt}{\dd{}t}

\newcommand{\EV}[1]{\vert\HFi\vert}

\newcommand{\HFi}{\bm{k}}

\newcommand{\bra}[1]{\left\langle#1\right\vert}
\newcommand{\ket}[1]{\left\vert{#1}\right\rangle}

\newcommand{\x}{x}
\DeclareFontFamily{U}{mathx}{\hyphenchar\font45}
\DeclareFontShape{U}{mathx}{m}{n}{<-> mathx10}{}
\DeclareSymbolFont{mathx}{U}{mathx}{m}{n}
\DeclareMathAccent{\widebar}{0}{mathx}{"73}

\newcommand{\cs}{c_s}

\begin{document}
\title{Primordial Magnetogenesis in a Bouncing Universe}

\author{E. Frion$^{1}$ \footnote{\href{mailto:emmanuel.frion@cosmo-ufes.org}{emmanuel.frion@cosmo-ufes.org}},  N. Pinto-Neto$^{1,2}$ \footnote{\href{mailto:nelsonpn@cbpf.br}{nelsonpn@cbpf.br}}, S.~D.~P. Vitenti$^{3}$ \footnote{\href{mailto:vitenti@uel.br}{vitenti@uel.br}} and S.~E. Perez Bergliaffa$^{4}$ \footnote{\href{mailto:sepbergliaffa@gmail.comg}{sepbergliaffa@gmail.com}}}

\affiliation{\vspace{0.7cm} $^{1}$PPGCosmo, CCE, Universidade Federal do Esp\'{i}rito Santo,Vit\'{o}ria, 29075-910, Esp\'{i}rito Santo, Brazil }

\affiliation{\vspace{0.7cm} $^{2}$Centro Brasileiro de Pesquisas F\'{i}sicas, Rua Dr.~Xavier Sigaud 150, Urca,  CEP 22290-180, Rio de Janeiro, RJ, Brazil}

\affiliation{\vspace{0.7cm} $^{3}$Departamento de F\'{i}sica, Universidade Estadual
        de Londrina, Rod. Celso Garcia Cid, Km 380, 86057-970,
        Londrina, Paran\'{a}, Brazil.}
        
\affiliation{\vspace{0.7cm} $^{4}$Departamento  de  F\'{i}sica  Te\'{o}rica,  Instituto  de  F\'{i}sica, Universidade  do  Estado  de  Rio  de  Janeiro,  CEP  20550-013,  Rio  de  Janeiro,  Brazil \vspace{0.5cm}}

\begin{abstract}
We investigate primordial magnetogenesis and the evolution of the electromagnetic field through a quantum bounce \cite{Peter:2006hx}, in a model that starts in the far past from a contracting phase where only dust is present and the electromagnetic field is in the adiabatic quantum vacuum state. By including a coupling between curvature and electromagnetism of the form $RF_{\mu \nu}F^{\mu \nu}$, we find acceptable magnetic field seeds within the current observational constraints at 1 mega-parsec (Mpc), and that the magnetic power spectrum evolves as a power-law with spectral index $n_B=6$. It is also shown that the electromagnetic backreaction is not an issue in the model under scrutiny. 
\end{abstract}


\date{\today}

\maketitle

\section{Introduction}

The existence of magnetic fields
in a variety of scales in the Universe (see for instance \cite{Durrer2013,beck2012magnetic,Beck:2013bxa})
calls the question of their origin.
In particular, there are several 
observations 
consistent with weak $\sim 10^{-16}$ Gauss fields in the intergalactic medium, coherent on Mpc scales: the 21-cm hydrogen line \cite{Minoda:2018gxj}, the anisotropy of ultra-high energy cosmic rays 
 \cite{Bray:2018ipq}, 
CMB distortions \cite{Ade:2015cva,Chluba:2019kpb}, B-mode polarization measurements \cite{Zucca:2016iur,Pogosian:2018vfr}, magnetic reheating \cite{Saga:2017wwr}, 
Big Bang Nucleosynthesis (BBN) \cite{Kawasaki:2012va}, and
$\gamma$-rays \cite{Barai:2018msb}, among others. Since such fields 
remained largely undisturbed during the cosmological evolution (as opposed to those in the presence of structure), they offer a window to  their origin, which is generally assumed to be primordial.

Primordial seed fields 
(which may be amplified later by the dynamo mechanism
\cite{Brandenburg2004})
are 
generated
before structure formation, for instance out of the expansion
 of the universe, either during inflation \cite{Martin:2007ue,Ratra:1991bn,Davis:1995mv,Berera:1998hv,Kandus:1999st,Bassett:1999ta,Battaner:2000kf,Davis:2000zp,Tornkvist:2000js,Davis:2005ih,Anber:2006xt,Jimenez:2010hu,Das:2010ywa,BeltranJimenez:2011vn,Bonvin:2011dt,Elizalde:2012kz,Bonvin:2013tba,Caprini:2014mja,Choudhury:2015jaa,Sharma:2017eps,Caprini:2017vnn,Sharma:2018kgs,Kamarpour:2018ckk, Savchenko:2018pdr,Sobol:2018djj,Subramanian:2019jyd,Patel:2019isj,Sobol:2019xls,Fujita:2019pmi,Shakeri:2019mnt,Kobayashi:2019uqs,Shtanov:2019civ,Sharma:2019jtb}, or in cosmological models with a bounce
\cite{Battefeld:2004cd,Salim:2006nw,Membiela2013,Sriramkumar:2015yza,Chowdhury2016,Qian2016,Koley:2016jdw,Chen:2017cjx,Leite:2018bbo,
Chowdhury:2018blx,Barrie:2020kpt}.\footnote{Cosmological magnetic fields may also be produced during phase transitions, see for instance \cite{Grasso:2000wj}, or through the generation of vortical currents \cite{Carrilho:2019qlb}.}
However, since minimally-coupled electromagnetism is conformally invariant, the expansion cannot affect its vacuum state. Hence such invariance must be broken 
in order to generate seed magnetic fields.

Conformal invariance can be broken in several ways: through the addition of a mass term \cite{Enqvist:2004yy}, by  coupling the electromagnetic (EM) field to a massless charged scalar field
\cite{Emami2009} or the axion \cite{Adshead:2016iae}, and by a non-minimal coupling with gravity. The last option has been
widely studied in the case of inflationary models (see  \cite{Turner1988,
Bamba2006,
Campanelli2008,
Kunze2009, Kunze2012, Savchenko2018}, among others). However, inflationary magnetogenesis is not free of problems. Among these, we can mention an exponential 
sensitivity of the amplitude of the generated magnetic field with the parameters of the inflationary model \cite{Subramanian:2015lua}, the strong coupling problem 
\cite{Demozzi2009}, and 
the limits in the 
magnetic field strength coming from the gravitational backreaction of
the electric fields that are produced simultaneously with the magnetic fields
\cite{Green2015}.
Hence, instead of an inflationary model, a nonsingular cosmological model (see \cite{Novello2008} for a review)
in conjunction with 
a coupling of the type $RF_{\mu\nu}F^{\mu\nu}$ will be used here to 
study the production of seed magnetic fields. 
Nonsingular models are likely to ease both the problem of the exponential sensitivity of the result and the strong coupling problem, since they expand slower than inflationary models. 
Moreover, we shall see below that  
backreaction is not an issue for the model 
chosen here.

It is worth remarking that magnetogenesis in nonsingular cosmological models has
been studied before, always in the presence of a scalar field. The models
already studied may be divided into two classes, depending on whether the
coupling of the EM field with the scalar field is fixed on theoretical grounds
(see for instance \cite{Salim2004, Salim:2006nw}), or chosen in a convenient way in
terms of the expansion factor (see \cite{Membiela2013,Qian2016,Chowdhury2016} ).
The coupling between the Ricci scalar and the EM field to be adopted in this
work, which is theoretically motivated by the 
vacuum polarization described 
quantum electrodynamics (QED)  in a curved
background \cite{Drummond:1979pp}, introduces a mass scale to be fixed by
observations.

We shall start in Sec.~\ref{backg} with a brief summary of the background model
that will be used in what follows. In Sec.~\ref{emsector}, the equations
governing the behavior of the 
perturbations of the 
electromagnetic field in a curved
background will be reviewed. We show in Sec.~\ref{Aresults} the analytic
solutions for the gauge field and its momentum. These results are used to
understand  the numerical solutions in Sec.~\ref{results}. 
The comparison of the results with observations
is presented in 
Sec.~\ref{discussion}. 
The fact that backreaction does not affect the background dynamics is shown in Sec.~\ref{backreaction}.  Also, we show in appendix \ref{sec:adiab} how to obtain  appropriate
initial conditions for the electromagnetic field from an adiabatic vacuum.

\section{The background}
\label{backg}

Cosmological models displaying a bounce solve the singularity problem by construction. They join a contracting phase, in which the Universe was initially very large and almost flat, to a subsequent expanding phase. In such models, the curvature scale tends to infinity in the asymptotic past. As a consequence,  vacuum initial conditions for cosmological perturbations can be imposed in the dust-dominated contracting phase
\footnote{In models in which radiation is important initially, thermal fluctuations may dominate over quantum fluctuations, see 
\cite{Cai2008,Cai2009, Bhattacharya2013}.
}, leading to a scale invariant spectrum \cite{Peter:2008qz}. The bounce can be either generated classically (see \textit{e.g.} \cite{Wands:2008tv,Ijjas:2016tpn,Cubero:2019lxw,Galkina:2019pir}) or by quantum effects (see \textit{e.g.} \cite{Peter:2008qz,Almeida:2018xvj,Bacalhau:2017hja,Frion:2018oij}).

The cosmological model that will be used here as background was obtained in \cite{Peter:2006hx} by solving the Wheeler-deWitt equation in the presence of a single perfect fluid. The solution was obtained in the minisuperspace approximation, and in the framework of the theory of de Broglie and Bohm
(dBB) \cite{Bohm:1951xw,Bohm:1951xx}. The reason behind this choice is that the dBB interpretation is very well suited for  cosmology, since it needs no external classical apparatus, as opposed to the Copenhagen interpretation. 

The expression of the scale factor in the case of a flat spatial section obtained in \cite{Peter:2006hx} is given by
\begin{equation}
\label{adet}
a(T)=a_b\left(1+\frac{T^2}{T_b^2}
\right)^{\frac{1}{3(1-w)}} \;,
\end{equation}
where $w$ is the equation of state of the fluid.\footnote{Note that a scale
	factor of this form was introduced by hand 
	in \cite{Sriramkumar:2015yza} to generate scale
	invariant magnetic fields, while it emerges naturally from quantum effects
	here.} All quantities appearing hereafter with the subscript $b$ are evaluated
at the bounce (with the exception of $T_b$, which fixes the
	bounce timescale), while quantities with the subscript $0$ are evaluated today.
The spacetime geometry associated with \eqref{adet} is nonsingular, and the
scale factor tends to the classical evolution for $\vert
	T\vert \gg T_b$. The relation between $T$ and the cosmic time $t$ is given by
\begin{align}
    dt=a^{3w}dT \;.
\end{align}

From now on, we shall set $w=0$, leading to a scale invariant spectrum for the curvature perturbations, and allowing us to set $t=T$. It will also be useful to express the scale factor as $a(t)\equiv a_0 Y(t)$, with
\begin{align}
    Y(t)= \frac{1}{x_b} \left( 1+\frac{t^2}{t_b^2}
\right)^{1/3} \;,
    \label{yt}
\end{align}
where we have defined $x\equiv a_0/a$ and $t_b \equiv 2\ell_b$, 
with $\ell_b$ the curvature scale at the bounce ($\ell_b \equiv 1/\sqrt{\vert R(0)\vert}$ where $R$ is the four-dimensional Ricci scalar) satisfying $10^3\:t_{\rm Planck}<t_b<10^{40}\:t_{\rm Planck}$.\footnote{The lower bound is set 
by imposing the validity of the Wheeler-DeWitt equation, \textit{i.e.},
by restricting 
the curvature
to values such that possible discreteness of the spacetime geometry is negligible, while quantum effects are still relevant \cite{Peter:2006hx}. Since $t_\text{Planck}\simeq 10^{-44}s$ and recalling that BBN happened around $10^{4}s$, the upper bound simply reflects the latest time at which the bounce can occur.}

For the subsequent calculations, it is convenient to define parameters that are directly related to observations. Let us first write down the Friedmann equation
\begin{align}
    H^2 = \frac{8\pi G}{3} \frac{\rho_m}{a^3} \;,
    \label{friedrho}
\end{align}
with $\rho_m$ the dark matter density energy. The ratio between Eq.~\eqref{friedrho} at some time $t$ and the same equation evaluated today leads to
\begin{align}
    H^2 = H^{2}_{0} \Omega_m x^3 \;,
    \label{fried}
\end{align}
with $\Omega_m$ the dimensionless dark matter density today. Note that at $x=1$ we have $H^2 = H_0^2\Omega_m$, this means that in the contraction phase, at the same scale as today $a = a_0$, the Hubble factor is $-H_0\sqrt{\Omega_m}$ due to the lack of other matter components. Then, from the expansion of $a(t)$ for large values of $t$, it follows that 
\begin{align}
    H^2\approx  \frac{4}{9t_b^2}\left( \frac{x}{x_b}\right)^3 \;.
    \label{larget}
\end{align}
Now, using $H_0=70 \; \textup{km\;s}^{-1}\textup{\;Mpc}^{-1}$ and the lower bound on $t_b$, it is straightforward to derive an upper limit on $x_b$ by equating Eqs.~\eqref{fried} and \eqref{larget},
\begin{align}
    \Omega_m = \frac 4 9 \frac{1}{t_b^2 x_b^3 H_0^2} \; \implies \x_b < \frac{10^{38}}{\Omega_m^{1/3}} \; .
    \label{limit xb}
\end{align}

For later convenience, we define
 $R_{H_0} \equiv H_0^{-1}$,
$t_s \equiv {t}/{R_{H_0}}$, and 
$
\alpha \equiv {R_{H_0}}/{t_b}$,
and rewrite $Y(t)$
as
\begin{align}
    Y(t_s) = \frac{1}{x_b} \left(1+\alpha^2t_s^2\right)^{1/3} \;,
\end{align}
with
\begin{align}
\alpha = \frac 3 2 \sqrt{\Omega_mx_b^3} \;.
\label{alpha}
\end{align}
We will see in the next section how to relate the previous quantities to the electromagnetic power spectrum, and what constraints can be derived on the parameters of the model.

\section{The electromagnetic sector}
\label{emsector}

To describe electromagnetism we shall adopt the Lagrangian
\begin{equation}
\label{lagr}    
{\cal L}= -f
F_{\mu\nu}F^{\mu\nu} \;,
\end{equation}
where
\begin{align}
    f \equiv \frac{1}{4}+\frac{R}{m_{\star}^2} \;,
    \label{coupling}
\end{align}
and $m_{\star}$ is a mass scale to be determined by observations. As mentioned in the Introduction, the nonminimal coupling in this Lagrangian breaks conformal invariance, and paves the way to the production of primordial electromagnetic fields.

The equations of motion for the electromagnetic field that follow from Eq.~\eqref{lagr} are
\begin{equation}
\label{eqmov}
\partial_\mu(\sqrt{-g}\:f\:F^{\mu\nu})=0 \;,
\end{equation}
where the field $F_{\mu\nu}$ is expressed in terms of the gauge potential $A_\mu$ as $F_{\mu\nu}=\partial_\mu A_\nu-\partial_\nu A_\mu$. To quantize the electromagnetic field, we expand the operator associated to the spatial part of the vector potential as
\begin{equation}
\label{decomp}
\hat{A}_i(t,\mathbf{x})=\sum_{\sigma=1,2}\int \frac{d^3k}{(2\pi)^{3/2}}\left[\epsilon_{i,\sigma}(\mathbf{k})\hat{a}_{\mathbf{k},\sigma} A_{k,\sigma} (t)e^{i\mathbf{k}\cdot\mathbf{x}}+H.C.\right] \;,  
\end{equation}
where $\epsilon_{i,\sigma}(\mathbf{k})$ are two orthonormal and transverse
vectors which are constant across spatial sheets (they have zero Lie derivative with respect to the spatial foliation vector field) and $H.C.$ stands for the Hermitian conjugate. The operators $\hat{a}_{\mathbf{k},\sigma}$
and $\hat{a}^\dagger_{\mathbf{k},\sigma}$ are respectively the annihilation and
creation operators. They satisfy %
$[\hat{a}_{\mathbf{k},\sigma}, \hat{a}^\dagger_{\mathbf{k'},\sigma'}] =
\delta_{\sigma\sigma'}\delta(\mathbf{k} - \mathbf{k'})$, %
$[\hat{a}_{\mathbf{k},\sigma}, \hat{a}_{\mathbf{k'},\sigma'}]=0$, %
and %
$[\hat{a}^\dagger_{\mathbf{k},\sigma},
\hat{a}^\dagger_{\mathbf{k'},\sigma'}]=0$. Note that in the equations above we adopted the Coulomb gauge with respect to the cosmic time foliation ($A_0=0$ and $\partial_i
A^i=0$). The time-dependent coefficients
	$A_{k,\sigma}(t)$ and their associated momenta $\Pi_{k,\sigma} \equiv
	4afA^\prime_{k,\sigma}(t)$ must satisfy
\begin{equation}\label{vacuumnorma}
A_{k,\sigma}(t)\Pi^*_{k,\sigma}(t)
-A^*_{k,\sigma}(t)\Pi_{k,\sigma}(t) = i,
\end{equation}	
	 for each $k$ and $\sigma$. It should be emphasized that the quantization of
the gauge-fixed electromagnetic field in the absence of charges is equivalent
to that of two free real scalar fields. Consequently, the choice of vacuum for
each polarization $\sigma$ corresponds to the choice of vacuum of each scalar
degree of freedom. However, using the fact that we are dealing with an
isotropic background, there is no reason to make different choices of vacuum
for different polarizations. For this reason, we choose a single time-dependent
coefficient to describe both polarizations, i.e., $A_{k,1} = A_{k,2} \equiv{A_{k}} $. 
Therefore, the same vacuum is chosen for both polarizations. Now, inserting this decomposition in
Eq.~\eqref{eqmov}, we get the equation governing the evolution of the modes
$A_k(t)$
\begin{align}
\label{pot}
\ddot{A}_k+\left(\frac{\dot a}{a}+\frac{\dot f}{f}
\right)\dot{A}_k +\frac{k^2}{a^2}A_k=0 \;.
\end{align}
Defining
\begin{align}
    k_s \equiv kR_H,\;\;
\;\;A_{sk}(t_s) \equiv \frac{A_k(t_s)}{\sqrt{x_bR_{H_0}}} \;,
\end{align}
where $R_H=R_{H_0}/a_0$ is the comoving Hubble radius today, 
the differential Eq.~\eqref{pot} can be written as
\begin{align}
A_{sk}^{\prime\prime}+\left(\frac{ Y^\prime}{Y}+\frac{f^\prime}{f}
\right){A}_{sk}^\prime +\frac{k_s^2}{Y^2}A_{sk}=0 \;,
\label{diffeq}
\end{align}
where a prime denotes the derivative with respect to $t_s$. The coupling \eqref{coupling} then takes the form 
\begin{align}
    f=\frac 1 4 \left[ 1+C^2\frac{\alpha^2 t_s^2+3}{(\alpha^2t_s^2+1)^2}
\right] \;; \;\;\;\;\; \textup{with}\;\; C^2 \equiv \frac{4}{3} \frac{\ell_*^2}{t_b^2} \;,
\;\;\;\;\;\ell_* \equiv \frac{1}{m_*} \;.
\label{cc}
\end{align}
An upper limit on $C$ can be straightforwardly derived from Eq.~\eqref{cc}. Since any contribution to the usual Maxwell's equations at BBN must be negligible, we impose the second term in Eq.~\eqref{cc} to be smaller than $10^{-2}$ at BBN. Together with the fact that $\alpha^2 t_s^2 \gg 1$ at this time, we get
\begin{align}
    C < 10^{-19} x_b^{3/2} \;.
    \label{limit c}
\end{align}
The energy densities of the electric and magnetic fields are respectively given by
\begin{align}
    \rho_E &=\frac{f}{8\pi}g^{ij}A_i^\prime A_j^\prime\;, \\
    \rho_B &=\frac{f}{16\pi}g^{ij}g^{lm}(\partial_j A_m-\partial_m A_j) (\partial_i A_l-\partial_l A_i) \;,
\end{align}
where $g^{ij}=\delta^{ij}/a^2$ are the spatial components of the inverse metric. To find the spectral energy densities, we first insert expansion \eqref{decomp} into $\rho_E$ and $\rho_B$. The resulting operators $\hat{\rho}_E$ and $\hat{\rho}_B$ upon quantization are
\begin{align}
    \hat{\rho}_B &= \frac{f}{2\pi^2 R_{H_0}^4 Y^{4}} \int \dd{\ln{k}} \;  \vert A_{sk}\vert^2 k^5 \;, \label{magnetic spectral density} \\
    \hat{\rho}_E &= \frac{f}{2\pi^2 R_{H_0}^4 Y^{2}} \int \dd{\ln{k}} \; \vert A^{\prime}_{sk}\vert^2 k^3 \;. \label{electric spectral density}
\end{align}
We now evaluate the expectation value of the two densities in vacuum, defined by 
$\hat{a}_{\mathbf{k},\sigma} \ket{0}=0$, and define the spectra as
\begin{align}
    {{\cal P}_{i}} \equiv \frac{\textup{d}\bra{0} \hat{\rho}_i \ket{0}}{\dd{\ln{k}}} \;,  \quad i=E,B \;.
\end{align}
This yields the magnetic and electric spectra, respectively
\begin{align}
    {\cal P_{B}} &\equiv B^2_\lambda = \frac{f}{2\pi^2 R_{H_0}^4}\frac{\vert A_{sk}\vert^2}{Y^4}k^5 \;, 
    \label{magpow} \\ 
    {\cal P_{E}} &\equiv E_{\lambda}^2=\frac{f}{2\pi^2 R_{H_0}^4}\frac{\vert A_{sk}^\prime\vert^2}{Y^2}k^3 = \frac{1}{2\pi^2 R_{H_0}^4}\frac{\vert \Pi_{sk}\vert^2}{fY^4}k^3 \;.
        \label{elecpow}
\end{align}
In the last line, we also expressed ${\cal P_{E}}$ in terms of the momentum canonically conjugate to the gauge field $\Pi_{sk} = Yf A_{sk}^\prime$ (see Appendix ~\ref{sec:adiab}), which is nothing but the electric field mode itself.

Finally, we can express the magnetic and electric fields, $B_{\lambda}$ and $E_{\lambda}$, using $H_0^2=1.15\times 10^{-64}$ G
\begin{align}
    \label{est}
B_\lambda &=\sqrt{\frac{f}{2\pi^2}} 
\frac{\vert A_{sk}\vert }{Y^2}k^{5/2}\:1.15\times 10^{-64} {\rm G} \;, \\
E_\lambda &=\sqrt{\frac{1}{2 \pi^2 f}} 
\frac{\vert  \Pi_{sk}\vert }{Y^2}k^{3/2}\:1.15\times 10^{-64} {\rm G}.
\end{align}

\section{Analytical results}
\label{Aresults}

In this section, we obtain analytically the time behavior and spectra of $A_{k}$ satisfying Eq.~\eqref{diffeq} (from now on the index $s$ on the time variable and wavenumber will be omitted), and its canonical momentum $\Pi_k$, in the different stages of the cosmic evolution. In the sequel, this analysis will be compared with the numerical results.

As shown in Appendix \ref{sec:adiab}, the adiabatic vacuum is a consistent choice for the EM field initial conditions. The modes in vacuum are 
\begin{equation}
\begin{split}
\vert A_k\vert  &= \sqrt{\frac{2}{k}} +\dots \;, \\
\vert \Pi_k\vert  &= \sqrt{\frac{k}{8}} +\dots. \;,
\end{split}
\end{equation}
and both the field and its canonical momentum are constant in this regime.
Now that the initial conditions for the EM field have been defined, we can move on to the analysis of the evolution of the electric and magnetic modes from the far past up to the present day. 

Three important characteristic times related to the evolution of the modes are worthy of note
. The first is the time limit of the adiabatic regime, $\vert t_c\vert $, defined in Eq.~\eqref{xi2}. The second one is the time where quantum effects leading to the bounce take place, \textit{i.e.} $\vert t_b\vert  = 1/\alpha$. Consequently,  the bounce phase takes place for $t$ such that  $-1/\alpha < t < 1/\alpha$. The third one is the characteristic time when the evolution of $f$  becomes important. Examining Eq.~\eqref{cc}, one gets the time $\vert t_f\vert  = C/\alpha$, up to $\vert t_b\vert $, which means that the evolution of $f$ is important when $-C/\alpha < t < -1/\alpha$, and  $1/\alpha < t < C/\alpha$. The domain of physically allowed parameters imposes that
\begin{equation}
\label{times}
\vert t_c\vert  \gg \vert t_f\vert  \gg \vert t_b\vert  \;.
\end{equation}

For $\vert t\vert <\vert t_c\vert $, the solution leaves the frequency-dominated
region. In this case, one can perform the usual expansion in $\nu^2$ derived
from the Hamilton Eqs.~\eqref{eq-hamiltonian} through iterative substitutions:
\begin{equation}
\label{iterations1}
\begin{split}
\Pi_k(t) &= -\int^t m(t_1)\nu^2(t_1) A_k(t_1)\dd t_1 + A_2(k) = mA^\prime_k(t)\Rightarrow\\
A_k(t) &= -\int^t \frac{\dd t_2}{m(t_2)}\int^{t_2}m(t_1)\nu^2(t_1) A_k(t_1) \dd t_1 + A_2(k)\int^t \frac{\dd t_1}{m(t_1)} + A_1(k)\Rightarrow\\
A_k(t) &= A_1(k)\left(1-\int^t \frac{\dd t_2}{m(t_2)}\int^{t_2}m(t_1)\nu^2(t_1) \dd t_1\right) + \\ 
& A_2(k)\left(\int^t \frac{\dd t_1}{m(t_1)} - \int^t \frac{\dd t_2}{m(t_2)}\int^{t_2}m(t_1)\nu^2(t_1)\dd t_1\int^{t_1} \frac{\dd t_3}{m(t_3)}\right) +\dots \; ,
\end{split}
\end{equation} 
where $A_1(k)$ and $A_2(k)$ are constants in time depending only on $k$, leading to the momentum expression
\begin{equation}
\label{iterations2}
\Pi_k(t) = -A_1(k)\int^{t}m(t_1)\nu^2(t_1) \dd t_1 + A_2(k)\left(1 - \int^t m(t_1)\nu^2(t_1)\dd t_1\int^{t_1} \frac{\dd t_2}{m(t_2)}\right)+\dots \;.
\end{equation}
We can now evaluate the time evolution and spectra in the different phases of the cosmic evolution. 

\subsection{The contracting phase and the bounce}

In the case of $A_k(t)$, all time-dependent terms are decaying in the contracting era up to the end of the bounce. As a consequence, $A_k(t)=A_1(k)$ is constant during all this phase. By continuity with the adiabatic phase, we conclude that
\begin{equation}
\label{A1}
A_1(k)\propto k^{-1/2} \;.
\end{equation}

The time-dependent terms of the momentum $\Pi_k(t)$ are also decaying, except for the one multiplying $A_1(k)$, which grows as $t^{-5/3}$ for $-C/\alpha < t < -1/\alpha$, since $f\propto 1/t^2$ in this region. Then, for $t< -C/\alpha$, $\Pi_k(t)=A_2(k)$ which, by continuity with the adiabatic phase, implies that 
\begin{equation}
\label{A2}
A_2(k)\propto k^{1/2}.
\end{equation}
In the period  $-C/\alpha < t < -1/\alpha$, the term multiplying $A_1(k)$ eventually surpasses the constant mode at a time $t_\pi$, and $\Pi_k(t)$ grows.

At the bounce itself $Y$ and $f$ are almost constant, therefore the modes will not evolve during this phase.

\subsection{The expanding phase}

In the expanding phase, the most important growing function related to $A_k(t)$ is the first one multiplying $A_2(k)$, which grows as fast as $t^{7/3}$ starting from some time $t_A$ in the interval $1/\alpha < t < C/\alpha$, and as $t^{1/3}$ for $C/\alpha < t < t_c$. 

In the case of $\Pi_k(t)$, as the integral multiplying $A_1(k)$ strongly decreases as $t^{-5/3}$ when $1/\alpha < t < C/\alpha$, the value of $\Pi_k(t)$ saturates in the value it gets by the end of the bounce, $t\approx 1/\alpha$. Also, $\Pi_k(t)$ acquires a $k^2$ dependence through the $\nu^2$ term. Combined with the $k$ dependence of $A_1(k)$, we obtain $\Pi_k(t) \propto k^{3/2}$.

After $t_c$, both $A_k(t)$ and $\Pi_k(t)$ begin to oscillate.

\subsection{Summary}

For the $A$-field, the spectra and time dependence in the different cosmic evolution phases is:
\begin{equation}
\label{historyA}
\begin{split}
-\infty < t < t_A \; &: \vert A_k(t)\vert  \propto k^{-1/2} \;, \\
t_A < t < C/\alpha \; &: \vert A_k(t)\vert  \propto k^{1/2}t^{7/3} \;, \\
C/\alpha < t < k^{-3} \; &: \vert A_k(t)\vert  \propto k^{1/2}t^{1/3}  \;, \\
t > k^{-3} \; &: \vert A_k(t)\vert  \propto k^{1/2} \; \times {\textrm{(oscillatory factors)}},
\end{split}
\end{equation}
where $t_A \in (1/\alpha,C/\alpha)$.

For the $\Pi$-field, we have:
\begin{equation}
\label{historyP}
\begin{split}
-\infty < t < t_\pi \; &: \vert \Pi_k(t)\vert \propto k^{1/2} \;,\\
t_\pi < t < -1/\alpha \; &: \vert \Pi_k(t)\vert  \propto k^{3/2} t^{-5/3} \;,\\-1/\alpha< t < k^{-3} \; &: \vert \Pi_k(t)\vert  \propto k^{3/2} \;,\\
t > k^{-3} \; &: \vert \Pi_k(t)\vert  \propto k^{3/2} \; \times {\textrm{(oscillatory factors)}},
\end{split}
\end{equation}
where $t_\pi \in (-C/\alpha,-1/\alpha)$.

Note that both the final spectrum of ${\cal P_{B}}$ and ${\cal P_{E}}$ (given in Eqs. \eqref{magpow} and \eqref{elecpow}) go as $k^6$.

After these analytical considerations, let us now turn to the numerical calculations, which confirm the behaviors presented in this section, and 
allow the calculation of  
the amplitudes.

\section{Numerical results}
\label{results}

\begin{figure}
    \centering
    \includegraphics[scale=0.41]{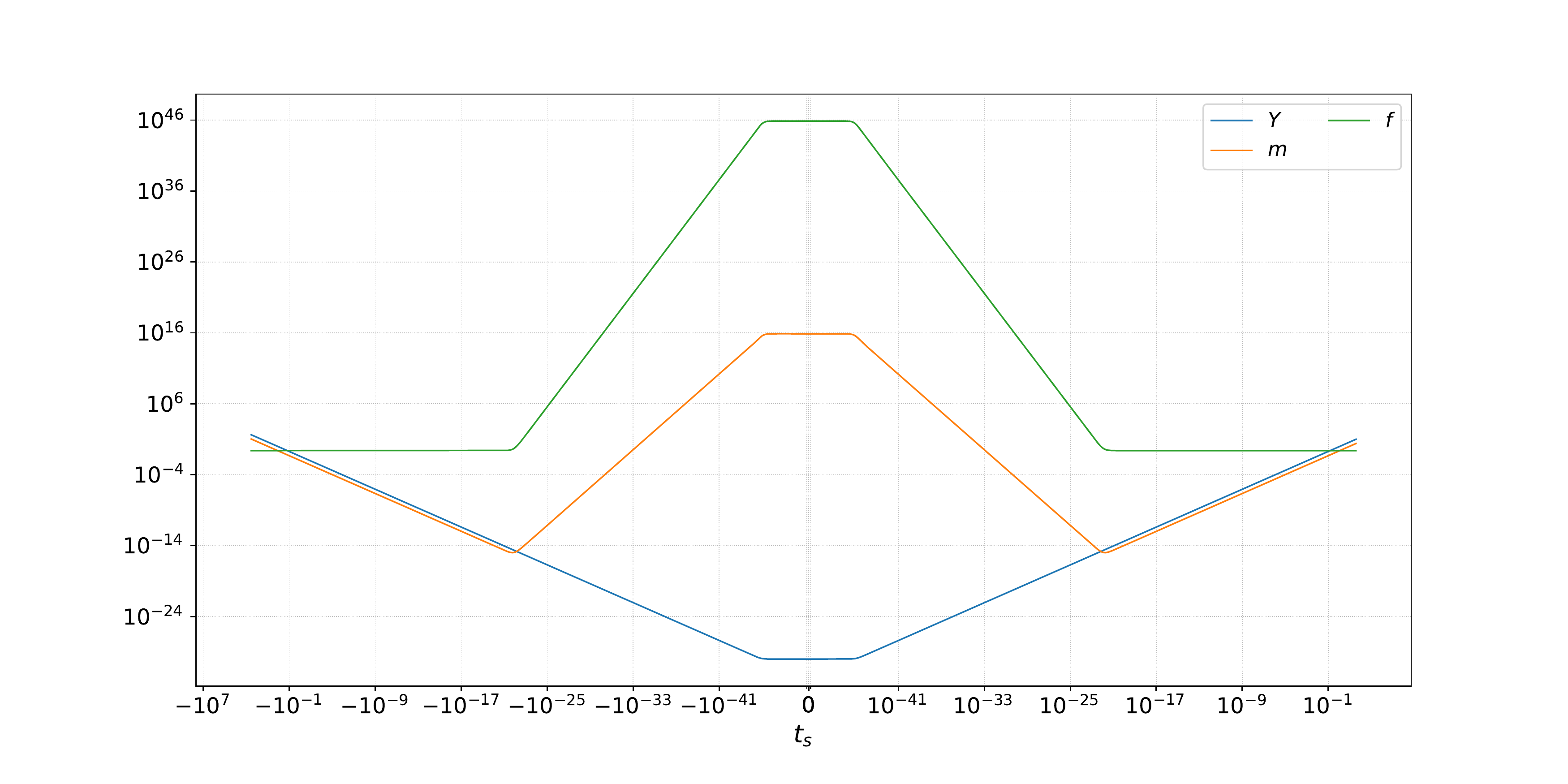}
    \caption{Evolution of the coupling $f$, the scale factor $Y$ normalised today, and the mass $m=af$ with time. We have used $C=10^{23}$ and $x_b=10^{30}$.}
    \label{ymf}
\end{figure}

\begin{figure}
    \includegraphics[scale=0.41]{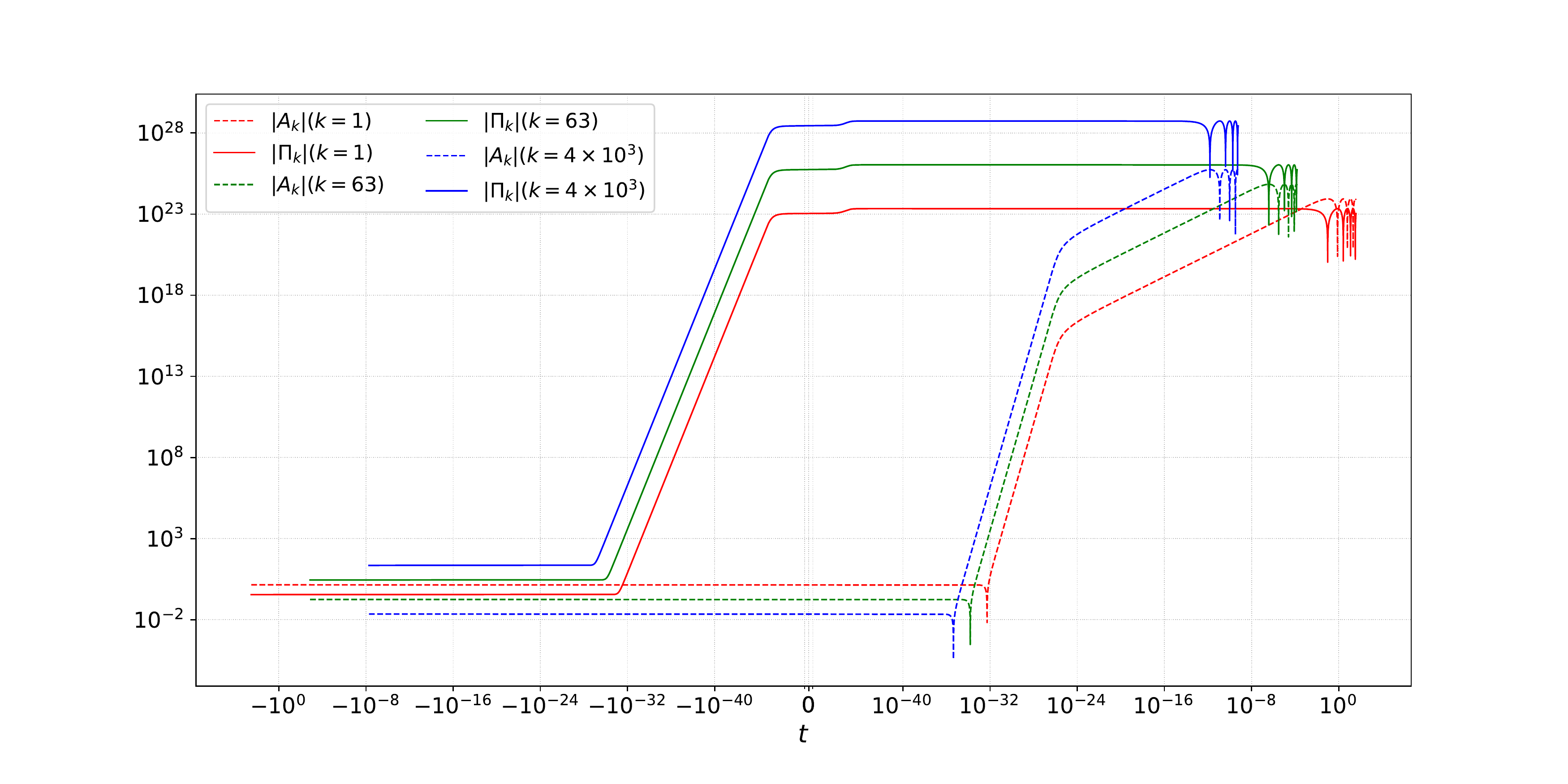}
    \includegraphics[scale=0.41]{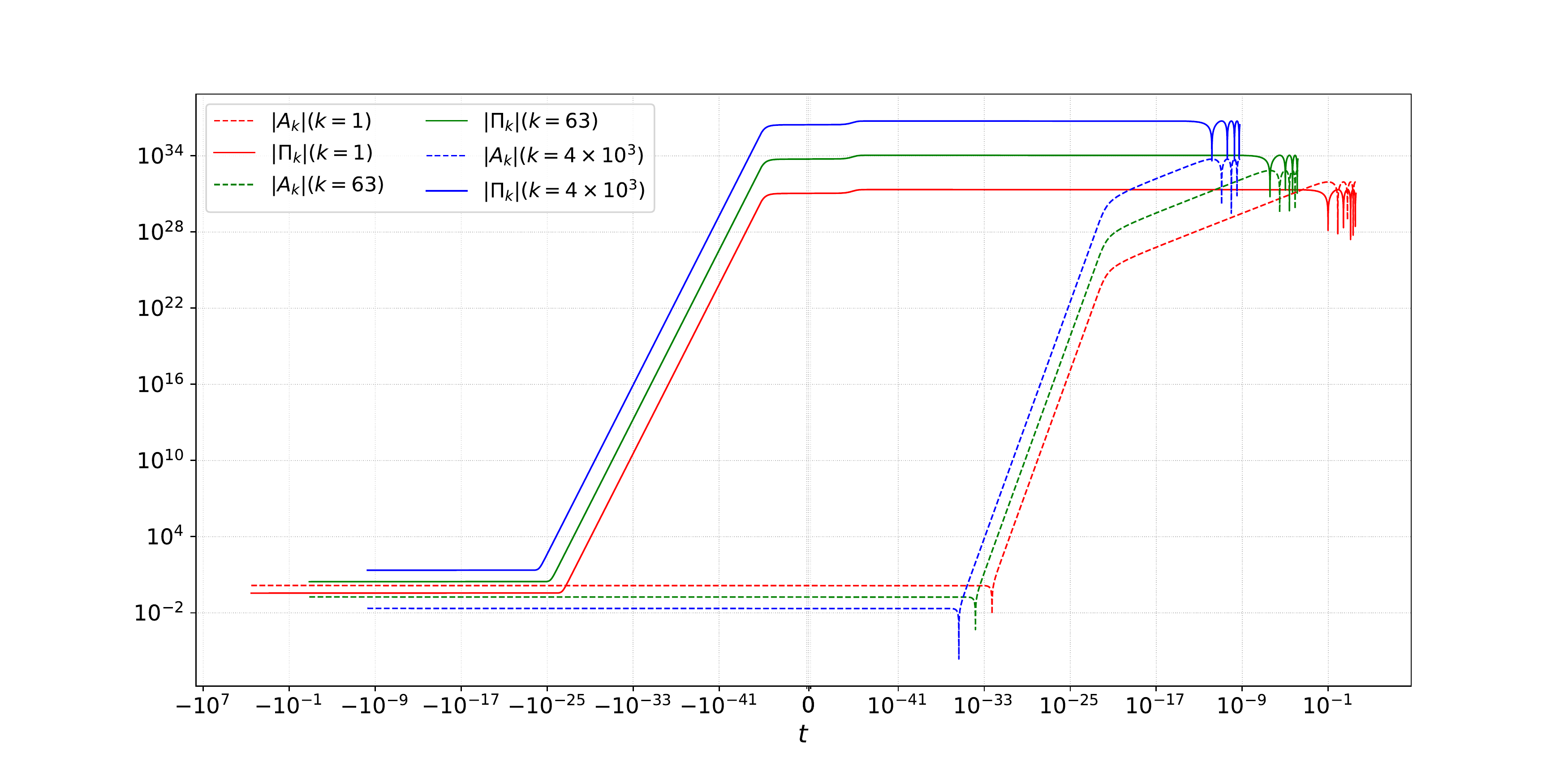}
    \caption{Evolution of the absolute values of the magnetic modes ($A_{k}$) and their momentum ($\Pi_k $) through the bounce in a dust background for $C=10^{19}$ and $x_b=10^{30}$ (top), and for $C=10^{23}$ and $x_b=10^{30}$ (bottom). The same colour for the gauge field and its momentum evolution is chosen for a given $k_s$. We see that larger values of $C$ lead to a higher final amplitude.}
    \label{mode evolution cc}
\end{figure}

\begin{figure}
    \centering
    \includegraphics[scale=0.41]{MagDustModeEvol_cc23_xb30.pdf}
    \includegraphics[scale=0.41]{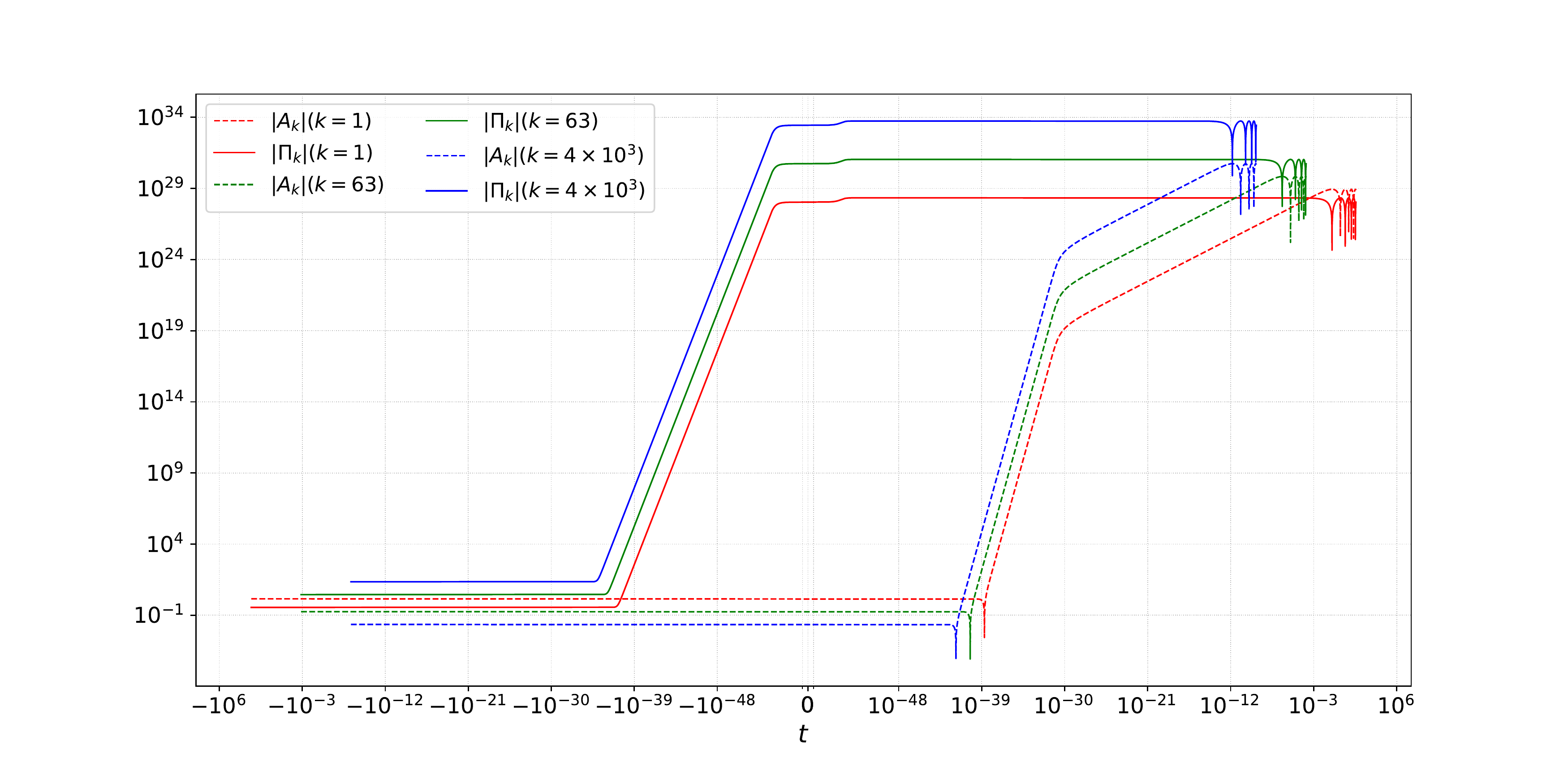}
    \caption{Same as Fig.~\ref{mode evolution cc} for $C=10^{23}$ and $x_b=10^{30}$ (top), and for $C=10^{23}$ and $x_b=10^{36}$ (bottom). We see that larger values of  $x_b$ lead to a quicker evolution of the modes.}
    \label{mode evolution xb}
\end{figure}

\begin{figure}
    \centering
    \includegraphics[scale=0.4]{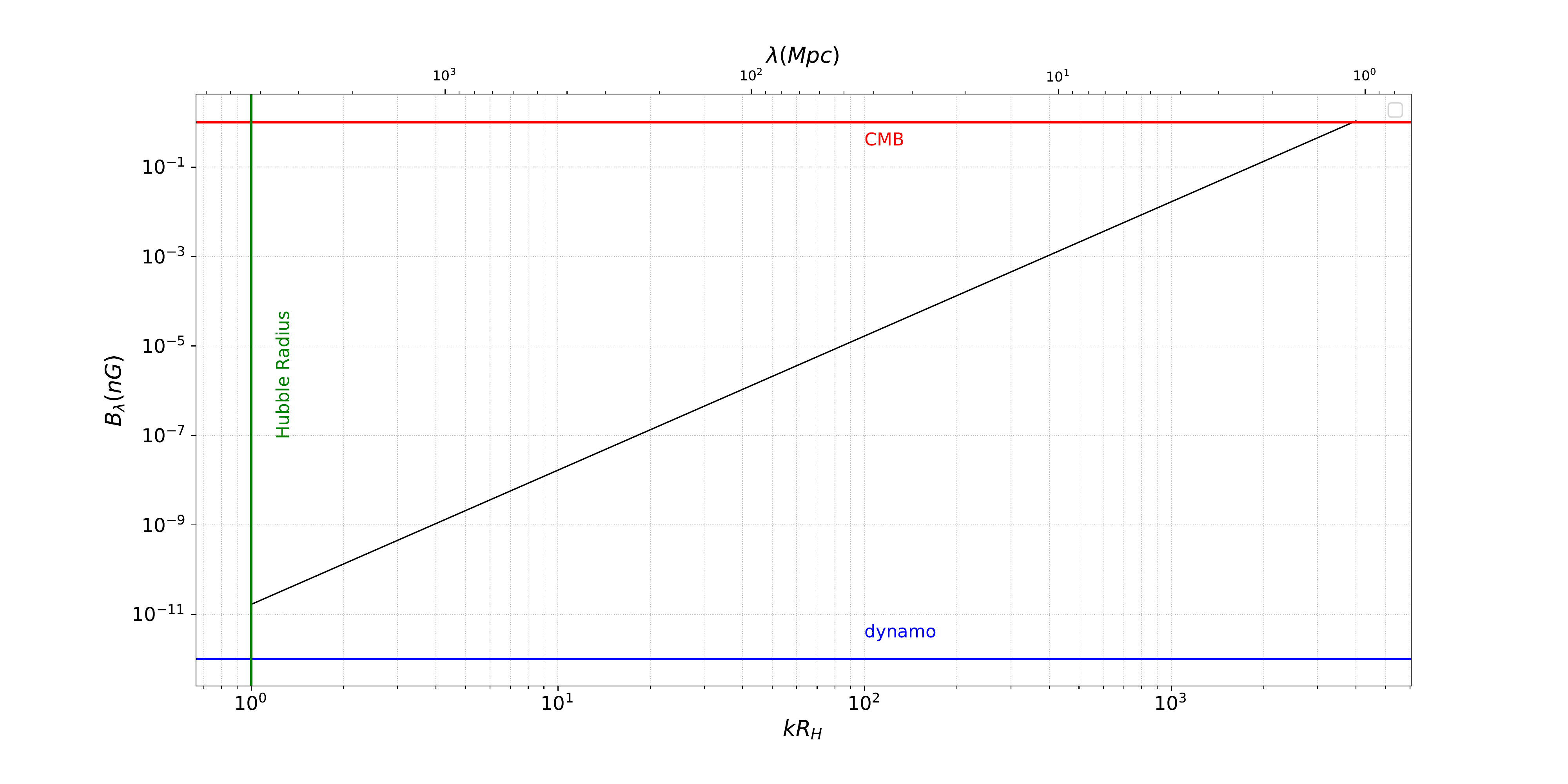}
    \includegraphics[scale=0.4]{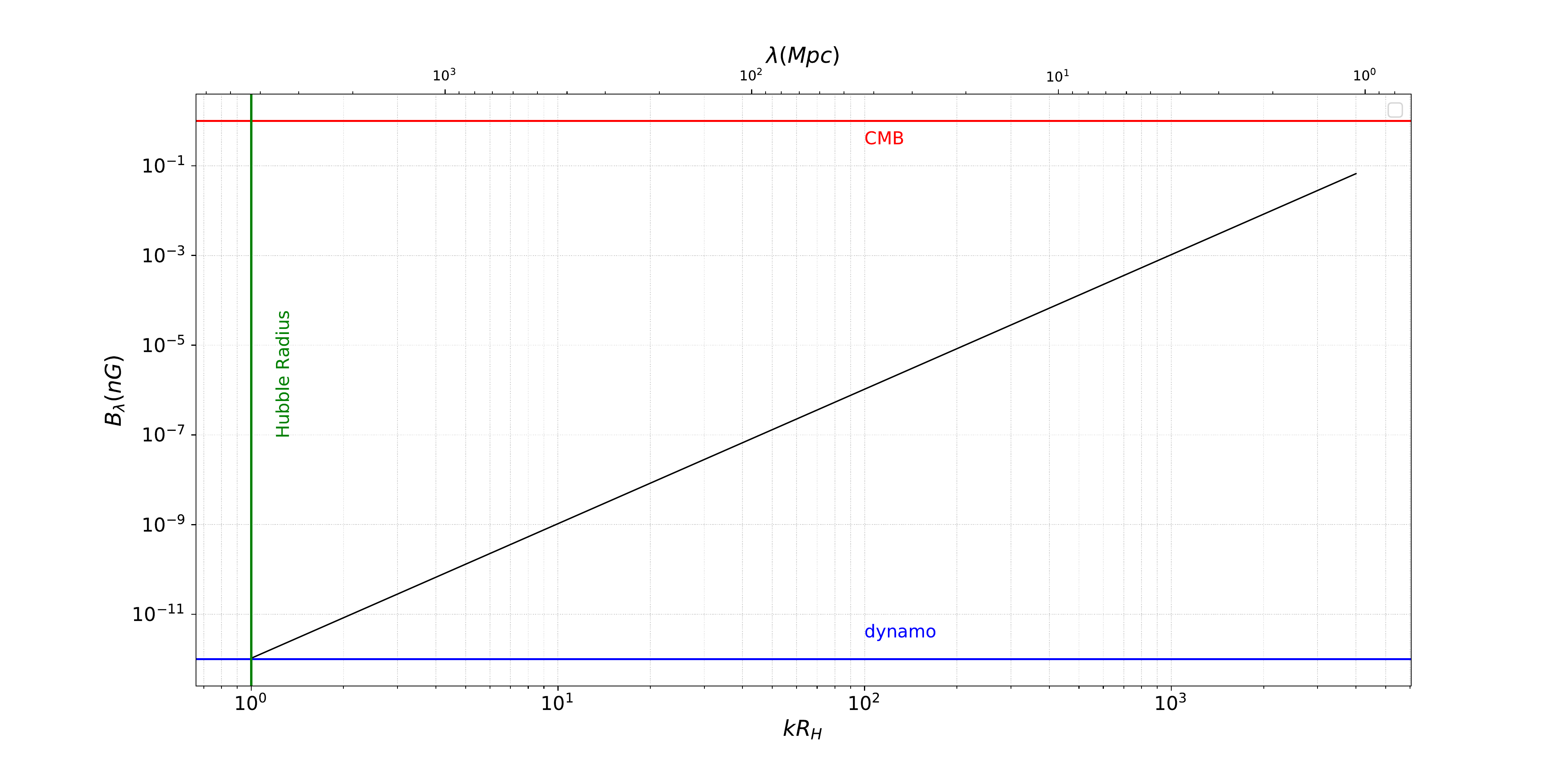}
    \caption{Magnetic field amplitude for $C=2.6 \times 10^{26}$ and $x_b=10^{38}$ (top), and $C=6.5 \times 10^{25}$ and $x_b=10^{38}$ (bottom). For these values, the seed field is sufficient to trigger the dynamo mecanism at large scales. The amplitude today is larger at all scales for larger values of $C$.}
    \label{magfield}
\end{figure}

\begin{figure}
    \centering
    \includegraphics[scale=0.4]{MagDustB_cc6-5e25_xb38.pdf}
    \includegraphics[scale=0.4]{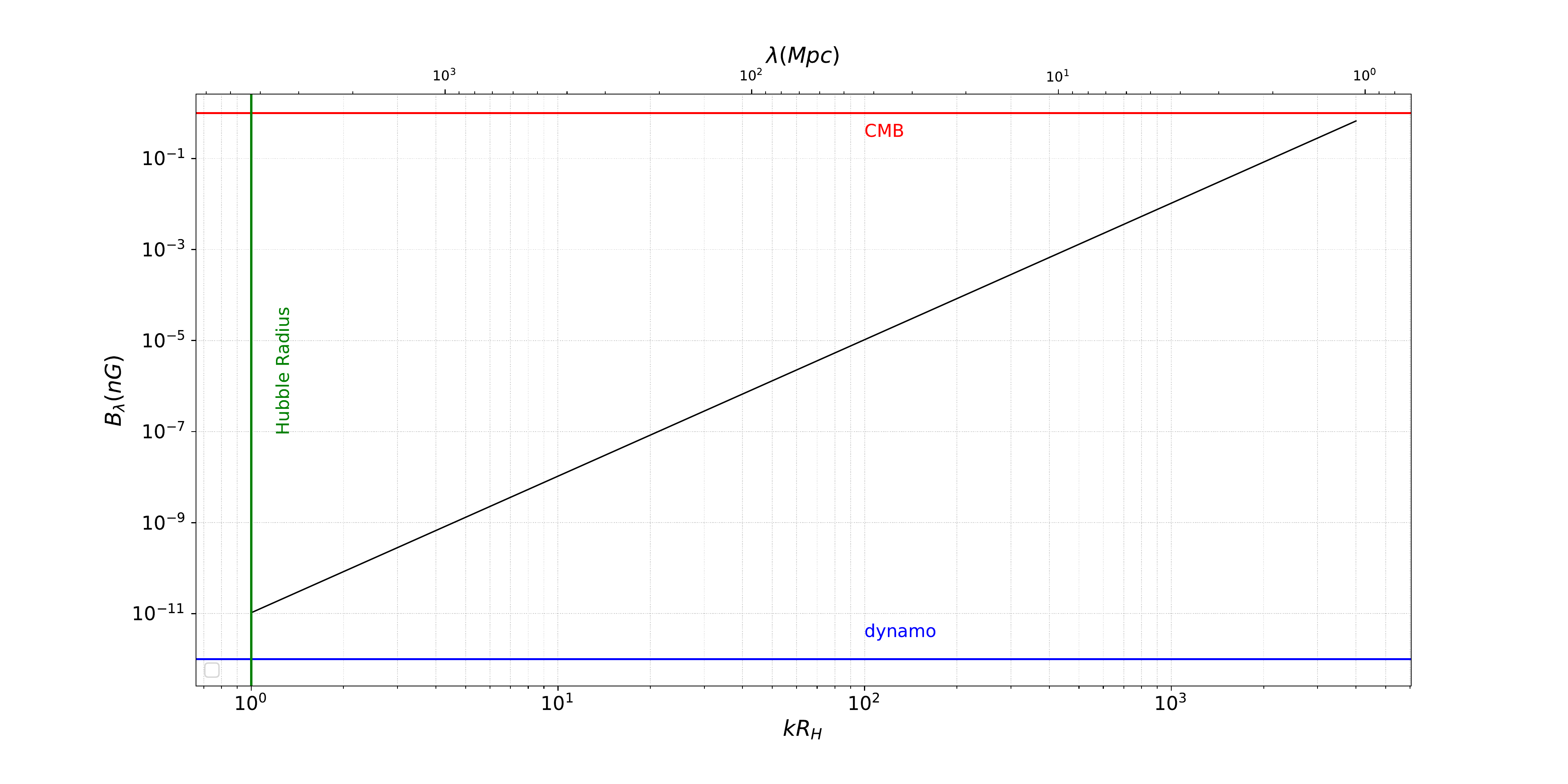}
    \caption{Magnetic field amplitude for $C=6.5 \times 10^{25}$ and $x_b=10^{38}$ (top) and for $C=6.5 \times 10^{25}$ and $x_b=10^{36}$ (bottom). The amplitude today is bigger at all scales when $x_b$ is smaller.}
    \label{magfield2}
\end{figure}

We start this section by showing in Fig.~\ref{ymf} the time behavior of the
coupling $f$ given in Eq.~\eqref{coupling}, the scale factor $Y=a/a_0$ from
Eq.~\eqref{yt}, and the mass $m=Yf$. From the definition of $|t_f|$ and $|t_b|$ in the previous section, and choosing $C=10^{23}$ and $x_b=10^{30}$, we obtain respectively $|t_f|\simeq 10^{-22}$ and $|t_b|\simeq 10^{-45}$. This is consistent with the behavior shown in the figure. %
The numerical evolution of the gauge field $A_{k}$ and its momentum $\Pi_k$ is shown next. In Fig.~\ref{mode evolution cc}, the influence of the parameter $C$ on the evolution of the modes is shown explicitly for $C=10^{19}$ and $C=10^{23}$ with $x_b=10^{30}$, while the influence of $x_b$ is shown in Fig.~\ref{mode evolution xb} for $x_b=10^{30}$ and $C=10^{36}$ with $C=10^{23}$.\footnote{We choose the values of $C$ and $x_b$  to be well inside the allowed parameter space at 1 Mpc, as can be seen in Fig.~\ref{parameter space 1Mpc}. We will use the same set of values throughout this section, except for Figs~\ref{magfield} and \ref{magfield2}.} Note that in these figures, as well as in the following ones, we performed the computation for $1<k<4000$, since $k=4000$ implies a physical wavelength of about 1 Mpc (remember that $k$ is in units of Hubble radius). One can verify in these figures all time and $k$ dependence described in
Sec.~\ref{Aresults}, summarized in Eqs.~\eqref{historyA} and \eqref{historyP}.

\begin{figure}
    \centering
    \includegraphics[scale=0.41]{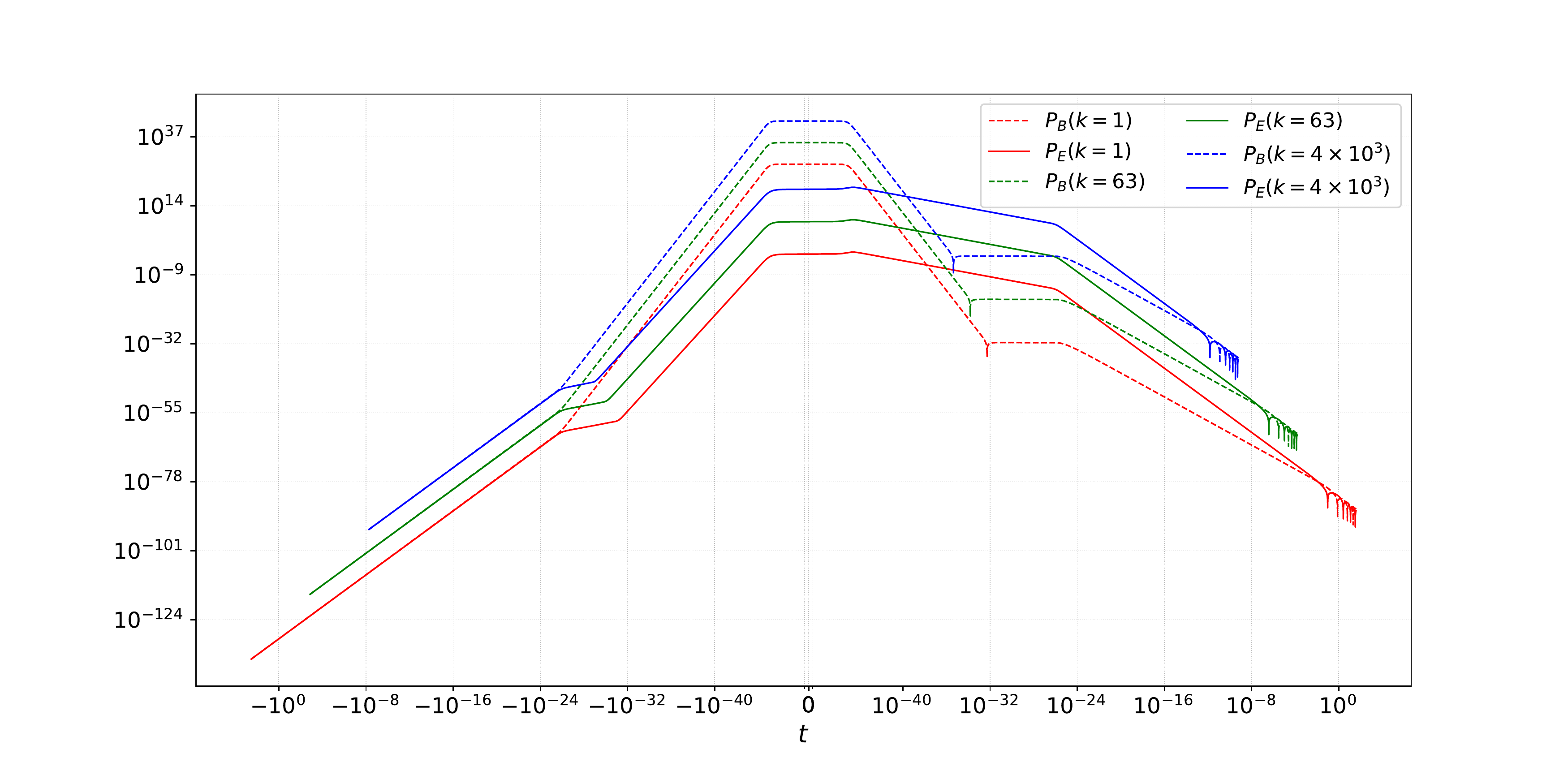}
    \includegraphics[scale=0.41]{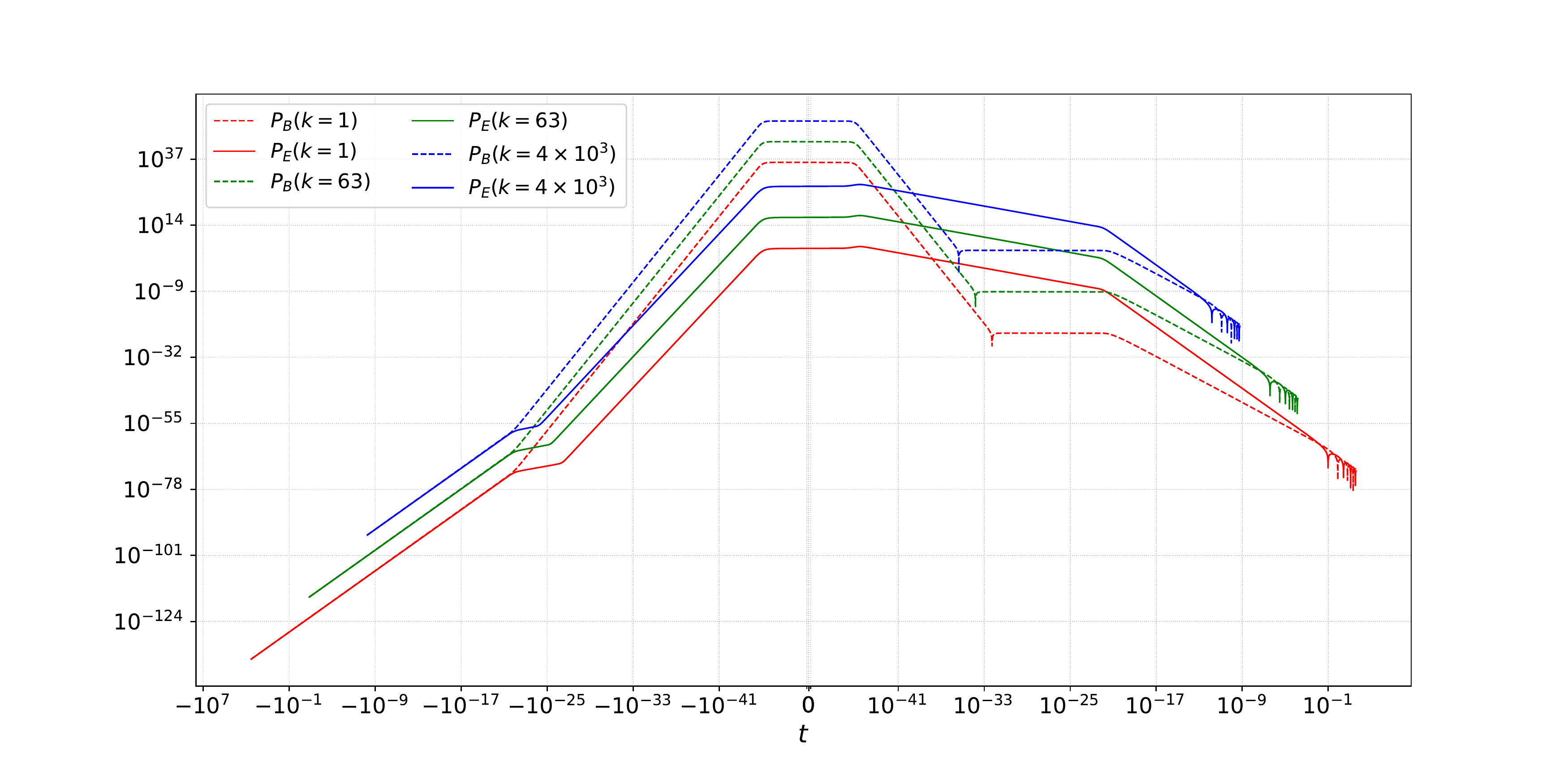}
    \caption{Evolution of the magnetic (dashed lines) and electric (continuous line) power spectra for $C=10^{19}$ and $x_b=10^{30}$ (top), and for $C=10^{23}$ and $x_b=10^{30}$ (bottom). We see that with larger $C$'s, the decrease of the electric contribution at late times happens later, and the total electromagnetic power spectrum is more important.}
    \label{spectra evolution cc}
\end{figure}

\begin{figure}
    \centering
    \includegraphics[scale=0.41]{MagDustPBPE_cc23_xb30.pdf}
    \includegraphics[scale=0.41]{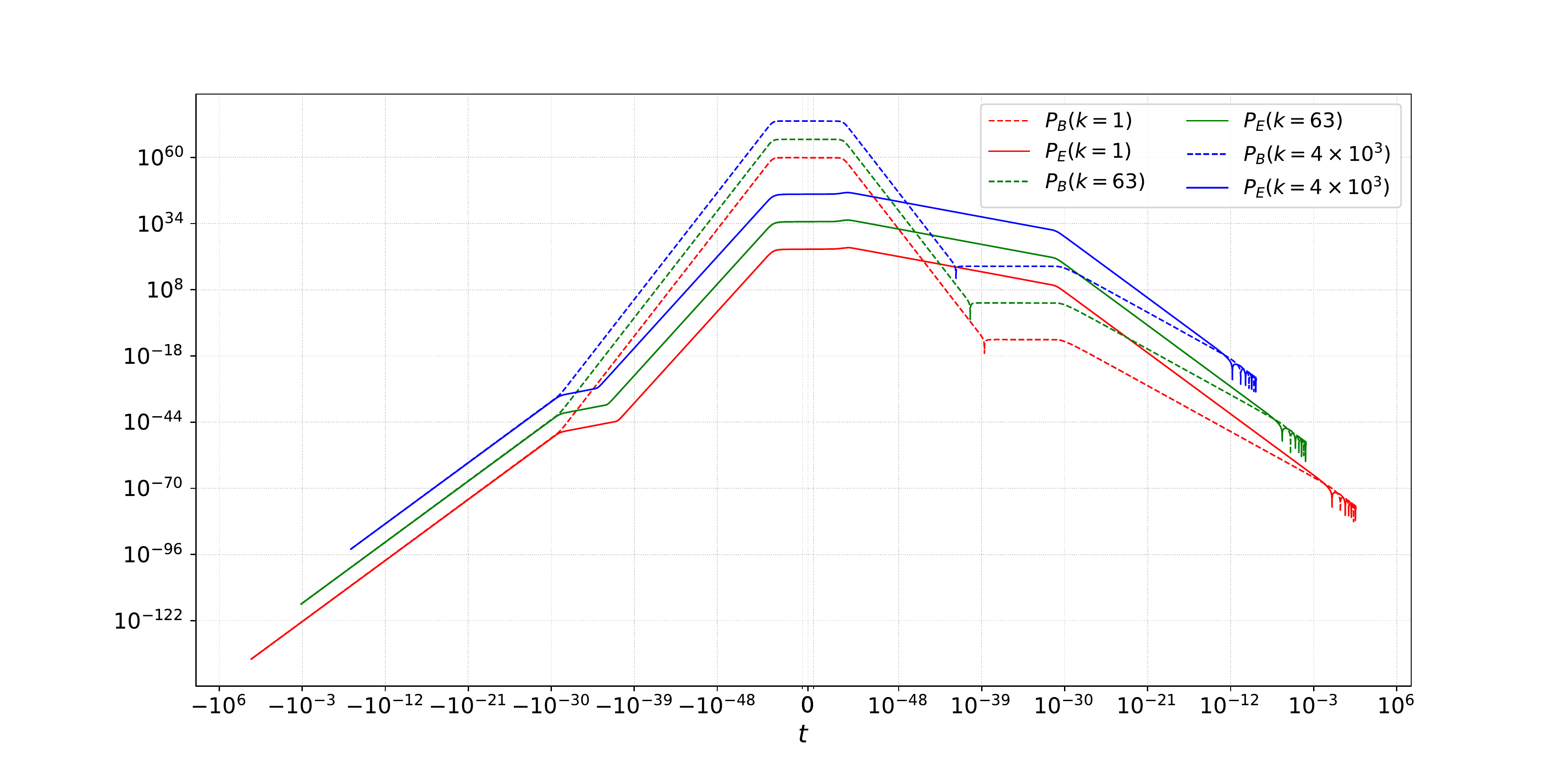}
    \caption{Same as Fig.~\ref{spectra evolution cc} for $C=10^{23}$ and $x_b=10^{30}$ (top), and for $C=10^{23}$ and $x_b=10^{36}$ (bottom). Higher values of $x_b$ imply an overall stronger total electromagnetic power spectrum, but with a stronger decrease rate at late times.}
    \label{spectra evolution xb}
\end{figure}

Now that the evolution of the modes has been described, we can use the shape of the spectra that follows from the results in Figs.~\ref{mode evolution cc} and \ref{mode evolution xb}, and Eqs~\eqref{magpow} and \eqref{elecpow}, the last one expressed in terms of the momentum, to fathom the time evolution of the magnetic and electric power spectra shown in Figs~\ref{spectra evolution cc} and \ref{spectra evolution xb}. At the beginning of the  evolution, modes are not excited. Only vacuum fluctuations are present, with the usual $k^4$ spectrum, increasing as $Y^{-4}$ due to contraction. When the coupling $f$ becomes relevant, the magnetic field power spectrum begins to increase faster, since $f$ is a growing function in the contracting phase, while the electric field power spectrum presents a slower increment, up to the time when $\Pi_k$ also begins to increase. After the bounce the situation is reversed, because $f$ is a decaying function of time in the expanding phase: the electric power spectrum decreases much slower than the magnetic one. Using Eq.~\eqref{historyP}, one can see that the decay is mild, going as $t^{-2/3}$, when $1/\alpha < t < C/\alpha$, opening a window in time where the electric spectrum has a significantly higher contribution than the magnetic one.

\begin{figure}
    \centering
    \includegraphics[scale=0.37]{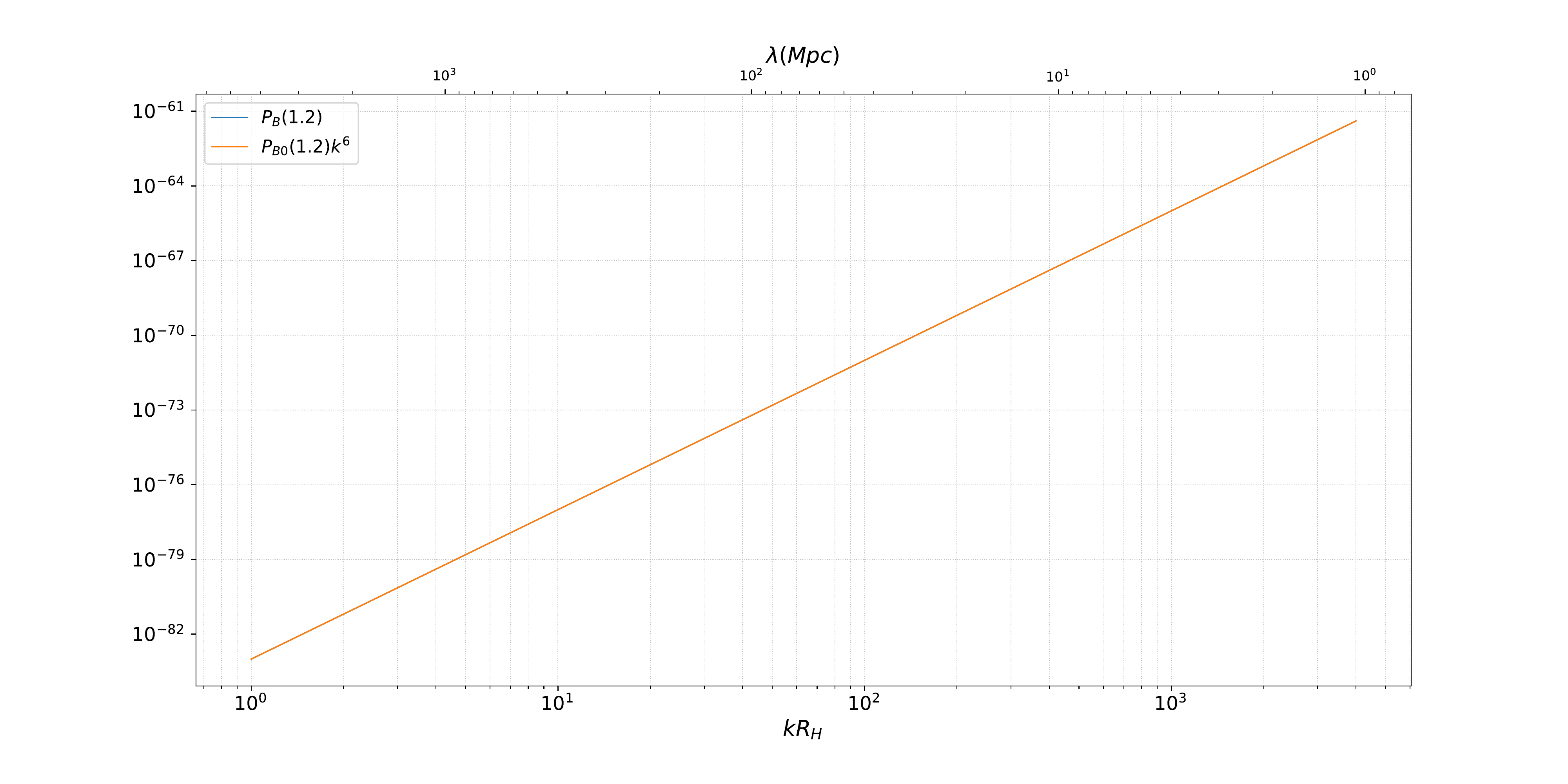}
    \includegraphics[scale=0.37]{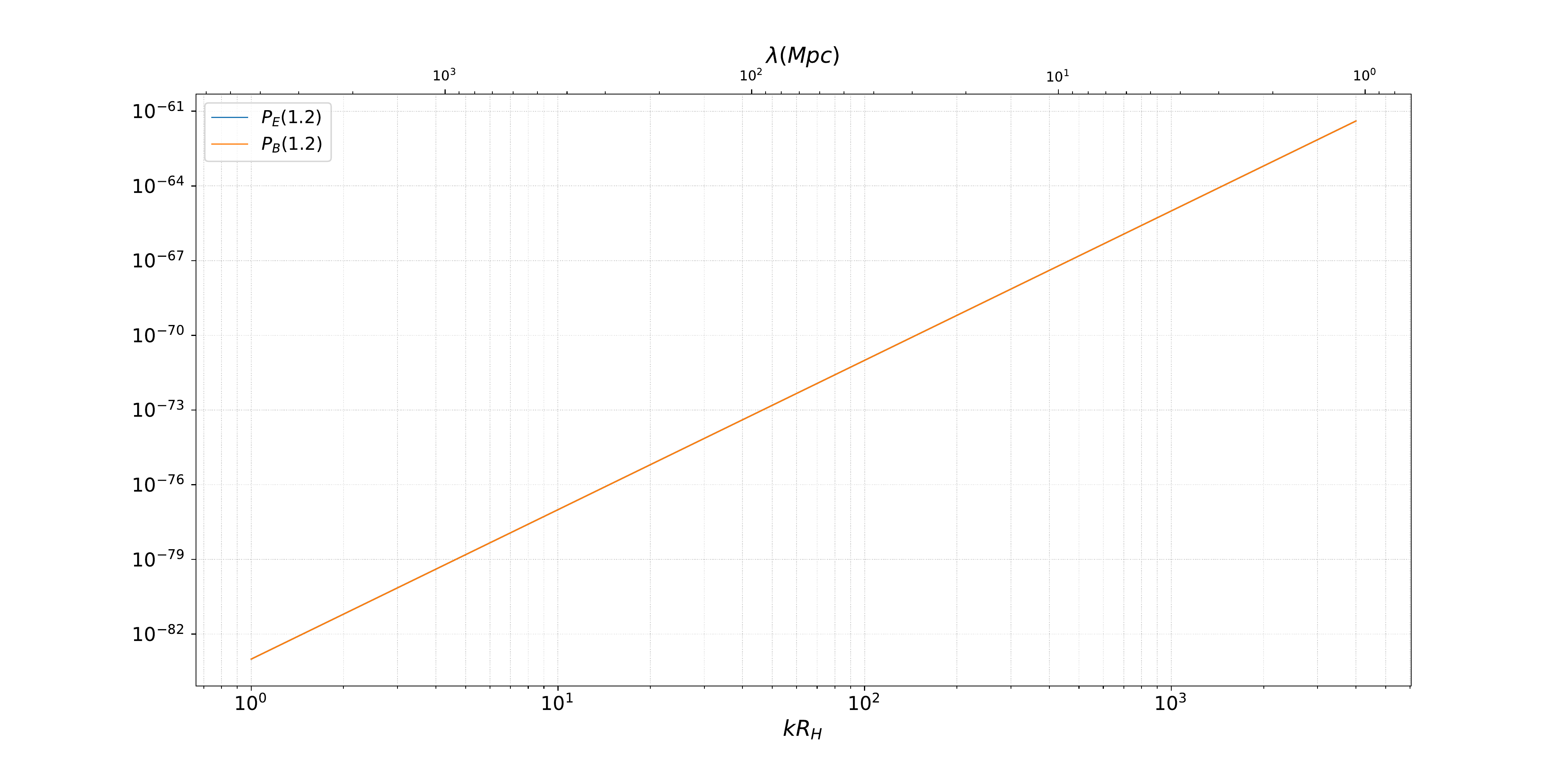}
    \caption{Behavior of the magnetic power spectrum today from \eqref{magpow} (blue) for $C=10^{23}$ and $x_b=10^{30}$. It is perfectly compatible with a power-law (top figure) with spectral index $n_B=6$ (orange). Note that $P_{B0}\equiv P_B(kR_H=1)$. We also show that the electric power spectrum behaves in the same fashion (bottom).}
    \label{powerspectrum}
\end{figure}

Another interesting aspect of the magnetic and electric power
spectra is their dependence in terms of $k$, shown 
in Fig.~\ref{powerspectrum}. As predicted in Sec.~\ref{Aresults}, we obtain the spectral index $n_B=6$. This
is typical of non-helicoidal and causally generated magnetic fields, as noted by
Caprini and Durrer~\cite{Caprini:2001nb, Durrer:2003ja}. %

From the power spectrum, we are able to get the amplitude of the magnetic field
\eqref{est} as a function of the scale, which is shown in Figs.~\ref{magfield}
and \ref{magfield2}.  Fig.~\ref{magfield} shows that a larger $x_b$, or
equivalently a lower scale factor at the bounce ($a_b$),  results in a lower
amplitude of the field. Thus a deeper bounce tends to generate weaker magnetic
fields. This is because electric and magnetic fields are generated when $f$
effectively changes in time, which happens for $-C/\alpha < t < C/\alpha$
(except for the short period of the bounce). Since $\alpha \propto
x_b^{3/2}$, a larger $x_b$ implies a shorter period in which the non-minimal
coupling is effective. For the same reason, a larger value of $C$ leads to a
larger amplitude of the magnetic field.

In the next section, we discuss how observations and theoretical limits can be used to constrain the parameters of our models.

\section{Discussion}
\label{discussion}

We now wish to confront the results of the previous section with observational
and theoretical limits found in the literature. Limits coming from several
physical processes can be invoked, as recalled in the introduction. However, it
is worth noting that many of them focus on specific models with considerable
uncertainties, or use specific priors leading to confusion on the possible upper
and lower bounds.\footnote{For instance, see
	\cite{Zucca:2016iur,Pogosian:2018vfr} for a discussion about the suppressed
	apparent limit  on the magnetic spectral index $n_B$, when assuming a different
	prior from Planck 2015 \cite{Ade:2015cva}} Since there is no unanimously accepted
limit on the spectral index, we will focus on the bounds derived considering
$n_B$ as a free parameter. Thus, we shall consider an upper bound around
$B_{\lambda}<10^{-9}G$,\footnote{See~\cite{Bray:2018ipq} and
	\cite{Safarzadeh:2019kyq} for recent limits using ultra-high-energy cosmic rays
	anisotropy and ultra-faint dwarf galaxies, respectively. See also
	\cite{Broderick:2018nqf} for a stronger upper limit of $B_{\lambda}<10^{-15}G$,
	putting detections of intergalactic magnetic fields with $\gamma$-ray under
	pressure.} and a first lower bound of around $B_{\lambda}>10^{-17}G$.\footnote{This limit comes from the non-detection of secondary GeV $\gamma$-rays
	around TeV blazars. However, there is still an ongoing debate on whether this
	lower limit should be trusted. See for example
	\cite{Broderick:2011av,Subramanian:2019jyd}.} The second lower limit we
consider concerns the minimum seed field in galaxies that would be amplified via
dynamo mechanism \cite{subramanian1994}, namely $B_{\lambda}>10^{-21}G$.

\begin{figure}
    \centering
    \includegraphics[scale=0.6]{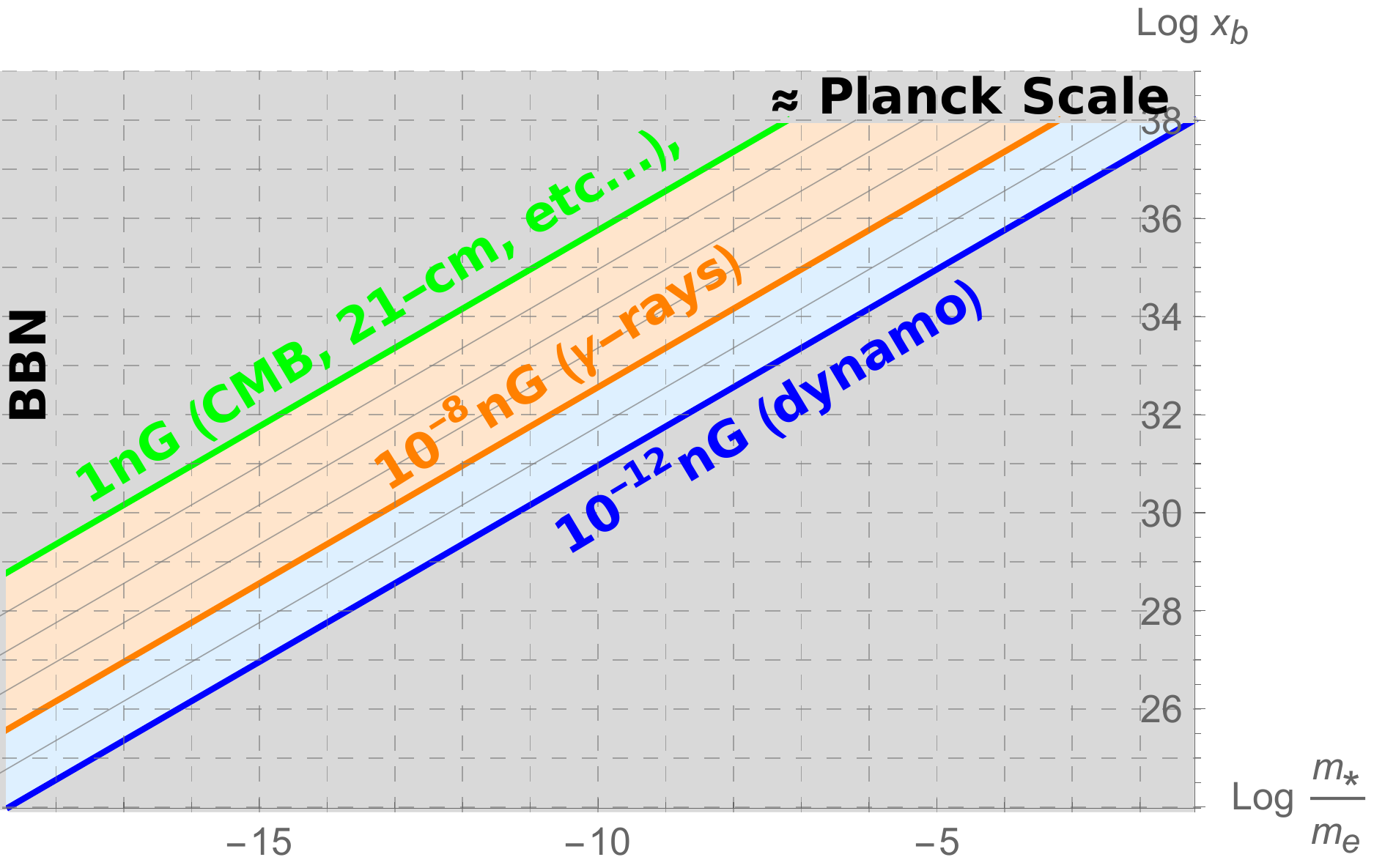}
    \caption{Parameter space with magnetic field amplitudes consistent with current limits at 1Mpc. The blue region represents the allowed values to initiate the dynamo effect, with the blue line a theoretical lower limit \cite{subramanian1994,Martin:2007ue}. The orange region represents allowed values by observations at large scales in voids, with the orange line a lower limit derived by blazars observations \cite{Taylor:2011bn} and the green line an upper limit derived using Ultra-High-Energy Cosmic Rays, Ultra-Faint Dwarf galaxies, 21-cm hydrogen lines, etc. \cite{Bray:2018ipq,Safarzadeh:2019kyq,Minoda:2018gxj}. Note the orange and blue regions are overlapping. The grey shaded region represents excluded values of the magnetic field. Each oblique grey line gives an amplitude for the magnetic field a hundred times higher than the lower line.}
    \label{parameter space 1Mpc}
\end{figure}

These theoretical and observational limits are used in Figure~\ref{parameter space 1Mpc} to constrain the region in parameter space for which consistent values of magnetic seed fields,\footnote{Within the commonly invoked limits, as discussed earlier.} evaluated today, are obtained at 1 Mpc. The upper value $x_b\lesssim10^{38}$ comes from Eq.~\eqref{limit xb} reflecting the earliest possible time for the bounce to occur. It is denoted ``Planck Scale'' in the graph. There is another limit set to preserve nucleosynthesis denoted ``BBN''. This can be derived by plugging Eq.~\eqref{limit c} into Eq.~\eqref{mass} presented below, giving $m_{\star}=10^{-19}m_e$.

In order to infer the allowed mass scales for the minimal coupling, one can use, for instance, the relation between $C$ and $m_{\star}$ coming from Eq.~\eqref{cc} to show
\begin{equation}
\label{mass}
\frac{m_{\star}}{m_e} = \frac{\alpha}{10^{38} C},
\end{equation}
where $m_e$ is the electron mass. The maximum mass allowed in this model is then $0.1 m_e$. Therefore, the value of the electron mass for $m_{\star}$ is not allowed by our model, a feature shared with power-law inflationary models
\cite{Turner1988, Campanelli2008}.

\section{Backreaction}
\label{backreaction}

When dealing with primordial magnetogenesis, a recurrent issue one must be aware of is the backreaction of the electromagnetic contribution on the background. When the electromagnetic energy density becomes higher than the background energy density, the background dynamics is modified and anisotropies can appear \cite{Kanno:2009ei}.

We define the matter and radiation energy densities, respectively, as
\begin{align}
    \rho_m \equiv \frac{\Omega_m}{Y^3}\;, \quad \rho_r \equiv \int \dd{\ln{k}}\; \left(\mathcal{P}_{E,0}+\mathcal{P}_{B,0}\right) \left(\frac{Y_0}{Y}\right)^{4} \;.
\end{align}
As pointed out in previous works on magnetogenesis in bouncing models, see
\textit{e.g.} \cite{Sriramkumar:2015yza}, the vanishing of the Hubble rate at
the bounce leads, via the Friedmann Eq.~\eqref{friedrho}, to $\rho_m=0$.
However, this is not the case here. The classical Friedmann equations are not
valid around the bounce, which is dominated by quantum cosmological effects, and
$\rho_m \propto Y^{-3}$ always. However, this does not guarantee that the model is free from backreaction. Let
us examine this point in more detail in this section.

As the electromagnetic power goes as $Y^{-4}$, and $\rho_m \propto Y^{-3}$, the first obvious critical point to investigate the issue of backreaction is at the bounce itself. As shown in the previous section, we have near the bounce that $\vert A_k \vert  \propto k^{-1/2}$ and $\vert \Pi_k\vert  \propto k^{3/2}$. Furthermore, $\vert A_k\vert $ does not depend on $x_b$ and $C$, and $\vert \Pi_k\vert  \propto C^2/\sqrt{x_b}$. This can be seen by inspecting the integral appearing in the first term of Eq.~\eqref{iterations2}, where after integration, and evaluating at the bounce, we get the constants $C^2 x_b/\alpha = C^2/x_b^{1/2}$. 

After integrating the magnetic and electric energy densities at the bounce, see Eqs.~\eqref{magnetic spectral density} and \eqref{electric spectral density}, and denoting the cut-off scale as $k_f$ (which we will refer to galactic scales, where this simple treatment may cease to be valid due to short range interactions leading to dissipation and other effects), we obtain
\begin{align}
    \rho_{B,b} = \frac{3C^2 x_b^4}{32 \pi^2 R_{H_0}^4} k_f^4 \;, \quad \rho_{E,b} = \frac{C^2 x_b^3}{9 \pi^2 R_{H_0}^4} k_f^6 \;.
\end{align}
The ratio of magnetic energy density over electric energy density is then simply
\begin{align}
    \frac{\rho_{B,b}}{\rho_{E,b}} \approx \frac{x_b}{k_f^2} \;,
\end{align}
and the magnetic field is dominant when $\rho_{B,b} \gg \rho_{E,b}$, or $\sqrt{x_b}\gg k_f$. As $x_b\gg 1$, this condition is always satisfied.

In units of Hubble radius, the matter energy density reads
\begin{align}
    \rho_{m,b} = \frac{7.8 \Omega_m 10^{120}}{R_{H_0}^4 Y^3} \;.
\end{align}
At the bounce, the matter energy density is given by
\begin{align}
    \rho_{m,b} = \frac{7.8 \Omega_m 10^{120} x_b^3}{R_{H_0}^4} \;.
\end{align}
Then, comparing the magnetic density to the matter density, and requiring the ratio be small enough gives
\begin{align}
    \frac{\rho_{B,b}}{\rho_{m,b}} < 10^{-4}  \quad \implies C^2 x_b k_f^4 < 10^{118} \;. 
    \label{magnetic backreaction}
\end{align}
Choosing the galactic scale (tens of kiloparsecs), $k\approx 10^5$, gives $C^2 x_b < 10^{98}$. The values given in Fig.~\ref{parameter space 1Mpc} all respect this constraint. In conclusion, there is no electromagnetic backreaction at the bounce.

As we have seen in Figs.~\ref{spectra evolution cc} and \ref{spectra evolution xb}, and discussed when commenting them, the electric density overcomes the magnetic density after the bounce for some time during the period $1/\alpha < t < C/\alpha$. The coupling behaves as
\begin{align}
    f \propto t^{-2} \;, \quad \frac{1}{\alpha} < t < \frac{C}{\alpha} \;,
\end{align}
and the scale factor as $Y \propto t^{\frac{2}{3}}$ in this region. This can be also be seen in Fig.~\ref{ymf}. Then, the electric density goes as $\rho_E \propto t^{-2/3}$. This is to be compared to the matter density $\rho_m \propto t^{-2}$, giving the ratio evolution
\begin{align}
    \frac{\rho_E}{\rho_m} \propto t^{\frac{4}{3}} \;.
\end{align}
To get an estimate of the electric backreaction, let us evaluate the initial conditions at the bounce and evolve this ratio in the considered time range. Performing a procedure similar to the one leading to \eqref{magnetic backreaction}, we obtain
\begin{align}
    \frac{\rho_{E,b}}{\rho_{m,b}} = 10^{-122}\:C^2 k_f^6  . 
\end{align}
Then, the ratio will evolve as
\begin{align}
    \frac{\rho_{E}}{\rho_{m}} = 10^{-122}\:C^2 k_f^6  \left(\frac{t_f}{t_i} \right)^{\frac{4}{3}} \;.
\end{align}
Choosing the initial time $t_i\equiv1/\alpha$ and the final time $t_f\equiv C/\alpha$ and imposing once again that the backreaction be small, we finally obtain
\begin{align}
    \frac{\rho_{E}}{\rho_{m}} < 10^{-4} \quad \implies C^{\frac{10}{3}} k_f^6 < 10^{118} \;. 
\end{align}
Once again, $k_f \approx 10^{5}$ is compatible with the maximum value $C\approx10^{26.3}$ allowed in Fig.~\ref{parameter space 1Mpc}. Again, there is no backreaction problem in our model
\footnote{To discuss the (absence of) backreaction in our model, we have shown that the electromagnetic energy density is always smaller than the matter energy density. In other models of bounce,
such as those based on the Lee-Wick theory \cite{Cai2008}, there are mechanisms preventing \emph{ab initio}
the uncontrolled growth of the electromagnetic energy density.}.

\section{Conclusions}

We presented in this work the generation of primordial magnetic fields in the context of a cosmological bounce, through a coupling between curvature and electromagnetism, predicted by 
QED in curved spacetimes \cite{Drummond:1979pp}. A homogeneous and isotropic background filled with pressureless (dark) matter in the contracting phase, followed by a bounce and an expanding phase has been considered. The bounce is produced by quantum effects described in the de Broglie-Bohm interpretation of quantum mechanics, motivated by the inconsistency of using standard quantum mechanics in quantum cosmology \cite{Pinto-Neto:2013toa}. Moreover, one of the advantages of bouncing magnetogenesis is the absence of the strong coupling problem. The model is characterized by three parameters, namely the presureless (dark) matter density today, $\Omega_m$, the scale factor at which the bounce happens, $x_b$, and the mass scale of the coupling $m_{\star}$.

We showed that an adiabatic vacuum can be defined as initial condition for the electromagnetic field in the far past of the contracting phase. 
Having defined the vacuum, we were able to explain analytically the behavior of the electric and magnetic modes, summarized in Eqs. \eqref{historyA} and \eqref{historyP}. We then confronted these analytical results with a numerical integration of the modes, given in Figs.~\ref{mode evolution cc} and \ref{mode evolution xb}, and presented in Figs.~\ref{spectra evolution cc}
and \ref{spectra evolution xb} the time evolution of the magnetic and electric power spectra. We illustrated the scale dependence of both spectra  in Fig~\ref{powerspectrum}, finding they behaved as a power-law with the same spectral index $n_E=n_B=6$. This result is reminiscent of
non-helicoidal, causally generated magnetic fields from phase transitions in the
early Universe \cite{Durrer:2003ja}. In Figs.~\ref{magfield} and \ref{magfield2}, we showed the amplitude of the magnetic field today was found to be strong enough on a wide range of scales to pass the current limits from observations.\footnote{It is worthwhile noting that the same coupling, when
	considered in the context of power-law inflation, does not generate large enough
	magnetic fields, see \cite{Campanelli2008}.}At the scale of 1~Mpc, we have
derived constraints on $x_b$ and $m_{\star}$, summarized in Fig.~\ref{parameter space
	1Mpc}. Finally, we also demonstrated that backreaction is not a problem in our model.

Though the results from our analysis are quite promising, we have omitted several possible effects that could constrain our results further. First, the presence of an electromagnetic energy density in spacetime should induce a stochastic background of
gravitational waves, even moreso since the magnetic fields generated have a very blue spectrum. Thus, the inclusion of theoretical  limits on gravitational waves production \cite{Caprini:2001nb} will be investigated in the future. This will be even more relevant with the upcoming detections from LISA \cite{Caprini:2009yp,Caprini:2018mtu,Saga:2018ont,Pol:2019yex}.

A second point of interest would be to take into account other possible backreaction effects. It has been shown recently that the vacuum polarisation in a dielectric medium, the so-called Schwinger effect, increases the medium conductivity and subsequently stops the magnetic field production \cite{Sobol:2018djj,Sobol:2019xls,Sharma:2017eps,Sharma:2018kgs,Shakeri:2019mnt}. This would lead to weaker magnetic fields than expected, and could constrain further our model. 

As a possible extension of our work, other non-minimal couplings between the
electromagnetic and the gravitational field (involving the Ricci and Riemann
tensors) could be considered in the generation of primordial magnetic fields.\footnote{Such couplings were considered in the framework of power-law inflation
	in \cite{Kunze2009}. } 
Also of importance is the parity-violating coupling $RF\tilde F$
\footnote{ The authors wish to thank the referee for calling their attention to this possibility.}
, which may be associated to the generation of helical magnetic fields.
	We leave these open questions for future work.

\begin{acknowledgments}

The authors wish to thank Samuel Colin for his important contributions in earlier stages of this work, particularly for helping them to discard a previous proposal for the coupling. EF acknowledges support by the \emph{Coordena\c{c}\~ao de Aperfei\c{c}oamento de Pessoal de N\'{i}vel Superior} - Brazil (CAPES) - Finance Code 001. EF thanks the \emph{Centro Brasileiro de Pesquisas Físicas} (CBPF) for its hospitality. NPN acknowledges support of CNPq of Brazil under grant PQ-IB 309073/2017-0. 

\end{acknowledgments}

\appendix

\section{Adiabatic vacuum initial conditions}
\label{sec:adiab}

First, we need to impose initial conditions for the EM field. To this end, we follow the adiabatic vacuum prescription implemented in Ref.~\cite{sandro_vacuum}. Even though we are dealing with vector degrees of freedom, since the time-dependent coefficient $A_k(t)$ satisfies the normalization condition~\eqref{vacuumnorma}, it follows that $A_k(t)$ has a behavior similar to the one of the coefficient one would obtain when quantizing a single free scalar field. Let us then consider the Hamiltonian
\begin{equation}
\label{hamiltonian}
\mathcal{H} = \frac{\Pi_k^2}{2m} + \frac{m\nu^2 A_k^2}{2} \;,
\end{equation}
where $m$ and $\nu$ can be functions of time. The Hamilton equations of motion
\begin{equation}
\label{eq-hamiltonian}
A^\prime_k=\frac{\Pi_k}{m}; \quad \Pi^\prime_k = -m\nu^2 A_k 
\end{equation}
lead to Eq.~\eqref{diffeq} if one identifies $m = Yf$ and $\nu = k / Y$.

A convenient choice is to express $A_{k}$ and $\Pi_k$ as the components of a particular eigenvector of the complex structure matrix (see Ref.~\cite{sandro_vacuum} for the mathematical and physical reasons to implement this choice),
\begin{equation}
\label{v:components}
\begin{split}
A_k &\equiv \frac{1}{2}\exp{(-\gamma_k/2)} \left[\exp{(\chi_k/2)}-\ci\exp{(-\chi_k/2)}\right] \;,\\ 
\Pi_k &\equiv -\frac{1}{2}\exp{(\gamma_k/2)} \left[\exp{(\chi_k/2)}+\ci\exp{(-\chi_k/2)} \right)] \;.
\end{split}
\end{equation}
The variables $\chi_k$ and $\gamma_k$ are real time-dependent functions, and can be used to represent the aforementioned matrix as
\begin{equation}
\label{M:components}
M_a{}^b = \left( \begin{array}{cc} \sinh{\chi_k} & \cosh{\chi_k} \exp({-\gamma_k}) \\ -\cosh{\chi_k} \exp({\gamma_k}) & -\sinh{\chi_k} \end{array} \right).
\end{equation}
Latin indices ($a,b,c, $ \dots) refer to the phase space vector components
defined by $v_a \equiv \left(A_k,\Pi_k\right)$, which are raised and lowered using the symplectic matrix as defined in Ref.~\cite{sandro_vacuum}. The phase space vectors $v_a$  satisfying the normalization condition~\eqref{vacuumnorma} (modulo a global time-dependent phase) have an one-to-one correspondence with matrices of the form shown in Eq.~\eqref{M:components}, and consequently with a pair $(\chi_k,\;\gamma_k)$. For this reason, we will denote interchangeably  $\left(A_k,\Pi_k\right)$ and $\left(\chi_k,\;\gamma_k\right)$ with the same symbol $v_a$.

The Hamilton Eqs.~\eqref{eq-hamiltonian} induce the dynamics of the matrix $M_a{}^b$, which reads
\begin{equation}
\label{set:equations}
\begin{split}
\chi_k^\prime&= -2\nu\sinh(\gamma_k-\xi) \;,\\
\gamma_k^\prime &= +2\nu\cosh(\gamma_k-\xi)\tanh(\chi_k) \;,
\end{split}
\end{equation}
where $\xi \equiv \ln (m\nu)$. The complex structure matrix satisfies
\begin{equation}
\label{M2:identity}
M_a{}^c M_c{}^b = -\delta_a{}^b \;,
\end{equation}
and, the comparison of two different vacuum definitions, given respectively by $v_a$ and $u_a$, yields the Bogoliubov coefficients
\begin{equation}
\label{beta-M}
\vert \beta_{v,u}\vert^2 = -\frac{1}{4} {\rm Tr} \left[{\rm\bf I} + {\bf M}(v) {\bf M}(u)\right] \;,
\end{equation}
with ${\rm Tr}$ the trace operator, ${\rm\bf I}$ the identity matrix and ${\bf M}(v)$ (${\bf M}(u)$) is the matrix  associated with the components $M_a{}^b(v)$ ($M_a{}^b(u)$) defined by the vector components $v_a$ ($u_a$). In this framework, a vacuum choice translates into a choice of functions $v^V_a \equiv \left(\chi_k^V(t),\;\gamma_k^{V}(t)\right)$ defined locally (with a finite number of time derivatives of the background variables), which do not necessarily satisfy the equations of motion~\eqref{set:equations} but give an approximation close enough to a solution. Moreover, the vacuum must be fixed by choosing a time $t_0$ where the variables satisfy 
$$v_a(t_0) = v^V_a(t_0), \quad\implies\quad \left(\chi_k(t_0), \gamma_k(t_0)\right) = \left(\chi^V_k(t_0), \gamma^V_k(t_0)\right).$$ In other words, if $v_a^V$ is stable in the sense that $$\Delta v_a \equiv \left(\delta\chi_k,\;\delta\gamma_k \right) = \left(\chi_k(t) - \chi^V_k(t), \gamma_k(t) - \gamma^V_k(t)\right)$$ remains small for a finite time interval, then  particle creation will also be small in this interval. This characterizes the so-called adiabatic vacuum. Hence, we find the adiabatic vacuum by finding the critical points of the system \eqref{set:equations}. When $\xi$ is constant in time, the critical points of the system \eqref{set:equations} are obvious: $\chi^V_k = 0$ and $\gamma^V_k = \xi$, a choice satisfying the condition of being locally defined in terms of the background. Then, substituting into Eq.~\eqref{v:components}, and using it as initial conditons for the system~\eqref{eq-hamiltonian}, yields the following solution
\begin{equation}
\label{WKB}
\begin{split}
A_k &= \frac{e^{-\ci\pi/4}}{\sqrt{2m\nu}}
\exp\left[-\ci\int_{t_0}^t\nu\ddt\right] \;,\\
\Pi_k &= -\ci e^{-\ci\pi/4}\sqrt{\frac{m\nu}{2}}
\exp\left[-\ci\int_{t_0}^t\nu\ddt\right] \;.
\end{split}
\end{equation}
In this case, the vacuum is perfectly stable, there is never particle production because $\chi_k(t) = \chi^V_k(t) = 0$ and $\gamma_k(t) = \gamma^V_k(t) = \xi$ for any time $t$, and consequently $\vert\beta_{v,v^V}\vert^2=0$, see Eq.~\eqref{beta-M}. We have a perfect adiabatic vacuum, which coincides with the WKB solution.

In the case where $\xi$ changes in time, there is one well-known situation where adiabatic vacua can be defined: when the mode frequencies dominate the dynamics. Let us define $$F_n \equiv \left(\frac{1}{2\nu}\frac{\dd}{\dd t}\right)^n \xi \;,$$ where $F_0 = \xi$, the function $F_1$ is the ratio between the time derivative of $\xi$ and $\theta \equiv \int2\nu\dd t$, $F_2$ the ratio between the time derivative of $F_1$ and $\theta$ and so forth.  Then, in the case $1 \gg F_1 \gg\dots\gg F_n>\dots$, which means that $\xi$ slowly varies in cosmic time when compared with the variation of $\theta$, one can still find approximate critical points (\textit{i.e.} adiabatic vacua), which can be reached through successive approximations, as explained in Ref.~\cite{sandro_vacuum}. Up to second order, the approximate critical points read
\begin{align}
\label{variables2}
\chi^V_k&= F_1 \;,\nonumber \\
\gamma^V_k&= F_0 -F_2\;.
\end{align}
If they are inserted in Eq.~\eqref{v:components}, they lead to the usual WKB expansion (modulo a time-dependent phase). As discussed in~\cite{sandro_vacuum}, around these functions, the variables $\Delta v_a$ satisfy a forced harmonic oscillator equation of motion with force of order $\mathcal{O}(F_3)$.

In our case, we have $m\nu=kf$. In the far past of the contracting phase one gets, for $f$ given in Eq.~\eqref{cc},
\begin{equation}
\label{xi1}
\left\vert\frac{\dd\xi}{\dd\theta}\right\vert \approx \frac{C^2}{x_b^3k\vert t\vert^{7/3}} \ll 1 \;,
\end{equation}
which implies that $$\vert t\vert \gg \vert t_a\vert \equiv\left(\frac{C^2}{x_b^3k}\right)^{3/7}.$$ As the physically relevant parameter space we consider satisfies $C^2/x_b^3 \ll 1$, then $\vert t_a\vert  \ll 1$, and this condition is easily satisfied.

However, the other adiabaticity conditions impose a more stringent constraint on $\vert t\vert$. Indeed, 

\begin{equation}
\label{xi2}
\left\vert\frac{\dd^2\xi}{\dd\theta^2}\right\vert \ll \left\vert\frac{\dd\xi}{\dd\theta}\right\vert \Rightarrow \vert t\vert  \gg \vert t_c\vert \equiv
\left(\frac{7}{3 k}\right)^{3}\approx k^{-3} \;.
\end{equation}
One can easily verify that all other conditions yield, apart numerical factors of order $1$,\footnote{It starts around $1$ and grows slowly with $n$, this is a natural feature of an asymptotic expansion. In other words, for a fixed time and mode $k$ there is a maximum order $n$ from which the series starts to be a bad approximation.} the same condition \eqref{xi2}. Hence, the adiabaticity condition reads

\begin{equation}
\label{adiabatic}
\vert t\vert  \gg \vert t_c\vert  \approx k^{-3}.
\end{equation}
This means that modes with the size of the Hubble radius today leave (enter) the adiabatic regime in the contracting (expanding) phase for times of the order the Hubble time today, independently of the parameters $x_b$ and $C$. Smaller wavelengths leave (enter) the adiabatic regime later (earlier) than the present Hubble time, following the rule $k^{-3}$.

To summarize, one can impose adiabatic vacuum initial conditions for the electromagnetic field in the contracting phase of the present bouncing model when $\vert t\vert  \gg \vert t_c\vert  \approx k^{-3}$. In this regime, the modes read, at leading order,
\begin{equation}
\label{initial-vacuum}
\begin{split}
A_k &= \frac{e^{-\ci\pi/4}}{\sqrt{2kf}}\exp\left(-\ci k\eta\right)+\dots \;,\\
\Pi_k &= -\ci e^{-\ci\pi/4}\sqrt{\frac{kf}{2}}
\exp\left(-\ci k\eta\right)+\dots \;,
\end{split}
\end{equation}
where $\eta$ is the conformal time $dt = Yd\eta$.

Since $f\approx 1/4$ for $\vert t\vert  \gg \vert t_c\vert $ , it follows that 
\begin{equation}
\label{initial-modulus}
\begin{split}
\vert A_k\vert  &= \sqrt{\frac{2}{k}} +\dots \;, \\
\vert \Pi_k\vert  &= \sqrt{\frac{k}{8}} +\dots. \;,
\end{split}
\end{equation}
and both the field and its canonical momentum are constant in this regime.

\bibliographystyle{apsrev4-2.bst}
\bibliography{bibliography} 

\begin{thebibliography}{102}%
\makeatletter
\providecommand \@ifxundefined [1]{%
 \@ifx{#1\undefined}
}%
\providecommand \@ifnum [1]{%
 \ifnum #1\expandafter \@firstoftwo
 \else \expandafter \@secondoftwo
 \fi
}%
\providecommand \@ifx [1]{%
 \ifx #1\expandafter \@firstoftwo
 \else \expandafter \@secondoftwo
 \fi
}%
\providecommand \natexlab [1]{#1}%
\providecommand \enquote  [1]{``#1''}%
\providecommand \bibnamefont  [1]{#1}%
\providecommand \bibfnamefont [1]{#1}%
\providecommand \citenamefont [1]{#1}%
\providecommand \href@noop [0]{\@secondoftwo}%
\providecommand \href [0]{\begingroup \@sanitize@url \@href}%
\providecommand \@href[1]{\@@startlink{#1}\@@href}%
\providecommand \@@href[1]{\endgroup#1\@@endlink}%
\providecommand \@sanitize@url [0]{\catcode `\\12\catcode `\$12\catcode
  `\&12\catcode `\#12\catcode `\^12\catcode `\_12\catcode `\%12\relax}%
\providecommand \@@startlink[1]{}%
\providecommand \@@endlink[0]{}%
\providecommand \url  [0]{\begingroup\@sanitize@url \@url }%
\providecommand \@url [1]{\endgroup\@href {#1}{\urlprefix }}%
\providecommand \urlprefix  [0]{URL }%
\providecommand \Eprint [0]{\href }%
\providecommand \doibase [0]{https://doi.org/}%
\providecommand \selectlanguage [0]{\@gobble}%
\providecommand \bibinfo  [0]{\@secondoftwo}%
\providecommand \bibfield  [0]{\@secondoftwo}%
\providecommand \translation [1]{[#1]}%
\providecommand \BibitemOpen [0]{}%
\providecommand \bibitemStop [0]{}%
\providecommand \bibitemNoStop [0]{.\EOS\space}%
\providecommand \EOS [0]{\spacefactor3000\relax}%
\providecommand \BibitemShut  [1]{\csname bibitem#1\endcsname}%
\let\auto@bib@innerbib\@empty
\bibitem [{\citenamefont {Peter}\ \emph {et~al.}(2007)\citenamefont {Peter},
  \citenamefont {Pinho},\ and\ \citenamefont {Pinto-Neto}}]{Peter:2006hx}%
  \BibitemOpen
  \bibfield  {author} {\bibinfo {author} {\bibfnamefont {P.}~\bibnamefont
  {Peter}}, \bibinfo {author} {\bibfnamefont {E.~J.}\ \bibnamefont {Pinho}},\
  and\ \bibinfo {author} {\bibfnamefont {N.}~\bibnamefont {Pinto-Neto}},\
  }\href {https://doi.org/10.1103/PhysRevD.75.023516} {\bibfield  {journal}
  {\bibinfo  {journal} {Phys.\ Rev.\ D}\ }\textbf {\bibinfo {volume} {75}},\
  \bibinfo {pages} {023516} (\bibinfo {year} {2007})},\ \Eprint
  {https://arxiv.org/abs/hep-th/0610205} {arXiv:hep-th/0610205 [hep-th]}
  \BibitemShut {NoStop}%
\bibitem [{\citenamefont {Durrer}\ and\ \citenamefont
  {Neronov}(2013)}]{Durrer2013}%
  \BibitemOpen
  \bibfield  {author} {\bibinfo {author} {\bibfnamefont {R.}~\bibnamefont
  {Durrer}}\ and\ \bibinfo {author} {\bibfnamefont {A.}~\bibnamefont
  {Neronov}},\ }\href {https://doi.org/10.1007/s00159-013-0062-7} {\bibfield
  {journal} {\bibinfo  {journal} {Astron. Astrophys. Rev.}\ }\textbf {\bibinfo
  {volume} {21}},\ \bibinfo {pages} {62} (\bibinfo {year} {2013})},\ \Eprint
  {https://arxiv.org/abs/1303.7121} {arXiv:1303.7121 [astro-ph.CO]}
  \BibitemShut {NoStop}%
\bibitem [{\citenamefont {Beck}(2012)}]{beck2012magnetic}%
  \BibitemOpen
  \bibfield  {author} {\bibinfo {author} {\bibfnamefont {R.}~\bibnamefont
  {Beck}},\ }\href {https://doi.org/10.1007/s11214-011-9782-z} {\bibfield
  {journal} {\bibinfo  {journal} {Space Science Reviews}\ }\textbf {\bibinfo
  {volume} {166}},\ \bibinfo {pages} {215} (\bibinfo {year}
  {2012})}\BibitemShut {NoStop}%
\bibitem [{\citenamefont {Beck}\ and\ \citenamefont
  {Wielebinski}(2013)}]{Beck:2013bxa}%
  \BibitemOpen
  \bibfield  {author} {\bibinfo {author} {\bibfnamefont {R.}~\bibnamefont
  {Beck}}\ and\ \bibinfo {author} {\bibfnamefont {R.}~\bibnamefont
  {Wielebinski}},\ }\href {https://doi.org/10.1007/978-94-007-5612-0\_13}
  {\bibfield  {journal} {\bibinfo  {journal} {Planets, Stars and Stellar
  Systems}\ ,\ \bibinfo {pages} {641}} (\bibinfo {year} {2013})},\ \Eprint
  {https://arxiv.org/abs/1302.5663} {arXiv:1302.5663 [astro-ph.GA]}
  \BibitemShut {NoStop}%
\bibitem [{\citenamefont {Minoda}\ \emph {et~al.}(2019)\citenamefont {Minoda},
  \citenamefont {Tashiro},\ and\ \citenamefont {Takahashi}}]{Minoda:2018gxj}%
  \BibitemOpen
  \bibfield  {author} {\bibinfo {author} {\bibfnamefont {T.}~\bibnamefont
  {Minoda}}, \bibinfo {author} {\bibfnamefont {H.}~\bibnamefont {Tashiro}},\
  and\ \bibinfo {author} {\bibfnamefont {T.}~\bibnamefont {Takahashi}},\ }\href
  {https://doi.org/10.1093/mnras/stz1860} {\bibfield  {journal} {\bibinfo
  {journal} {Mon.\ Not.\ Roy.\ Astron.\ Soc.}\ }\textbf {\bibinfo {volume}
  {488}},\ \bibinfo {pages} {2001} (\bibinfo {year} {2019})},\ \Eprint
  {https://arxiv.org/abs/1812.00730} {arXiv:1812.00730 [astro-ph.CO]}
  \BibitemShut {NoStop}%
\bibitem [{\citenamefont {Bray}\ and\ \citenamefont
  {Scaife}(2018)}]{Bray:2018ipq}%
  \BibitemOpen
  \bibfield  {author} {\bibinfo {author} {\bibfnamefont {J.}~\bibnamefont
  {Bray}}\ and\ \bibinfo {author} {\bibfnamefont {A.}~\bibnamefont {Scaife}},\
  }\href {https://doi.org/10.3847/1538-4357/aac777} {\bibfield  {journal}
  {\bibinfo  {journal} {Astrophys.\ J.}\ }\textbf {\bibinfo {volume} {861}},\
  \bibinfo {pages} {3} (\bibinfo {year} {2018})},\ \Eprint
  {https://arxiv.org/abs/1805.07995} {arXiv:1805.07995 [astro-ph.CO]}
  \BibitemShut {NoStop}%
\bibitem [{\citenamefont {Ade}\ \emph {et~al.}(2016)\citenamefont {Ade} \emph
  {et~al.}}]{Ade:2015cva}%
  \BibitemOpen
  \bibfield  {author} {\bibinfo {author} {\bibfnamefont {P.}~\bibnamefont
  {Ade}} \emph {et~al.} (\bibinfo {collaboration} {Planck}),\ }\href
  {https://doi.org/10.1051/0004-6361/201525821} {\bibfield  {journal} {\bibinfo
   {journal} {Astron.\ Astrophys.}\ }\textbf {\bibinfo {volume} {594}},\
  \bibinfo {pages} {A19} (\bibinfo {year} {2016})},\ \Eprint
  {https://arxiv.org/abs/1502.01594} {arXiv:1502.01594 [astro-ph.CO]}
  \BibitemShut {NoStop}%
\bibitem [{\citenamefont {Chluba}\ \emph {et~al.}(2019)\citenamefont {Chluba}
  \emph {et~al.}}]{Chluba:2019kpb}%
  \BibitemOpen
  \bibfield  {author} {\bibinfo {author} {\bibfnamefont {J.}~\bibnamefont
  {Chluba}} \emph {et~al.},\ }\href@noop {} {\bibfield  {journal} {\bibinfo
  {journal} {Bull.\ Am.\ Astron.\ Soc.}\ }\textbf {\bibinfo {volume} {51}},\
  \bibinfo {pages} {184} (\bibinfo {year} {2019})},\ \Eprint
  {https://arxiv.org/abs/1903.04218} {arXiv:1903.04218 [astro-ph.CO]}
  \BibitemShut {NoStop}%
\bibitem [{\citenamefont {Zucca}\ \emph {et~al.}(2017)\citenamefont {Zucca},
  \citenamefont {Li},\ and\ \citenamefont {Pogosian}}]{Zucca:2016iur}%
  \BibitemOpen
  \bibfield  {author} {\bibinfo {author} {\bibfnamefont {A.}~\bibnamefont
  {Zucca}}, \bibinfo {author} {\bibfnamefont {Y.}~\bibnamefont {Li}},\ and\
  \bibinfo {author} {\bibfnamefont {L.}~\bibnamefont {Pogosian}},\ }\href
  {https://doi.org/10.1103/PhysRevD.95.063506} {\bibfield  {journal} {\bibinfo
  {journal} {Phys. Rev.}\ }\textbf {\bibinfo {volume} {D95}},\ \bibinfo {pages}
  {063506} (\bibinfo {year} {2017})},\ \Eprint
  {https://arxiv.org/abs/1611.00757} {arXiv:1611.00757 [astro-ph.CO]}
  \BibitemShut {NoStop}%
\bibitem [{\citenamefont {Pogosian}\ and\ \citenamefont
  {Zucca}(2018)}]{Pogosian:2018vfr}%
  \BibitemOpen
  \bibfield  {author} {\bibinfo {author} {\bibfnamefont {L.}~\bibnamefont
  {Pogosian}}\ and\ \bibinfo {author} {\bibfnamefont {A.}~\bibnamefont
  {Zucca}},\ }\href {https://doi.org/10.1088/1361-6382/aac398} {\bibfield
  {journal} {\bibinfo  {journal} {Class. Quant. Grav.}\ }\textbf {\bibinfo
  {volume} {35}},\ \bibinfo {pages} {124004} (\bibinfo {year} {2018})},\
  \Eprint {https://arxiv.org/abs/1801.08936} {arXiv:1801.08936 [astro-ph.CO]}
  \BibitemShut {NoStop}%
\bibitem [{\citenamefont {Saga}\ \emph
  {et~al.}(2018{\natexlab{a}})\citenamefont {Saga}, \citenamefont {Tashiro},\
  and\ \citenamefont {Yokoyama}}]{Saga:2017wwr}%
  \BibitemOpen
  \bibfield  {author} {\bibinfo {author} {\bibfnamefont {S.}~\bibnamefont
  {Saga}}, \bibinfo {author} {\bibfnamefont {H.}~\bibnamefont {Tashiro}},\ and\
  \bibinfo {author} {\bibfnamefont {S.}~\bibnamefont {Yokoyama}},\ }\href
  {https://doi.org/10.1093/mnrasl/slx195} {\bibfield  {journal} {\bibinfo
  {journal} {Mon.\ Not.\ Roy.\ Astron.\ Soc.}\ }\textbf {\bibinfo {volume}
  {474}},\ \bibinfo {pages} {L52} (\bibinfo {year} {2018}{\natexlab{a}})},\
  \Eprint {https://arxiv.org/abs/1708.08225} {arXiv:1708.08225 [astro-ph.CO]}
  \BibitemShut {NoStop}%
\bibitem [{\citenamefont {Kawasaki}\ and\ \citenamefont
  {Kusakabe}(2012)}]{Kawasaki:2012va}%
  \BibitemOpen
  \bibfield  {author} {\bibinfo {author} {\bibfnamefont {M.}~\bibnamefont
  {Kawasaki}}\ and\ \bibinfo {author} {\bibfnamefont {M.}~\bibnamefont
  {Kusakabe}},\ }\href {https://doi.org/10.1103/PhysRevD.86.063003} {\bibfield
  {journal} {\bibinfo  {journal} {Phys.\ Rev.\ D}\ }\textbf {\bibinfo {volume}
  {86}},\ \bibinfo {pages} {063003} (\bibinfo {year} {2012})},\ \Eprint
  {https://arxiv.org/abs/1204.6164} {arXiv:1204.6164 [astro-ph.CO]}
  \BibitemShut {NoStop}%
\bibitem [{\citenamefont {Barai}\ and\ \citenamefont {de~Gouveia
  Dal~Pino}(2018)}]{Barai:2018msb}%
  \BibitemOpen
  \bibfield  {author} {\bibinfo {author} {\bibfnamefont {P.}~\bibnamefont
  {Barai}}\ and\ \bibinfo {author} {\bibfnamefont {E.~M.}\ \bibnamefont
  {de~Gouveia Dal~Pino}} (\bibinfo {collaboration} {CTA Consortium})\
  }(\bibinfo {year} {2018})\ \Eprint {https://arxiv.org/abs/1811.06025}
  {arXiv:1811.06025 [astro-ph.HE]} \BibitemShut {NoStop}%
\bibitem [{\citenamefont {Brandenburg}\ and\ \citenamefont
  {Subramanian}(2005)}]{Brandenburg2004}%
  \BibitemOpen
  \bibfield  {author} {\bibinfo {author} {\bibfnamefont {A.}~\bibnamefont
  {Brandenburg}}\ and\ \bibinfo {author} {\bibfnamefont {K.}~\bibnamefont
  {Subramanian}},\ }\href {https://doi.org/10.1016/j.physrep.2005.06.005}
  {\bibfield  {journal} {\bibinfo  {journal} {Phys. Rept.}\ }\textbf {\bibinfo
  {volume} {417}},\ \bibinfo {pages} {1} (\bibinfo {year} {2005})},\ \Eprint
  {https://arxiv.org/abs/astro-ph/0405052} {arXiv:astro-ph/0405052 [astro-ph]}
  \BibitemShut {NoStop}%
\bibitem [{\citenamefont {Martin}\ and\ \citenamefont
  {Yokoyama}(2008)}]{Martin:2007ue}%
  \BibitemOpen
  \bibfield  {author} {\bibinfo {author} {\bibfnamefont {J.}~\bibnamefont
  {Martin}}\ and\ \bibinfo {author} {\bibfnamefont {J.}~\bibnamefont
  {Yokoyama}},\ }\href {https://doi.org/10.1088/1475-7516/2008/01/025}
  {\bibfield  {journal} {\bibinfo  {journal} {JCAP}\ }\textbf {\bibinfo
  {volume} {01}},\ \bibinfo {pages} {025}},\ \Eprint
  {https://arxiv.org/abs/0711.4307} {arXiv:0711.4307 [astro-ph]} \BibitemShut
  {NoStop}%
\bibitem [{\citenamefont {Ratra}(1992)}]{Ratra:1991bn}%
  \BibitemOpen
  \bibfield  {author} {\bibinfo {author} {\bibfnamefont {B.}~\bibnamefont
  {Ratra}},\ }\href {https://doi.org/10.1086/186384} {\bibfield  {journal}
  {\bibinfo  {journal} {Astrophys.\ J.}\ }\textbf {\bibinfo {volume} {391}},\
  \bibinfo {pages} {L1} (\bibinfo {year} {1992})}\BibitemShut {NoStop}%
\bibitem [{\citenamefont {Davis}\ and\ \citenamefont
  {Dimopoulos}(1997)}]{Davis:1995mv}%
  \BibitemOpen
  \bibfield  {author} {\bibinfo {author} {\bibfnamefont {A.-C.}\ \bibnamefont
  {Davis}}\ and\ \bibinfo {author} {\bibfnamefont {K.}~\bibnamefont
  {Dimopoulos}},\ }\href {https://doi.org/10.1103/PhysRevD.55.7398} {\bibfield
  {journal} {\bibinfo  {journal} {Phys. Rev.}\ }\textbf {\bibinfo {volume}
  {D55}},\ \bibinfo {pages} {7398} (\bibinfo {year} {1997})},\ \Eprint
  {https://arxiv.org/abs/astro-ph/9506132} {arXiv:astro-ph/9506132 [astro-ph]}
  \BibitemShut {NoStop}%
\bibitem [{\citenamefont {Berera}\ \emph {et~al.}(1999)\citenamefont {Berera},
  \citenamefont {Kephart},\ and\ \citenamefont {Wick}}]{Berera:1998hv}%
  \BibitemOpen
  \bibfield  {author} {\bibinfo {author} {\bibfnamefont {A.}~\bibnamefont
  {Berera}}, \bibinfo {author} {\bibfnamefont {T.~W.}\ \bibnamefont
  {Kephart}},\ and\ \bibinfo {author} {\bibfnamefont {S.~D.}\ \bibnamefont
  {Wick}},\ }\href {https://doi.org/10.1103/PhysRevD.59.043510} {\bibfield
  {journal} {\bibinfo  {journal} {Phys. Rev.}\ }\textbf {\bibinfo {volume}
  {D59}},\ \bibinfo {pages} {043510} (\bibinfo {year} {1999})},\ \Eprint
  {https://arxiv.org/abs/hep-ph/9809404} {arXiv:hep-ph/9809404 [hep-ph]}
  \BibitemShut {NoStop}%
\bibitem [{\citenamefont {Kandus}\ \emph {et~al.}(2000)\citenamefont {Kandus},
  \citenamefont {Calzetta}, \citenamefont {Mazzitelli},\ and\ \citenamefont
  {Wagner}}]{Kandus:1999st}%
  \BibitemOpen
  \bibfield  {author} {\bibinfo {author} {\bibfnamefont {A.}~\bibnamefont
  {Kandus}}, \bibinfo {author} {\bibfnamefont {E.~A.}\ \bibnamefont
  {Calzetta}}, \bibinfo {author} {\bibfnamefont {F.~D.}\ \bibnamefont
  {Mazzitelli}},\ and\ \bibinfo {author} {\bibfnamefont {C.~E.~M.}\
  \bibnamefont {Wagner}},\ }\href
  {https://doi.org/10.1016/S0370-2693(99)01389-1} {\bibfield  {journal}
  {\bibinfo  {journal} {Phys. Lett.}\ }\textbf {\bibinfo {volume} {B472}},\
  \bibinfo {pages} {287} (\bibinfo {year} {2000})},\ \Eprint
  {https://arxiv.org/abs/hep-ph/9908524} {arXiv:hep-ph/9908524 [hep-ph]}
  \BibitemShut {NoStop}%
\bibitem [{\citenamefont {Bassett}\ \emph {et~al.}(2000)\citenamefont
  {Bassett}, \citenamefont {Gordon}, \citenamefont {Maartens},\ and\
  \citenamefont {Kaiser}}]{Bassett:1999ta}%
  \BibitemOpen
  \bibfield  {author} {\bibinfo {author} {\bibfnamefont {B.~A.}\ \bibnamefont
  {Bassett}}, \bibinfo {author} {\bibfnamefont {C.}~\bibnamefont {Gordon}},
  \bibinfo {author} {\bibfnamefont {R.}~\bibnamefont {Maartens}},\ and\
  \bibinfo {author} {\bibfnamefont {D.~I.}\ \bibnamefont {Kaiser}},\ }\href
  {https://doi.org/10.1103/PhysRevD.61.061302} {\bibfield  {journal} {\bibinfo
  {journal} {Phys. Rev.}\ }\textbf {\bibinfo {volume} {D61}},\ \bibinfo {pages}
  {061302} (\bibinfo {year} {2000})},\ \Eprint
  {https://arxiv.org/abs/hep-ph/9909482} {arXiv:hep-ph/9909482 [hep-ph]}
  \BibitemShut {NoStop}%
\bibitem [{\citenamefont {Battaner}\ and\ \citenamefont
  {Lesch}(2000)}]{Battaner:2000kf}%
  \BibitemOpen
  \bibfield  {author} {\bibinfo {author} {\bibfnamefont {E.}~\bibnamefont
  {Battaner}}\ and\ \bibinfo {author} {\bibfnamefont {H.}~\bibnamefont
  {Lesch}},\ }\href@noop {} {\  (\bibinfo {year} {2000})},\ \Eprint
  {https://arxiv.org/abs/astro-ph/0003370} {arXiv:astro-ph/0003370 [astro-ph]}
  \BibitemShut {NoStop}%
\bibitem [{\citenamefont {Davis}\ \emph {et~al.}(2001)\citenamefont {Davis},
  \citenamefont {Dimopoulos}, \citenamefont {Prokopec},\ and\ \citenamefont
  {Tornkvist}}]{Davis:2000zp}%
  \BibitemOpen
  \bibfield  {author} {\bibinfo {author} {\bibfnamefont {A.-C.}\ \bibnamefont
  {Davis}}, \bibinfo {author} {\bibfnamefont {K.}~\bibnamefont {Dimopoulos}},
  \bibinfo {author} {\bibfnamefont {T.}~\bibnamefont {Prokopec}},\ and\
  \bibinfo {author} {\bibfnamefont {O.}~\bibnamefont {Tornkvist}},\ }\href
  {https://doi.org/10.1016/S0370-2693(01)00138-1} {\bibfield  {journal}
  {\bibinfo  {journal} {Phys.\ Lett.\ B}\ }\textbf {\bibinfo {volume} {501}},\
  \bibinfo {pages} {165} (\bibinfo {year} {2001})},\ \Eprint
  {https://arxiv.org/abs/astro-ph/0007214} {arXiv:astro-ph/0007214 [astro-ph]}
  \BibitemShut {NoStop}%
\bibitem [{\citenamefont {Törnkvist}\ \emph {et~al.}(2001)\citenamefont
  {Törnkvist}, \citenamefont {Davis}, \citenamefont {Dimopoulos},\ and\
  \citenamefont {Prokopec}}]{Tornkvist:2000js}%
  \BibitemOpen
  \bibfield  {author} {\bibinfo {author} {\bibfnamefont {O.}~\bibnamefont
  {Törnkvist}}, \bibinfo {author} {\bibfnamefont {A.-C.}\ \bibnamefont
  {Davis}}, \bibinfo {author} {\bibfnamefont {K.}~\bibnamefont {Dimopoulos}},\
  and\ \bibinfo {author} {\bibfnamefont {T.}~\bibnamefont {Prokopec}},\ }\href
  {https://doi.org/10.1063/1.1363559} {\bibfield  {journal} {\bibinfo
  {journal} {AIP Conf.\ Proc.}\ }\textbf {\bibinfo {volume} {555}},\ \bibinfo
  {pages} {443} (\bibinfo {year} {2001})},\ \Eprint
  {https://arxiv.org/abs/astro-ph/0011278} {arXiv:astro-ph/0011278 [astrop-ph]}
  \BibitemShut {NoStop}%
\bibitem [{\citenamefont {Davis}\ and\ \citenamefont
  {Dimopoulos}(2005)}]{Davis:2005ih}%
  \BibitemOpen
  \bibfield  {author} {\bibinfo {author} {\bibfnamefont {A.-C.}\ \bibnamefont
  {Davis}}\ and\ \bibinfo {author} {\bibfnamefont {K.}~\bibnamefont
  {Dimopoulos}},\ }\href {https://doi.org/10.1103/PhysRevD.72.043517}
  {\bibfield  {journal} {\bibinfo  {journal} {Phys. Rev.}\ }\textbf {\bibinfo
  {volume} {D72}},\ \bibinfo {pages} {043517} (\bibinfo {year} {2005})},\
  \Eprint {https://arxiv.org/abs/hep-ph/0505242} {arXiv:hep-ph/0505242
  [hep-ph]} \BibitemShut {NoStop}%
\bibitem [{\citenamefont {Anber}\ and\ \citenamefont
  {Sorbo}(2006)}]{Anber:2006xt}%
  \BibitemOpen
  \bibfield  {author} {\bibinfo {author} {\bibfnamefont {M.~M.}\ \bibnamefont
  {Anber}}\ and\ \bibinfo {author} {\bibfnamefont {L.}~\bibnamefont {Sorbo}},\
  }\href {https://doi.org/10.1088/1475-7516/2006/10/018} {\bibfield  {journal}
  {\bibinfo  {journal} {JCAP}\ }\textbf {\bibinfo {volume} {10}},\ \bibinfo
  {pages} {018}},\ \Eprint {https://arxiv.org/abs/astro-ph/0606534}
  {arXiv:astro-ph/0606534 [astro-ph]} \BibitemShut {NoStop}%
\bibitem [{\citenamefont {Beltran~Jimenez}\ and\ \citenamefont
  {Maroto}(2011{\natexlab{a}})}]{Jimenez:2010hu}%
  \BibitemOpen
  \bibfield  {author} {\bibinfo {author} {\bibfnamefont {J.}~\bibnamefont
  {Beltran~Jimenez}}\ and\ \bibinfo {author} {\bibfnamefont {A.~L.}\
  \bibnamefont {Maroto}},\ }\href {https://doi.org/10.1103/PhysRevD.83.023514}
  {\bibfield  {journal} {\bibinfo  {journal} {Phys. Rev.}\ }\textbf {\bibinfo
  {volume} {D83}},\ \bibinfo {pages} {023514} (\bibinfo {year}
  {2011}{\natexlab{a}})},\ \Eprint {https://arxiv.org/abs/1010.3960}
  {arXiv:1010.3960 [astro-ph.CO]} \BibitemShut {NoStop}%
\bibitem [{\citenamefont {Das}\ and\ \citenamefont
  {Mohanty}(2012)}]{Das:2010ywa}%
  \BibitemOpen
  \bibfield  {author} {\bibinfo {author} {\bibfnamefont {M.}~\bibnamefont
  {Das}}\ and\ \bibinfo {author} {\bibfnamefont {S.}~\bibnamefont {Mohanty}},\
  }\href {https://doi.org/10.1142/S0217751X12500406} {\bibfield  {journal}
  {\bibinfo  {journal} {Int. J. Mod. Phys.}\ }\textbf {\bibinfo {volume}
  {A27}},\ \bibinfo {pages} {1250040} (\bibinfo {year} {2012})},\ \Eprint
  {https://arxiv.org/abs/1004.1927} {arXiv:1004.1927 [astro-ph.CO]}
  \BibitemShut {NoStop}%
\bibitem [{\citenamefont {Beltran~Jimenez}\ and\ \citenamefont
  {Maroto}(2011{\natexlab{b}})}]{BeltranJimenez:2011vn}%
  \BibitemOpen
  \bibfield  {author} {\bibinfo {author} {\bibfnamefont {J.}~\bibnamefont
  {Beltran~Jimenez}}\ and\ \bibinfo {author} {\bibfnamefont {A.~L.}\
  \bibnamefont {Maroto}},\ }\href
  {https://doi.org/10.1088/1742-6596/314/1/012089} {\bibfield  {journal}
  {\bibinfo  {journal} {J.\ Phys.\ Conf.\ Ser.}\ }\textbf {\bibinfo {volume}
  {314}},\ \bibinfo {pages} {012089} (\bibinfo {year} {2011}{\natexlab{b}})},\
  \Eprint {https://arxiv.org/abs/1101.1763} {arXiv:1101.1763 [astro-ph.CO]}
  \BibitemShut {NoStop}%
\bibitem [{\citenamefont {Bonvin}\ \emph {et~al.}(2012)\citenamefont {Bonvin},
  \citenamefont {Caprini},\ and\ \citenamefont {Durrer}}]{Bonvin:2011dt}%
  \BibitemOpen
  \bibfield  {author} {\bibinfo {author} {\bibfnamefont {C.}~\bibnamefont
  {Bonvin}}, \bibinfo {author} {\bibfnamefont {C.}~\bibnamefont {Caprini}},\
  and\ \bibinfo {author} {\bibfnamefont {R.}~\bibnamefont {Durrer}},\ }\href
  {https://doi.org/10.1103/PhysRevD.86.023519} {\bibfield  {journal} {\bibinfo
  {journal} {Phys. Rev.}\ }\textbf {\bibinfo {volume} {D86}},\ \bibinfo {pages}
  {023519} (\bibinfo {year} {2012})},\ \Eprint
  {https://arxiv.org/abs/1112.3901} {arXiv:1112.3901 [astro-ph.CO]}
  \BibitemShut {NoStop}%
\bibitem [{\citenamefont {Elizalde}\ and\ \citenamefont
  {Skalozub}(2012)}]{Elizalde:2012kz}%
  \BibitemOpen
  \bibfield  {author} {\bibinfo {author} {\bibfnamefont {E.}~\bibnamefont
  {Elizalde}}\ and\ \bibinfo {author} {\bibfnamefont {V.}~\bibnamefont
  {Skalozub}},\ }\href {https://doi.org/10.1140/epjc/s10052-012-1968-3}
  {\bibfield  {journal} {\bibinfo  {journal} {Eur. Phys. J.}\ }\textbf
  {\bibinfo {volume} {C72}},\ \bibinfo {pages} {1968} (\bibinfo {year}
  {2012})},\ \Eprint {https://arxiv.org/abs/1202.3895} {arXiv:1202.3895
  [hep-ph]} \BibitemShut {NoStop}%
\bibitem [{\citenamefont {Bonvin}\ \emph {et~al.}(2013)\citenamefont {Bonvin},
  \citenamefont {Caprini},\ and\ \citenamefont {Durrer}}]{Bonvin:2013tba}%
  \BibitemOpen
  \bibfield  {author} {\bibinfo {author} {\bibfnamefont {C.}~\bibnamefont
  {Bonvin}}, \bibinfo {author} {\bibfnamefont {C.}~\bibnamefont {Caprini}},\
  and\ \bibinfo {author} {\bibfnamefont {R.}~\bibnamefont {Durrer}},\ }\href
  {https://doi.org/10.1103/PhysRevD.88.083515} {\bibfield  {journal} {\bibinfo
  {journal} {Phys. Rev.}\ }\textbf {\bibinfo {volume} {D88}},\ \bibinfo {pages}
  {083515} (\bibinfo {year} {2013})},\ \Eprint
  {https://arxiv.org/abs/1308.3348} {arXiv:1308.3348 [astro-ph.CO]}
  \BibitemShut {NoStop}%
\bibitem [{\citenamefont {Caprini}\ and\ \citenamefont
  {Sorbo}(2014)}]{Caprini:2014mja}%
  \BibitemOpen
  \bibfield  {author} {\bibinfo {author} {\bibfnamefont {C.}~\bibnamefont
  {Caprini}}\ and\ \bibinfo {author} {\bibfnamefont {L.}~\bibnamefont
  {Sorbo}},\ }\href {https://doi.org/10.1088/1475-7516/2014/10/056} {\bibfield
  {journal} {\bibinfo  {journal} {JCAP}\ }\textbf {\bibinfo {volume}
  {1410}}\bibfield  {number} {\bibinfo  {number} { (10)},\ \bibinfo {pages}
  {056}},\ }\Eprint {https://arxiv.org/abs/1407.2809} {arXiv:1407.2809
  [astro-ph.CO]} \BibitemShut {NoStop}%
\bibitem [{\citenamefont {Choudhury}(2015)}]{Choudhury:2015jaa}%
  \BibitemOpen
  \bibfield  {author} {\bibinfo {author} {\bibfnamefont {S.}~\bibnamefont
  {Choudhury}},\ }\href {https://doi.org/10.1007/JHEP10(2015)095} {\bibfield
  {journal} {\bibinfo  {journal} {JHEP}\ }\textbf {\bibinfo {volume} {10}},\
  \bibinfo {pages} {095}},\ \Eprint {https://arxiv.org/abs/1504.08206}
  {arXiv:1504.08206 [astro-ph.CO]} \BibitemShut {NoStop}%
\bibitem [{\citenamefont {Sharma}\ \emph {et~al.}(2017)\citenamefont {Sharma},
  \citenamefont {Jagannathan}, \citenamefont {Seshadri},\ and\ \citenamefont
  {Subramanian}}]{Sharma:2017eps}%
  \BibitemOpen
  \bibfield  {author} {\bibinfo {author} {\bibfnamefont {R.}~\bibnamefont
  {Sharma}}, \bibinfo {author} {\bibfnamefont {S.}~\bibnamefont {Jagannathan}},
  \bibinfo {author} {\bibfnamefont {T.~R.}\ \bibnamefont {Seshadri}},\ and\
  \bibinfo {author} {\bibfnamefont {K.}~\bibnamefont {Subramanian}},\ }\href
  {https://doi.org/10.1103/PhysRevD.96.083511} {\bibfield  {journal} {\bibinfo
  {journal} {Phys. Rev.}\ }\textbf {\bibinfo {volume} {D96}},\ \bibinfo {pages}
  {083511} (\bibinfo {year} {2017})},\ \Eprint
  {https://arxiv.org/abs/1708.08119} {arXiv:1708.08119 [astro-ph.CO]}
  \BibitemShut {NoStop}%
\bibitem [{\citenamefont {Caprini}\ \emph {et~al.}(2018)\citenamefont
  {Caprini}, \citenamefont {Guzzetti},\ and\ \citenamefont
  {Sorbo}}]{Caprini:2017vnn}%
  \BibitemOpen
  \bibfield  {author} {\bibinfo {author} {\bibfnamefont {C.}~\bibnamefont
  {Caprini}}, \bibinfo {author} {\bibfnamefont {M.~C.}\ \bibnamefont
  {Guzzetti}},\ and\ \bibinfo {author} {\bibfnamefont {L.}~\bibnamefont
  {Sorbo}},\ }\href {https://doi.org/10.1088/1361-6382/aac143} {\bibfield
  {journal} {\bibinfo  {journal} {Class. Quant. Grav.}\ }\textbf {\bibinfo
  {volume} {35}},\ \bibinfo {pages} {124003} (\bibinfo {year} {2018})},\
  \Eprint {https://arxiv.org/abs/1707.09750} {arXiv:1707.09750 [astro-ph.CO]}
  \BibitemShut {NoStop}%
\bibitem [{\citenamefont {Sharma}\ \emph {et~al.}(2018)\citenamefont {Sharma},
  \citenamefont {Subramanian},\ and\ \citenamefont
  {Seshadri}}]{Sharma:2018kgs}%
  \BibitemOpen
  \bibfield  {author} {\bibinfo {author} {\bibfnamefont {R.}~\bibnamefont
  {Sharma}}, \bibinfo {author} {\bibfnamefont {K.}~\bibnamefont
  {Subramanian}},\ and\ \bibinfo {author} {\bibfnamefont {T.~R.}\ \bibnamefont
  {Seshadri}},\ }\href {https://doi.org/10.1103/PhysRevD.97.083503} {\bibfield
  {journal} {\bibinfo  {journal} {Phys. Rev.}\ }\textbf {\bibinfo {volume}
  {D97}},\ \bibinfo {pages} {083503} (\bibinfo {year} {2018})},\ \Eprint
  {https://arxiv.org/abs/1802.04847} {arXiv:1802.04847 [astro-ph.CO]}
  \BibitemShut {NoStop}%
\bibitem [{\citenamefont {Kamarpour}\ and\ \citenamefont
  {Sobol}(2018)}]{Kamarpour:2018ckk}%
  \BibitemOpen
  \bibfield  {author} {\bibinfo {author} {\bibfnamefont {M.}~\bibnamefont
  {Kamarpour}}\ and\ \bibinfo {author} {\bibfnamefont {O.}~\bibnamefont
  {Sobol}},\ }\href {https://doi.org/10.15407/ujpe63.8.673} {\bibfield
  {journal} {\bibinfo  {journal} {Ukr.\ J.\ Phys.}\ }\textbf {\bibinfo {volume}
  {63}},\ \bibinfo {pages} {673} (\bibinfo {year} {2018})}\BibitemShut
  {NoStop}%
\bibitem [{\citenamefont {Savchenko}\ and\ \citenamefont
  {Shtanov}(2018{\natexlab{a}})}]{Savchenko:2018pdr}%
  \BibitemOpen
  \bibfield  {author} {\bibinfo {author} {\bibfnamefont {O.}~\bibnamefont
  {Savchenko}}\ and\ \bibinfo {author} {\bibfnamefont {Y.}~\bibnamefont
  {Shtanov}},\ }\href {https://doi.org/10.1088/1475-7516/2018/10/040}
  {\bibfield  {journal} {\bibinfo  {journal} {JCAP}\ }\textbf {\bibinfo
  {volume} {1810}},\ \bibinfo {pages} {040}},\ \Eprint
  {https://arxiv.org/abs/1808.06193} {arXiv:1808.06193 [astro-ph.CO]}
  \BibitemShut {NoStop}%
\bibitem [{\citenamefont {Sobol}\ \emph {et~al.}(2018)\citenamefont {Sobol},
  \citenamefont {Gorbar}, \citenamefont {Kamarpour},\ and\ \citenamefont
  {Vilchinskii}}]{Sobol:2018djj}%
  \BibitemOpen
  \bibfield  {author} {\bibinfo {author} {\bibfnamefont {O.~O.}\ \bibnamefont
  {Sobol}}, \bibinfo {author} {\bibfnamefont {E.~V.}\ \bibnamefont {Gorbar}},
  \bibinfo {author} {\bibfnamefont {M.}~\bibnamefont {Kamarpour}},\ and\
  \bibinfo {author} {\bibfnamefont {S.~I.}\ \bibnamefont {Vilchinskii}},\
  }\href {https://doi.org/10.1103/PhysRevD.98.063534} {\bibfield  {journal}
  {\bibinfo  {journal} {Phys. Rev.}\ }\textbf {\bibinfo {volume} {D98}},\
  \bibinfo {pages} {063534} (\bibinfo {year} {2018})},\ \Eprint
  {https://arxiv.org/abs/1807.09851} {arXiv:1807.09851 [hep-ph]} \BibitemShut
  {NoStop}%
\bibitem [{\citenamefont {Subramanian}(2019)}]{Subramanian:2019jyd}%
  \BibitemOpen
  \bibfield  {author} {\bibinfo {author} {\bibfnamefont {K.}~\bibnamefont
  {Subramanian}},\ }\href {https://doi.org/10.3390/galaxies7020047} {\bibfield
  {journal} {\bibinfo  {journal} {Galaxies}\ }\textbf {\bibinfo {volume} {7}},\
  \bibinfo {pages} {47} (\bibinfo {year} {2019})},\ \Eprint
  {https://arxiv.org/abs/1903.03744} {arXiv:1903.03744 [astro-ph.CO]}
  \BibitemShut {NoStop}%
\bibitem [{\citenamefont {Patel}\ \emph {et~al.}(2020)\citenamefont {Patel},
  \citenamefont {Tashiro},\ and\ \citenamefont {Urakawa}}]{Patel:2019isj}%
  \BibitemOpen
  \bibfield  {author} {\bibinfo {author} {\bibfnamefont {T.}~\bibnamefont
  {Patel}}, \bibinfo {author} {\bibfnamefont {H.}~\bibnamefont {Tashiro}},\
  and\ \bibinfo {author} {\bibfnamefont {Y.}~\bibnamefont {Urakawa}},\ }\href
  {https://doi.org/10.1088/1475-7516/2020/01/043} {\bibfield  {journal}
  {\bibinfo  {journal} {JCAP}\ }\textbf {\bibinfo {volume} {01}}\bibfield
  {number} {\bibinfo  {number} { (01)},\ \bibinfo {pages} {043}},\ }\Eprint
  {https://arxiv.org/abs/1909.00288} {arXiv:1909.00288 [astro-ph.CO]}
  \BibitemShut {NoStop}%
\bibitem [{\citenamefont {Sobol}\ \emph {et~al.}(2019)\citenamefont {Sobol},
  \citenamefont {Gorbar},\ and\ \citenamefont {Vilchinskii}}]{Sobol:2019xls}%
  \BibitemOpen
  \bibfield  {author} {\bibinfo {author} {\bibfnamefont {O.~O.}\ \bibnamefont
  {Sobol}}, \bibinfo {author} {\bibfnamefont {E.~V.}\ \bibnamefont {Gorbar}},\
  and\ \bibinfo {author} {\bibfnamefont {S.~I.}\ \bibnamefont {Vilchinskii}},\
  }\href {https://doi.org/10.1103/PhysRevD.100.063523} {\bibfield  {journal}
  {\bibinfo  {journal} {Phys. Rev.}\ }\textbf {\bibinfo {volume} {D100}},\
  \bibinfo {pages} {063523} (\bibinfo {year} {2019})},\ \Eprint
  {https://arxiv.org/abs/1907.10443} {arXiv:1907.10443 [astro-ph.CO]}
  \BibitemShut {NoStop}%
\bibitem [{\citenamefont {Fujita}\ and\ \citenamefont
  {Durrer}(2019)}]{Fujita:2019pmi}%
  \BibitemOpen
  \bibfield  {author} {\bibinfo {author} {\bibfnamefont {T.}~\bibnamefont
  {Fujita}}\ and\ \bibinfo {author} {\bibfnamefont {R.}~\bibnamefont
  {Durrer}},\ }\href {https://doi.org/10.1088/1475-7516/2019/09/008} {\bibfield
   {journal} {\bibinfo  {journal} {JCAP}\ }\textbf {\bibinfo {volume} {1909}},\
  \bibinfo {pages} {008}},\ \Eprint {https://arxiv.org/abs/1904.11428}
  {arXiv:1904.11428 [astro-ph.CO]} \BibitemShut {NoStop}%
\bibitem [{\citenamefont {Shakeri}\ \emph {et~al.}(2019)\citenamefont
  {Shakeri}, \citenamefont {Gorji},\ and\ \citenamefont
  {Firouzjahi}}]{Shakeri:2019mnt}%
  \BibitemOpen
  \bibfield  {author} {\bibinfo {author} {\bibfnamefont {S.}~\bibnamefont
  {Shakeri}}, \bibinfo {author} {\bibfnamefont {M.~A.}\ \bibnamefont {Gorji}},\
  and\ \bibinfo {author} {\bibfnamefont {H.}~\bibnamefont {Firouzjahi}},\
  }\href {https://doi.org/10.1103/PhysRevD.99.103525} {\bibfield  {journal}
  {\bibinfo  {journal} {Phys. Rev.}\ }\textbf {\bibinfo {volume} {D99}},\
  \bibinfo {pages} {103525} (\bibinfo {year} {2019})},\ \Eprint
  {https://arxiv.org/abs/1903.05310} {arXiv:1903.05310 [hep-th]} \BibitemShut
  {NoStop}%
\bibitem [{\citenamefont {Kobayashi}\ and\ \citenamefont
  {Sloth}(2019)}]{Kobayashi:2019uqs}%
  \BibitemOpen
  \bibfield  {author} {\bibinfo {author} {\bibfnamefont {T.}~\bibnamefont
  {Kobayashi}}\ and\ \bibinfo {author} {\bibfnamefont {M.~S.}\ \bibnamefont
  {Sloth}},\ }\href {https://doi.org/10.1103/PhysRevD.100.023524} {\bibfield
  {journal} {\bibinfo  {journal} {Phys. Rev.}\ }\textbf {\bibinfo {volume}
  {D100}},\ \bibinfo {pages} {023524} (\bibinfo {year} {2019})},\ \Eprint
  {https://arxiv.org/abs/1903.02561} {arXiv:1903.02561 [astro-ph.CO]}
  \BibitemShut {NoStop}%
\bibitem [{\citenamefont {Shtanov}(2019)}]{Shtanov:2019civ}%
  \BibitemOpen
  \bibfield  {author} {\bibinfo {author} {\bibfnamefont {Y.}~\bibnamefont
  {Shtanov}},\ }\href {https://doi.org/10.1088/1475-7516/2019/10/008}
  {\bibfield  {journal} {\bibinfo  {journal} {JCAP}\ }\textbf {\bibinfo
  {volume} {1910}}\bibfield  {number} {\bibinfo  {number} { (10)},\ \bibinfo
  {pages} {008}},\ }\Eprint {https://arxiv.org/abs/1902.05894}
  {arXiv:1902.05894 [astro-ph.CO]} \BibitemShut {NoStop}%
\bibitem [{\citenamefont {Sharma}\ \emph {et~al.}(2019)\citenamefont {Sharma},
  \citenamefont {Subramanian},\ and\ \citenamefont
  {Seshadri}}]{Sharma:2019jtb}%
  \BibitemOpen
  \bibfield  {author} {\bibinfo {author} {\bibfnamefont {R.}~\bibnamefont
  {Sharma}}, \bibinfo {author} {\bibfnamefont {K.}~\bibnamefont
  {Subramanian}},\ and\ \bibinfo {author} {\bibfnamefont {T.~R.}\ \bibnamefont
  {Seshadri}},\ }\href@noop {} {\  (\bibinfo {year} {2019})},\ \Eprint
  {https://arxiv.org/abs/1912.12089} {arXiv:1912.12089 [astro-ph.CO]}
  \BibitemShut {NoStop}%
\bibitem [{\citenamefont {Battefeld}\ and\ \citenamefont
  {Brandenberger}(2004)}]{Battefeld:2004cd}%
  \BibitemOpen
  \bibfield  {author} {\bibinfo {author} {\bibfnamefont {T.~J.}\ \bibnamefont
  {Battefeld}}\ and\ \bibinfo {author} {\bibfnamefont {R.}~\bibnamefont
  {Brandenberger}},\ }\href {https://doi.org/10.1103/PhysRevD.70.121302}
  {\bibfield  {journal} {\bibinfo  {journal} {Phys. Rev.}\ }\textbf {\bibinfo
  {volume} {D70}},\ \bibinfo {pages} {121302} (\bibinfo {year} {2004})},\
  \Eprint {https://arxiv.org/abs/hep-th/0406180} {arXiv:hep-th/0406180
  [hep-th]} \BibitemShut {NoStop}%
\bibitem [{\citenamefont {Salim}\ \emph {et~al.}(2007)\citenamefont {Salim},
  \citenamefont {Souza}, \citenamefont {Perez~Bergliaffa},\ and\ \citenamefont
  {Prokopec}}]{Salim:2006nw}%
  \BibitemOpen
  \bibfield  {author} {\bibinfo {author} {\bibfnamefont {J.~M.}\ \bibnamefont
  {Salim}}, \bibinfo {author} {\bibfnamefont {N.}~\bibnamefont {Souza}},
  \bibinfo {author} {\bibfnamefont {S.~E.}\ \bibnamefont {Perez~Bergliaffa}},\
  and\ \bibinfo {author} {\bibfnamefont {T.}~\bibnamefont {Prokopec}},\ }\href
  {https://doi.org/10.1088/1475-7516/2007/04/011} {\bibfield  {journal}
  {\bibinfo  {journal} {JCAP}\ }\textbf {\bibinfo {volume} {0704}},\ \bibinfo
  {pages} {011}},\ \Eprint {https://arxiv.org/abs/astro-ph/0612281}
  {arXiv:astro-ph/0612281 [astro-ph]} \BibitemShut {NoStop}%
\bibitem [{\citenamefont {Membiela}(2014)}]{Membiela2013}%
  \BibitemOpen
  \bibfield  {author} {\bibinfo {author} {\bibfnamefont {F.~A.}\ \bibnamefont
  {Membiela}},\ }\href {https://doi.org/10.1016/j.nuclphysb.2014.05.018}
  {\bibfield  {journal} {\bibinfo  {journal} {Nucl. Phys.}\ }\textbf {\bibinfo
  {volume} {B885}},\ \bibinfo {pages} {196} (\bibinfo {year} {2014})},\ \Eprint
  {https://arxiv.org/abs/1312.2162} {arXiv:1312.2162 [astro-ph.CO]}
  \BibitemShut {NoStop}%
\bibitem [{\citenamefont {Sriramkumar}\ \emph {et~al.}(2015)\citenamefont
  {Sriramkumar}, \citenamefont {Atmjeet},\ and\ \citenamefont
  {Jain}}]{Sriramkumar:2015yza}%
  \BibitemOpen
  \bibfield  {author} {\bibinfo {author} {\bibfnamefont {L.}~\bibnamefont
  {Sriramkumar}}, \bibinfo {author} {\bibfnamefont {K.}~\bibnamefont
  {Atmjeet}},\ and\ \bibinfo {author} {\bibfnamefont {R.~K.}\ \bibnamefont
  {Jain}},\ }\href {https://doi.org/10.1088/1475-7516/2015/09/010} {\bibfield
  {journal} {\bibinfo  {journal} {JCAP}\ }\textbf {\bibinfo {volume}
  {1509}}\bibfield  {number} {\bibinfo  {number} { (09)},\ \bibinfo {pages}
  {010}},\ }\Eprint {https://arxiv.org/abs/1504.06853} {arXiv:1504.06853
  [astro-ph.CO]} \BibitemShut {NoStop}%
\bibitem [{\citenamefont {Chowdhury}\ \emph {et~al.}(2016)\citenamefont
  {Chowdhury}, \citenamefont {Sriramkumar},\ and\ \citenamefont
  {Jain}}]{Chowdhury2016}%
  \BibitemOpen
  \bibfield  {author} {\bibinfo {author} {\bibfnamefont {D.}~\bibnamefont
  {Chowdhury}}, \bibinfo {author} {\bibfnamefont {L.}~\bibnamefont
  {Sriramkumar}},\ and\ \bibinfo {author} {\bibfnamefont {R.~K.}\ \bibnamefont
  {Jain}},\ }\href {https://doi.org/10.1103/PhysRevD.94.083512} {\bibfield
  {journal} {\bibinfo  {journal} {Phys. Rev.}\ }\textbf {\bibinfo {volume}
  {D94}},\ \bibinfo {pages} {083512} (\bibinfo {year} {2016})},\ \Eprint
  {https://arxiv.org/abs/1604.02143} {arXiv:1604.02143 [gr-qc]} \BibitemShut
  {NoStop}%
\bibitem [{\citenamefont {Qian}\ \emph {et~al.}(2016)\citenamefont {Qian},
  \citenamefont {Cai}, \citenamefont {Easson},\ and\ \citenamefont
  {Guo}}]{Qian2016}%
  \BibitemOpen
  \bibfield  {author} {\bibinfo {author} {\bibfnamefont {P.}~\bibnamefont
  {Qian}}, \bibinfo {author} {\bibfnamefont {Y.-F.}\ \bibnamefont {Cai}},
  \bibinfo {author} {\bibfnamefont {D.~A.}\ \bibnamefont {Easson}},\ and\
  \bibinfo {author} {\bibfnamefont {Z.-K.}\ \bibnamefont {Guo}},\ }\href
  {https://doi.org/10.1103/PhysRevD.94.083524} {\bibfield  {journal} {\bibinfo
  {journal} {Phys. Rev.}\ }\textbf {\bibinfo {volume} {D94}},\ \bibinfo {pages}
  {083524} (\bibinfo {year} {2016})},\ \Eprint
  {https://arxiv.org/abs/1607.06578} {arXiv:1607.06578 [gr-qc]} \BibitemShut
  {NoStop}%
\bibitem [{\citenamefont {Koley}\ and\ \citenamefont
  {Samtani}(2017)}]{Koley:2016jdw}%
  \BibitemOpen
  \bibfield  {author} {\bibinfo {author} {\bibfnamefont {R.}~\bibnamefont
  {Koley}}\ and\ \bibinfo {author} {\bibfnamefont {S.}~\bibnamefont
  {Samtani}},\ }\href {https://doi.org/10.1088/1475-7516/2017/04/030}
  {\bibfield  {journal} {\bibinfo  {journal} {JCAP}\ }\textbf {\bibinfo
  {volume} {1704}}\bibfield  {number} {\bibinfo  {number} { (04)},\ \bibinfo
  {pages} {030}},\ }\Eprint {https://arxiv.org/abs/1612.08556}
  {arXiv:1612.08556 [gr-qc]} \BibitemShut {NoStop}%
\bibitem [{\citenamefont {Chen}\ \emph {et~al.}(2018)\citenamefont {Chen},
  \citenamefont {Li}, \citenamefont {Li},\ and\ \citenamefont
  {Zhu}}]{Chen:2017cjx}%
  \BibitemOpen
  \bibfield  {author} {\bibinfo {author} {\bibfnamefont {J.-W.}\ \bibnamefont
  {Chen}}, \bibinfo {author} {\bibfnamefont {C.-H.}\ \bibnamefont {Li}},
  \bibinfo {author} {\bibfnamefont {Y.-B.}\ \bibnamefont {Li}},\ and\ \bibinfo
  {author} {\bibfnamefont {M.}~\bibnamefont {Zhu}},\ }\href
  {https://doi.org/10.1007/s11433-018-9211-5} {\bibfield  {journal} {\bibinfo
  {journal} {Sci. China Phys. Mech. Astron.}\ }\textbf {\bibinfo {volume}
  {61}},\ \bibinfo {pages} {100411} (\bibinfo {year} {2018})},\ \Eprint
  {https://arxiv.org/abs/1711.10897} {arXiv:1711.10897 [physics.gen-ph]}
  \BibitemShut {NoStop}%
\bibitem [{\citenamefont {Leite}\ and\ \citenamefont
  {Pavlović}(2018)}]{Leite:2018bbo}%
  \BibitemOpen
  \bibfield  {author} {\bibinfo {author} {\bibfnamefont {N.}~\bibnamefont
  {Leite}}\ and\ \bibinfo {author} {\bibfnamefont {P.}~\bibnamefont
  {Pavlović}},\ }\href {https://doi.org/10.1088/1361-6382/aae2d6} {\bibfield
  {journal} {\bibinfo  {journal} {Class. Quant. Grav.}\ }\textbf {\bibinfo
  {volume} {35}},\ \bibinfo {pages} {215005} (\bibinfo {year} {2018})},\
  \Eprint {https://arxiv.org/abs/1805.06036} {arXiv:1805.06036 [gr-qc]}
  \BibitemShut {NoStop}%
\bibitem [{\citenamefont {Chowdhury}\ \emph {et~al.}(2019)\citenamefont
  {Chowdhury}, \citenamefont {Sriramkumar},\ and\ \citenamefont
  {Kamionkowski}}]{Chowdhury:2018blx}%
  \BibitemOpen
  \bibfield  {author} {\bibinfo {author} {\bibfnamefont {D.}~\bibnamefont
  {Chowdhury}}, \bibinfo {author} {\bibfnamefont {L.}~\bibnamefont
  {Sriramkumar}},\ and\ \bibinfo {author} {\bibfnamefont {M.}~\bibnamefont
  {Kamionkowski}},\ }\href {https://doi.org/10.1088/1475-7516/2019/01/048}
  {\bibfield  {journal} {\bibinfo  {journal} {JCAP}\ }\textbf {\bibinfo
  {volume} {1901}},\ \bibinfo {pages} {048}},\ \Eprint
  {https://arxiv.org/abs/1807.05530} {arXiv:1807.05530 [astro-ph.CO]}
  \BibitemShut {NoStop}%
\bibitem [{\citenamefont {Barrie}(2020)}]{Barrie:2020kpt}%
  \BibitemOpen
  \bibfield  {author} {\bibinfo {author} {\bibfnamefont {N.~D.}\ \bibnamefont
  {Barrie}},\ }\href@noop {} {\  (\bibinfo {year} {2020})},\ \Eprint
  {https://arxiv.org/abs/2001.04773} {arXiv:2001.04773 [hep-ph]} \BibitemShut
  {NoStop}%
\bibitem [{\citenamefont {Grasso}\ and\ \citenamefont
  {Rubinstein}(2001)}]{Grasso:2000wj}%
  \BibitemOpen
  \bibfield  {author} {\bibinfo {author} {\bibfnamefont {D.}~\bibnamefont
  {Grasso}}\ and\ \bibinfo {author} {\bibfnamefont {H.~R.}\ \bibnamefont
  {Rubinstein}},\ }\href {https://doi.org/10.1016/S0370-1573(00)00110-1}
  {\bibfield  {journal} {\bibinfo  {journal} {Phys.\ Rept.}\ }\textbf {\bibinfo
  {volume} {348}},\ \bibinfo {pages} {163} (\bibinfo {year} {2001})},\ \Eprint
  {https://arxiv.org/abs/astro-ph/0009061} {arXiv:astro-ph/0009061 [astro-ph]}
  \BibitemShut {NoStop}%
\bibitem [{\citenamefont {Carrilho}\ and\ \citenamefont
  {Malik}(2019)}]{Carrilho:2019qlb}%
  \BibitemOpen
  \bibfield  {author} {\bibinfo {author} {\bibfnamefont {P.}~\bibnamefont
  {Carrilho}}\ and\ \bibinfo {author} {\bibfnamefont {K.~A.}\ \bibnamefont
  {Malik}},\ }\href {https://doi.org/10.1088/1475-7516/2019/04/028} {\bibfield
  {journal} {\bibinfo  {journal} {JCAP}\ }\textbf {\bibinfo {volume} {1904}},\
  \bibinfo {pages} {028}},\ \Eprint {https://arxiv.org/abs/1902.00459}
  {arXiv:1902.00459 [astro-ph.CO]} \BibitemShut {NoStop}%
\bibitem [{\citenamefont {Enqvist}\ \emph {et~al.}(2004)\citenamefont
  {Enqvist}, \citenamefont {Jokinen},\ and\ \citenamefont
  {Mazumdar}}]{Enqvist:2004yy}%
  \BibitemOpen
  \bibfield  {author} {\bibinfo {author} {\bibfnamefont {K.}~\bibnamefont
  {Enqvist}}, \bibinfo {author} {\bibfnamefont {A.}~\bibnamefont {Jokinen}},\
  and\ \bibinfo {author} {\bibfnamefont {A.}~\bibnamefont {Mazumdar}},\ }\href
  {https://doi.org/10.1088/1475-7516/2004/11/001} {\bibfield  {journal}
  {\bibinfo  {journal} {JCAP}\ }\textbf {\bibinfo {volume} {0411}},\ \bibinfo
  {pages} {001}},\ \Eprint {https://arxiv.org/abs/hep-ph/0404269}
  {arXiv:hep-ph/0404269 [hep-ph]} \BibitemShut {NoStop}%
\bibitem [{\citenamefont {Emami}\ \emph {et~al.}(2010)\citenamefont {Emami},
  \citenamefont {Firouzjahi},\ and\ \citenamefont {Movahed}}]{Emami2009}%
  \BibitemOpen
  \bibfield  {author} {\bibinfo {author} {\bibfnamefont {R.}~\bibnamefont
  {Emami}}, \bibinfo {author} {\bibfnamefont {H.}~\bibnamefont {Firouzjahi}},\
  and\ \bibinfo {author} {\bibfnamefont {M.~S.}\ \bibnamefont {Movahed}},\
  }\href {https://doi.org/10.1103/PhysRevD.81.083526} {\bibfield  {journal}
  {\bibinfo  {journal} {Phys. Rev.}\ }\textbf {\bibinfo {volume} {D81}},\
  \bibinfo {pages} {083526} (\bibinfo {year} {2010})},\ \Eprint
  {https://arxiv.org/abs/0908.4161} {arXiv:0908.4161 [hep-th]} \BibitemShut
  {NoStop}%
\bibitem [{\citenamefont {Adshead}\ \emph {et~al.}(2016)\citenamefont
  {Adshead}, \citenamefont {Giblin}, \citenamefont {Scully},\ and\
  \citenamefont {Sfakianakis}}]{Adshead:2016iae}%
  \BibitemOpen
  \bibfield  {author} {\bibinfo {author} {\bibfnamefont {P.}~\bibnamefont
  {Adshead}}, \bibinfo {author} {\bibfnamefont {J.~T.}\ \bibnamefont {Giblin}},
  \bibinfo {author} {\bibfnamefont {T.~R.}\ \bibnamefont {Scully}},\ and\
  \bibinfo {author} {\bibfnamefont {E.~I.}\ \bibnamefont {Sfakianakis}},\
  }\href {https://doi.org/10.1088/1475-7516/2016/10/039} {\bibfield  {journal}
  {\bibinfo  {journal} {JCAP}\ }\textbf {\bibinfo {volume} {10}},\ \bibinfo
  {pages} {039}},\ \Eprint {https://arxiv.org/abs/1606.08474} {arXiv:1606.08474
  [astro-ph.CO]} \BibitemShut {NoStop}%
\bibitem [{\citenamefont {Turner}\ and\ \citenamefont
  {Widrow}(1988)}]{Turner1988}%
  \BibitemOpen
  \bibfield  {author} {\bibinfo {author} {\bibfnamefont {M.~S.}\ \bibnamefont
  {Turner}}\ and\ \bibinfo {author} {\bibfnamefont {L.~M.}\ \bibnamefont
  {Widrow}},\ }\href {https://doi.org/10.1103/PhysRevD.37.2743} {\bibfield
  {journal} {\bibinfo  {journal} {Phys. Rev. D}\ }\textbf {\bibinfo {volume}
  {37}},\ \bibinfo {pages} {2743} (\bibinfo {year} {1988})}\BibitemShut
  {NoStop}%
\bibitem [{\citenamefont {Bamba}\ and\ \citenamefont
  {Sasaki}(2007)}]{Bamba2006}%
  \BibitemOpen
  \bibfield  {author} {\bibinfo {author} {\bibfnamefont {K.}~\bibnamefont
  {Bamba}}\ and\ \bibinfo {author} {\bibfnamefont {M.}~\bibnamefont {Sasaki}},\
  }\href {https://doi.org/10.1088/1475-7516/2007/02/030} {\bibfield  {journal}
  {\bibinfo  {journal} {JCAP}\ }\textbf {\bibinfo {volume} {0702}},\ \bibinfo
  {pages} {030}},\ \Eprint {https://arxiv.org/abs/astro-ph/0611701}
  {arXiv:astro-ph/0611701 [astro-ph]} \BibitemShut {NoStop}%
\bibitem [{\citenamefont {Campanelli}\ \emph {et~al.}(2008)\citenamefont
  {Campanelli}, \citenamefont {Cea}, \citenamefont {Fogli},\ and\ \citenamefont
  {Tedesco}}]{Campanelli2008}%
  \BibitemOpen
  \bibfield  {author} {\bibinfo {author} {\bibfnamefont {L.}~\bibnamefont
  {Campanelli}}, \bibinfo {author} {\bibfnamefont {P.}~\bibnamefont {Cea}},
  \bibinfo {author} {\bibfnamefont {G.~L.}\ \bibnamefont {Fogli}},\ and\
  \bibinfo {author} {\bibfnamefont {L.}~\bibnamefont {Tedesco}},\ }\href
  {https://doi.org/10.1103/PhysRevD.77.123002} {\bibfield  {journal} {\bibinfo
  {journal} {Phys. Rev.}\ }\textbf {\bibinfo {volume} {D77}},\ \bibinfo {pages}
  {123002} (\bibinfo {year} {2008})},\ \Eprint
  {https://arxiv.org/abs/0802.2630} {arXiv:0802.2630 [astro-ph]} \BibitemShut
  {NoStop}%
\bibitem [{\citenamefont {Kunze}(2010)}]{Kunze2009}%
  \BibitemOpen
  \bibfield  {author} {\bibinfo {author} {\bibfnamefont {K.~E.}\ \bibnamefont
  {Kunze}},\ }\href {https://doi.org/10.1103/PhysRevD.81.043526} {\bibfield
  {journal} {\bibinfo  {journal} {Phys. Rev.}\ }\textbf {\bibinfo {volume}
  {D81}},\ \bibinfo {pages} {043526} (\bibinfo {year} {2010})},\ \Eprint
  {https://arxiv.org/abs/0911.1101} {arXiv:0911.1101 [astro-ph.CO]}
  \BibitemShut {NoStop}%
\bibitem [{\citenamefont {Kunze}(2013)}]{Kunze2012}%
  \BibitemOpen
  \bibfield  {author} {\bibinfo {author} {\bibfnamefont {K.~E.}\ \bibnamefont
  {Kunze}},\ }\href {https://doi.org/10.1103/PhysRevD.87.063505} {\bibfield
  {journal} {\bibinfo  {journal} {Phys. Rev.}\ }\textbf {\bibinfo {volume}
  {D87}},\ \bibinfo {pages} {063505} (\bibinfo {year} {2013})},\ \Eprint
  {https://arxiv.org/abs/1210.6899} {arXiv:1210.6899 [astro-ph.CO]}
  \BibitemShut {NoStop}%
\bibitem [{\citenamefont {Savchenko}\ and\ \citenamefont
  {Shtanov}(2018{\natexlab{b}})}]{Savchenko2018}%
  \BibitemOpen
  \bibfield  {author} {\bibinfo {author} {\bibfnamefont {O.}~\bibnamefont
  {Savchenko}}\ and\ \bibinfo {author} {\bibfnamefont {Y.}~\bibnamefont
  {Shtanov}},\ }\href {https://doi.org/10.1088/1475-7516/2018/10/040}
  {\bibfield  {journal} {\bibinfo  {journal} {JCAP}\ }\textbf {\bibinfo
  {volume} {1810}},\ \bibinfo {pages} {040}},\ \Eprint
  {https://arxiv.org/abs/1808.06193} {arXiv:1808.06193 [astro-ph.CO]}
  \BibitemShut {NoStop}%
\bibitem [{\citenamefont {Subramanian}(2016)}]{Subramanian:2015lua}%
  \BibitemOpen
  \bibfield  {author} {\bibinfo {author} {\bibfnamefont {K.}~\bibnamefont
  {Subramanian}},\ }\href {https://doi.org/10.1088/0034-4885/79/7/076901}
  {\bibfield  {journal} {\bibinfo  {journal} {Rept. Prog. Phys.}\ }\textbf
  {\bibinfo {volume} {79}},\ \bibinfo {pages} {076901} (\bibinfo {year}
  {2016})},\ \Eprint {https://arxiv.org/abs/1504.02311} {arXiv:1504.02311
  [astro-ph.CO]} \BibitemShut {NoStop}%
\bibitem [{\citenamefont {Demozzi}\ \emph {et~al.}(2009)\citenamefont
  {Demozzi}, \citenamefont {Mukhanov},\ and\ \citenamefont
  {Rubinstein}}]{Demozzi2009}%
  \BibitemOpen
  \bibfield  {author} {\bibinfo {author} {\bibfnamefont {V.}~\bibnamefont
  {Demozzi}}, \bibinfo {author} {\bibfnamefont {V.}~\bibnamefont {Mukhanov}},\
  and\ \bibinfo {author} {\bibfnamefont {H.}~\bibnamefont {Rubinstein}},\
  }\href {https://doi.org/10.1088/1475-7516/2009/08/025} {\bibfield  {journal}
  {\bibinfo  {journal} {JCAP}\ }\textbf {\bibinfo {volume} {0908}},\ \bibinfo
  {pages} {025}},\ \Eprint {https://arxiv.org/abs/0907.1030} {arXiv:0907.1030
  [astro-ph.CO]} \BibitemShut {NoStop}%
\bibitem [{\citenamefont {Green}\ and\ \citenamefont
  {Kobayashi}(2016)}]{Green2015}%
  \BibitemOpen
  \bibfield  {author} {\bibinfo {author} {\bibfnamefont {D.}~\bibnamefont
  {Green}}\ and\ \bibinfo {author} {\bibfnamefont {T.}~\bibnamefont
  {Kobayashi}},\ }\href {https://doi.org/10.1088/1475-7516/2016/03/010}
  {\bibfield  {journal} {\bibinfo  {journal} {JCAP}\ }\textbf {\bibinfo
  {volume} {1603}}\bibfield  {number} {\bibinfo  {number} { (03)},\ \bibinfo
  {pages} {010}},\ }\Eprint {https://arxiv.org/abs/1511.08793}
  {arXiv:1511.08793 [astro-ph.CO]} \BibitemShut {NoStop}%
\bibitem [{\citenamefont {Novello}\ and\ \citenamefont
  {Bergliaffa}(2008)}]{Novello2008}%
  \BibitemOpen
  \bibfield  {author} {\bibinfo {author} {\bibfnamefont {M.}~\bibnamefont
  {Novello}}\ and\ \bibinfo {author} {\bibfnamefont {S.~E.~P.}\ \bibnamefont
  {Bergliaffa}},\ }\href {https://doi.org/10.1016/j.physrep.2008.04.006}
  {\bibfield  {journal} {\bibinfo  {journal} {Phys. Rept.}\ }\textbf {\bibinfo
  {volume} {463}},\ \bibinfo {pages} {127} (\bibinfo {year} {2008})},\ \Eprint
  {https://arxiv.org/abs/0802.1634} {arXiv:0802.1634 [astro-ph]} \BibitemShut
  {NoStop}%
\bibitem [{\citenamefont {Salim}\ \emph {et~al.}(2005)\citenamefont {Salim},
  \citenamefont {Perez~Bergliaffa},\ and\ \citenamefont {Souza}}]{Salim2004}%
  \BibitemOpen
  \bibfield  {author} {\bibinfo {author} {\bibfnamefont {J.~M.}\ \bibnamefont
  {Salim}}, \bibinfo {author} {\bibfnamefont {S.~E.}\ \bibnamefont
  {Perez~Bergliaffa}},\ and\ \bibinfo {author} {\bibfnamefont {N.}~\bibnamefont
  {Souza}},\ }\href {https://doi.org/10.1088/0264-9381/22/6/006} {\bibfield
  {journal} {\bibinfo  {journal} {Class. Quant. Grav.}\ }\textbf {\bibinfo
  {volume} {22}},\ \bibinfo {pages} {975} (\bibinfo {year} {2005})},\ \Eprint
  {https://arxiv.org/abs/astro-ph/0410423} {arXiv:astro-ph/0410423 [astro-ph]}
  \BibitemShut {NoStop}%
\bibitem [{\citenamefont {Drummond}\ and\ \citenamefont
  {Hathrell}(1980)}]{Drummond:1979pp}%
  \BibitemOpen
  \bibfield  {author} {\bibinfo {author} {\bibfnamefont {I.}~\bibnamefont
  {Drummond}}\ and\ \bibinfo {author} {\bibfnamefont {S.}~\bibnamefont
  {Hathrell}},\ }\href {https://doi.org/10.1103/PhysRevD.22.343} {\bibfield
  {journal} {\bibinfo  {journal} {Phys.\ Rev.\ D}\ }\textbf {\bibinfo {volume}
  {22}},\ \bibinfo {pages} {343} (\bibinfo {year} {1980})}\BibitemShut
  {NoStop}%
\bibitem [{\citenamefont {Cai}\ \emph {et~al.}(2009{\natexlab{a}})\citenamefont
  {Cai}, \citenamefont {Qiu}, \citenamefont {Brandenberger},\ and\
  \citenamefont {Zhang}}]{Cai2008}%
  \BibitemOpen
  \bibfield  {author} {\bibinfo {author} {\bibfnamefont {Y.-F.}\ \bibnamefont
  {Cai}}, \bibinfo {author} {\bibfnamefont {T.-t.}\ \bibnamefont {Qiu}},
  \bibinfo {author} {\bibfnamefont {R.}~\bibnamefont {Brandenberger}},\ and\
  \bibinfo {author} {\bibfnamefont {X.-m.}\ \bibnamefont {Zhang}},\ }\href
  {https://doi.org/10.1103/PhysRevD.80.023511} {\bibfield  {journal} {\bibinfo
  {journal} {Phys.\ Rev.\ D}\ }\textbf {\bibinfo {volume} {80}},\ \bibinfo
  {pages} {023511} (\bibinfo {year} {2009}{\natexlab{a}})},\ \Eprint
  {https://arxiv.org/abs/0810.4677} {arXiv:0810.4677 [hep-th]} \BibitemShut
  {NoStop}%
\bibitem [{\citenamefont {Cai}\ \emph {et~al.}(2009{\natexlab{b}})\citenamefont
  {Cai}, \citenamefont {Xue}, \citenamefont {Brandenberger},\ and\
  \citenamefont {Zhang}}]{Cai2009}%
  \BibitemOpen
  \bibfield  {author} {\bibinfo {author} {\bibfnamefont {Y.-F.}\ \bibnamefont
  {Cai}}, \bibinfo {author} {\bibfnamefont {W.}~\bibnamefont {Xue}}, \bibinfo
  {author} {\bibfnamefont {R.}~\bibnamefont {Brandenberger}},\ and\ \bibinfo
  {author} {\bibfnamefont {X.-m.}\ \bibnamefont {Zhang}},\ }\href
  {https://doi.org/10.1088/1475-7516/2009/06/037} {\bibfield  {journal}
  {\bibinfo  {journal} {JCAP}\ }\textbf {\bibinfo {volume} {06}},\ \bibinfo
  {pages} {037}},\ \Eprint {https://arxiv.org/abs/0903.4938} {arXiv:0903.4938
  [hep-th]} \BibitemShut {NoStop}%
\bibitem [{\citenamefont {Bhattacharya}\ \emph {et~al.}(2013)\citenamefont
  {Bhattacharya}, \citenamefont {Cai},\ and\ \citenamefont
  {Das}}]{Bhattacharya2013}%
  \BibitemOpen
  \bibfield  {author} {\bibinfo {author} {\bibfnamefont {K.}~\bibnamefont
  {Bhattacharya}}, \bibinfo {author} {\bibfnamefont {Y.-F.}\ \bibnamefont
  {Cai}},\ and\ \bibinfo {author} {\bibfnamefont {S.}~\bibnamefont {Das}},\
  }\href {https://doi.org/10.1103/PhysRevD.87.083511} {\bibfield  {journal}
  {\bibinfo  {journal} {Phys.\ Rev.\ D}\ }\textbf {\bibinfo {volume} {87}},\
  \bibinfo {pages} {083511} (\bibinfo {year} {2013})},\ \Eprint
  {https://arxiv.org/abs/1301.0661} {arXiv:1301.0661 [hep-th]} \BibitemShut
  {NoStop}%
\bibitem [{\citenamefont {Peter}\ and\ \citenamefont
  {Pinto-Neto}(2008)}]{Peter:2008qz}%
  \BibitemOpen
  \bibfield  {author} {\bibinfo {author} {\bibfnamefont {P.}~\bibnamefont
  {Peter}}\ and\ \bibinfo {author} {\bibfnamefont {N.}~\bibnamefont
  {Pinto-Neto}},\ }\href {https://doi.org/10.1103/PhysRevD.78.063506}
  {\bibfield  {journal} {\bibinfo  {journal} {Phys.\ Rev.\ D}\ }\textbf
  {\bibinfo {volume} {78}},\ \bibinfo {pages} {063506} (\bibinfo {year}
  {2008})},\ \Eprint {https://arxiv.org/abs/0809.2022} {arXiv:0809.2022
  [gr-qc]} \BibitemShut {NoStop}%
\bibitem [{\citenamefont {Wands}(2009)}]{Wands:2008tv}%
  \BibitemOpen
  \bibfield  {author} {\bibinfo {author} {\bibfnamefont {D.}~\bibnamefont
  {Wands}},\ }\href {https://doi.org/10.1166/asl.2009.1026} {\bibfield
  {journal} {\bibinfo  {journal} {Adv.\ Sci.\ Lett.}\ }\textbf {\bibinfo
  {volume} {2}},\ \bibinfo {pages} {194} (\bibinfo {year} {2009})},\ \Eprint
  {https://arxiv.org/abs/0809.4556} {arXiv:0809.4556 [astro-ph]} \BibitemShut
  {NoStop}%
\bibitem [{\citenamefont {Ijjas}\ and\ \citenamefont
  {Steinhardt}(2016)}]{Ijjas:2016tpn}%
  \BibitemOpen
  \bibfield  {author} {\bibinfo {author} {\bibfnamefont {A.}~\bibnamefont
  {Ijjas}}\ and\ \bibinfo {author} {\bibfnamefont {P.~J.}\ \bibnamefont
  {Steinhardt}},\ }\href {https://doi.org/10.1103/PhysRevLett.117.121304}
  {\bibfield  {journal} {\bibinfo  {journal} {Phys.\ Rev.\ Lett.}\ }\textbf
  {\bibinfo {volume} {117}},\ \bibinfo {pages} {121304} (\bibinfo {year}
  {2016})},\ \Eprint {https://arxiv.org/abs/1606.08880} {arXiv:1606.08880
  [gr-qc]} \BibitemShut {NoStop}%
\bibitem [{\citenamefont {Cubero}\ and\ \citenamefont
  {Popławski}(2020)}]{Cubero:2019lxw}%
  \BibitemOpen
  \bibfield  {author} {\bibinfo {author} {\bibfnamefont {J.~L.}\ \bibnamefont
  {Cubero}}\ and\ \bibinfo {author} {\bibfnamefont {N.~J.}\ \bibnamefont
  {Popławski}},\ }\href {https://doi.org/10.1088/1361-6382/ab5cb9} {\bibfield
  {journal} {\bibinfo  {journal} {Class. Quant. Grav.}\ }\textbf {\bibinfo
  {volume} {37}},\ \bibinfo {pages} {025011} (\bibinfo {year} {2020})},\
  \Eprint {https://arxiv.org/abs/1906.11824} {arXiv:1906.11824 [gr-qc]}
  \BibitemShut {NoStop}%
\bibitem [{\citenamefont {Galkina}\ \emph {et~al.}(2019)\citenamefont
  {Galkina}, \citenamefont {Fabris}, \citenamefont {Falciano},\ and\
  \citenamefont {Pinto-Neto}}]{Galkina:2019pir}%
  \BibitemOpen
  \bibfield  {author} {\bibinfo {author} {\bibfnamefont {O.}~\bibnamefont
  {Galkina}}, \bibinfo {author} {\bibfnamefont {J.}~\bibnamefont {Fabris}},
  \bibinfo {author} {\bibfnamefont {F.}~\bibnamefont {Falciano}},\ and\
  \bibinfo {author} {\bibfnamefont {N.}~\bibnamefont {Pinto-Neto}},\ }\href
  {https://doi.org/10.1134/S0021364019200013} {\bibfield  {journal} {\bibinfo
  {journal} {JETP Lett.}\ }\textbf {\bibinfo {volume} {110}},\ \bibinfo {pages}
  {523} (\bibinfo {year} {2019})},\ \Eprint {https://arxiv.org/abs/1908.04258}
  {arXiv:1908.04258 [gr-qc]} \BibitemShut {NoStop}%
\bibitem [{\citenamefont {Almeida}\ \emph {et~al.}(2018)\citenamefont
  {Almeida}, \citenamefont {Bergeron}, \citenamefont {Gazeau},\ and\
  \citenamefont {Scardua}}]{Almeida:2018xvj}%
  \BibitemOpen
  \bibfield  {author} {\bibinfo {author} {\bibfnamefont {C.}~\bibnamefont
  {Almeida}}, \bibinfo {author} {\bibfnamefont {H.}~\bibnamefont {Bergeron}},
  \bibinfo {author} {\bibfnamefont {J.-P.}\ \bibnamefont {Gazeau}},\ and\
  \bibinfo {author} {\bibfnamefont {A.}~\bibnamefont {Scardua}},\ }\href
  {https://doi.org/10.1016/j.aop.2018.03.010} {\bibfield  {journal} {\bibinfo
  {journal} {Annals Phys.}\ }\textbf {\bibinfo {volume} {392}},\ \bibinfo
  {pages} {206} (\bibinfo {year} {2018})},\ \Eprint
  {https://arxiv.org/abs/1708.06422} {1708.06422 [quant-ph]} \BibitemShut
  {NoStop}%
\bibitem [{\citenamefont {Bacalhau}\ \emph {et~al.}(2018)\citenamefont
  {Bacalhau}, \citenamefont {Pinto-Neto},\ and\ \citenamefont {Dias
  Pinto~Vitenti}}]{Bacalhau:2017hja}%
  \BibitemOpen
  \bibfield  {author} {\bibinfo {author} {\bibfnamefont {A.~P.}\ \bibnamefont
  {Bacalhau}}, \bibinfo {author} {\bibfnamefont {N.}~\bibnamefont
  {Pinto-Neto}},\ and\ \bibinfo {author} {\bibfnamefont {S.}~\bibnamefont {Dias
  Pinto~Vitenti}},\ }\href {https://doi.org/10.1103/PhysRevD.97.083517}
  {\bibfield  {journal} {\bibinfo  {journal} {Phys.\ Rev.\ D}\ }\textbf
  {\bibinfo {volume} {97}},\ \bibinfo {pages} {083517} (\bibinfo {year}
  {2018})},\ \Eprint {https://arxiv.org/abs/1706.08830} {arXiv:1706.08830
  [gr-qc]} \BibitemShut {NoStop}%
\bibitem [{\citenamefont {Frion}\ and\ \citenamefont
  {Almeida}(2019)}]{Frion:2018oij}%
  \BibitemOpen
  \bibfield  {author} {\bibinfo {author} {\bibfnamefont {E.}~\bibnamefont
  {Frion}}\ and\ \bibinfo {author} {\bibfnamefont {C.}~\bibnamefont
  {Almeida}},\ }\href {https://doi.org/10.1103/PhysRevD.99.023524} {\bibfield
  {journal} {\bibinfo  {journal} {Phys.\ Rev.\ D}\ }\textbf {\bibinfo {volume}
  {99}},\ \bibinfo {pages} {023524} (\bibinfo {year} {2019})},\ \Eprint
  {https://arxiv.org/abs/1810.00707} {arXiv:1810.00707 [gr-qc]} \BibitemShut
  {NoStop}%
\bibitem [{\citenamefont {Bohm}(1952{\natexlab{a}})}]{Bohm:1951xw}%
  \BibitemOpen
  \bibfield  {author} {\bibinfo {author} {\bibfnamefont {D.}~\bibnamefont
  {Bohm}},\ }\href {https://doi.org/10.1103/PhysRev.85.166} {\bibfield
  {journal} {\bibinfo  {journal} {Phys.\ Rev.}\ }\textbf {\bibinfo {volume}
  {85}},\ \bibinfo {pages} {166} (\bibinfo {year}
  {1952}{\natexlab{a}})}\BibitemShut {NoStop}%
\bibitem [{\citenamefont {Bohm}(1952{\natexlab{b}})}]{Bohm:1951xx}%
  \BibitemOpen
  \bibfield  {author} {\bibinfo {author} {\bibfnamefont {D.}~\bibnamefont
  {Bohm}},\ }\href {https://doi.org/10.1103/PhysRev.85.180} {\bibfield
  {journal} {\bibinfo  {journal} {Phys.\ Rev.}\ }\textbf {\bibinfo {volume}
  {85}},\ \bibinfo {pages} {180} (\bibinfo {year}
  {1952}{\natexlab{b}})}\BibitemShut {NoStop}%
\bibitem [{\citenamefont {Caprini}\ and\ \citenamefont
  {Durrer}(2001)}]{Caprini:2001nb}%
  \BibitemOpen
  \bibfield  {author} {\bibinfo {author} {\bibfnamefont {C.}~\bibnamefont
  {Caprini}}\ and\ \bibinfo {author} {\bibfnamefont {R.}~\bibnamefont
  {Durrer}},\ }\href {https://doi.org/10.1103/PhysRevD.65.023517} {\bibfield
  {journal} {\bibinfo  {journal} {Phys.\ Rev.\ D}\ }\textbf {\bibinfo {volume}
  {65}},\ \bibinfo {pages} {023517} (\bibinfo {year} {2001})},\ \Eprint
  {https://arxiv.org/abs/astro-ph/0106244} {arXiv:astro-ph/0106244 [astro-ph]}
  \BibitemShut {NoStop}%
\bibitem [{\citenamefont {Durrer}\ and\ \citenamefont
  {Caprini}(2003)}]{Durrer:2003ja}%
  \BibitemOpen
  \bibfield  {author} {\bibinfo {author} {\bibfnamefont {R.}~\bibnamefont
  {Durrer}}\ and\ \bibinfo {author} {\bibfnamefont {C.}~\bibnamefont
  {Caprini}},\ }\href {https://doi.org/10.1088/1475-7516/2003/11/010}
  {\bibfield  {journal} {\bibinfo  {journal} {JCAP}\ }\textbf {\bibinfo
  {volume} {11}},\ \bibinfo {pages} {010}},\ \Eprint
  {https://arxiv.org/abs/astro-ph/0305059} {arXiv:astro-ph/0305059 [astro-ph]}
  \BibitemShut {NoStop}%
\bibitem [{\citenamefont {Safarzadeh}\ and\ \citenamefont
  {Loeb}(2019)}]{Safarzadeh:2019kyq}%
  \BibitemOpen
  \bibfield  {author} {\bibinfo {author} {\bibfnamefont {M.}~\bibnamefont
  {Safarzadeh}}\ and\ \bibinfo {author} {\bibfnamefont {A.}~\bibnamefont
  {Loeb}},\ }\href {https://doi.org/10.3847/2041-8213/ab2335} {\bibfield
  {journal} {\bibinfo  {journal} {Astrophys.\ J.}\ }\textbf {\bibinfo {volume}
  {877}},\ \bibinfo {pages} {L27} (\bibinfo {year} {2019})},\ \Eprint
  {https://arxiv.org/abs/1901.03341} {arXiv:1901.03341 [astro-ph.CO]}
  \BibitemShut {NoStop}%
\bibitem [{\citenamefont {Broderick}\ \emph {et~al.}(2018)\citenamefont
  {Broderick} \emph {et~al.}}]{Broderick:2018nqf}%
  \BibitemOpen
  \bibfield  {author} {\bibinfo {author} {\bibfnamefont {A.~E.}\ \bibnamefont
  {Broderick}} \emph {et~al.},\ }\href
  {https://doi.org/10.3847/1538-4357/aae5f2} {\bibfield  {journal} {\bibinfo
  {journal} {Astrophys.\ J.}\ }\textbf {\bibinfo {volume} {868}},\ \bibinfo
  {pages} {87} (\bibinfo {year} {2018})},\ \Eprint
  {https://arxiv.org/abs/1808.02959} {arXiv:1808.02959 [astro-ph.HE]}
  \BibitemShut {NoStop}%
\bibitem [{\citenamefont {Broderick}\ \emph {et~al.}(2012)\citenamefont
  {Broderick}, \citenamefont {Chang},\ and\ \citenamefont
  {Pfrommer}}]{Broderick:2011av}%
  \BibitemOpen
  \bibfield  {author} {\bibinfo {author} {\bibfnamefont {A.~E.}\ \bibnamefont
  {Broderick}}, \bibinfo {author} {\bibfnamefont {P.}~\bibnamefont {Chang}},\
  and\ \bibinfo {author} {\bibfnamefont {C.}~\bibnamefont {Pfrommer}},\ }\href
  {https://doi.org/10.1088/0004-637X/752/1/22} {\bibfield  {journal} {\bibinfo
  {journal} {Astrophys. J.}\ }\textbf {\bibinfo {volume} {752}},\ \bibinfo
  {pages} {22} (\bibinfo {year} {2012})},\ \Eprint
  {https://arxiv.org/abs/1106.5494} {arXiv:1106.5494 [astro-ph.CO]}
  \BibitemShut {NoStop}%
\bibitem [{\citenamefont {Subramanian}\ \emph {et~al.}(1994)\citenamefont
  {Subramanian}, \citenamefont {Narasimha},\ and\ \citenamefont
  {Chitre}}]{subramanian1994}%
  \BibitemOpen
  \bibfield  {author} {\bibinfo {author} {\bibfnamefont {K.}~\bibnamefont
  {Subramanian}}, \bibinfo {author} {\bibfnamefont {D.}~\bibnamefont
  {Narasimha}},\ and\ \bibinfo {author} {\bibfnamefont {S.}~\bibnamefont
  {Chitre}},\ }\href {https://doi.org/10.1093/mnras/271.1.L15} {\bibfield
  {journal} {\bibinfo  {journal} {Mon.\ Not.\ Roy.\ Astron.\ Soc.}\ }\textbf
  {\bibinfo {volume} {271}},\ \bibinfo {pages} {L15} (\bibinfo {year}
  {1994})}\BibitemShut {NoStop}%
\bibitem [{\citenamefont {Taylor}\ \emph {et~al.}(2011)\citenamefont {Taylor},
  \citenamefont {Vovk},\ and\ \citenamefont {Neronov}}]{Taylor:2011bn}%
  \BibitemOpen
  \bibfield  {author} {\bibinfo {author} {\bibfnamefont {A.}~\bibnamefont
  {Taylor}}, \bibinfo {author} {\bibfnamefont {I.}~\bibnamefont {Vovk}},\ and\
  \bibinfo {author} {\bibfnamefont {A.}~\bibnamefont {Neronov}},\ }\href
  {https://doi.org/10.1051/0004-6361/201116441} {\bibfield  {journal} {\bibinfo
   {journal} {Astron.\ Astrophys.}\ }\textbf {\bibinfo {volume} {529}},\
  \bibinfo {pages} {A144} (\bibinfo {year} {2011})},\ \Eprint
  {https://arxiv.org/abs/1101.0932} {arXiv:1101.0932 [astro-ph.HE]}
  \BibitemShut {NoStop}%
\bibitem [{\citenamefont {Kanno}\ \emph {et~al.}(2009)\citenamefont {Kanno},
  \citenamefont {Soda},\ and\ \citenamefont {Watanabe}}]{Kanno:2009ei}%
  \BibitemOpen
  \bibfield  {author} {\bibinfo {author} {\bibfnamefont {S.}~\bibnamefont
  {Kanno}}, \bibinfo {author} {\bibfnamefont {J.}~\bibnamefont {Soda}},\ and\
  \bibinfo {author} {\bibfnamefont {M.-a.}\ \bibnamefont {Watanabe}},\ }\href
  {https://doi.org/10.1088/1475-7516/2009/12/009} {\bibfield  {journal}
  {\bibinfo  {journal} {JCAP}\ }\textbf {\bibinfo {volume} {0912}},\ \bibinfo
  {pages} {009}},\ \Eprint {https://arxiv.org/abs/0908.3509} {arXiv:0908.3509
  [astro-ph.CO]} \BibitemShut {NoStop}%
\bibitem [{\citenamefont {Pinto-Neto}\ and\ \citenamefont
  {Fabris}(2013)}]{Pinto-Neto:2013toa}%
  \BibitemOpen
  \bibfield  {author} {\bibinfo {author} {\bibfnamefont {N.}~\bibnamefont
  {Pinto-Neto}}\ and\ \bibinfo {author} {\bibfnamefont {J.}~\bibnamefont
  {Fabris}},\ }\href {https://doi.org/10.1088/0264-9381/30/14/143001}
  {\bibfield  {journal} {\bibinfo  {journal} {Class.\ Quant.\ Grav.}\ }\textbf
  {\bibinfo {volume} {30}},\ \bibinfo {pages} {143001} (\bibinfo {year}
  {2013})},\ \Eprint {https://arxiv.org/abs/1306.0820} {arXiv:1306.0820
  [gr-qc]} \BibitemShut {NoStop}%
\bibitem [{\citenamefont {Caprini}\ \emph {et~al.}(2009)\citenamefont
  {Caprini}, \citenamefont {Durrer},\ and\ \citenamefont
  {Servant}}]{Caprini:2009yp}%
  \BibitemOpen
  \bibfield  {author} {\bibinfo {author} {\bibfnamefont {C.}~\bibnamefont
  {Caprini}}, \bibinfo {author} {\bibfnamefont {R.}~\bibnamefont {Durrer}},\
  and\ \bibinfo {author} {\bibfnamefont {G.}~\bibnamefont {Servant}},\ }\href
  {https://doi.org/10.1088/1475-7516/2009/12/024} {\bibfield  {journal}
  {\bibinfo  {journal} {JCAP}\ }\textbf {\bibinfo {volume} {0912}},\ \bibinfo
  {pages} {024}},\ \Eprint {https://arxiv.org/abs/0909.0622} {arXiv:0909.0622
  [astro-ph.CO]} \BibitemShut {NoStop}%
\bibitem [{\citenamefont {Caprini}\ and\ \citenamefont
  {Figueroa}(2018)}]{Caprini:2018mtu}%
  \BibitemOpen
  \bibfield  {author} {\bibinfo {author} {\bibfnamefont {C.}~\bibnamefont
  {Caprini}}\ and\ \bibinfo {author} {\bibfnamefont {D.~G.}\ \bibnamefont
  {Figueroa}},\ }\href {https://doi.org/10.1088/1361-6382/aac608} {\bibfield
  {journal} {\bibinfo  {journal} {Class. Quant. Grav.}\ }\textbf {\bibinfo
  {volume} {35}},\ \bibinfo {pages} {163001} (\bibinfo {year} {2018})},\
  \Eprint {https://arxiv.org/abs/1801.04268} {arXiv:1801.04268 [astro-ph.CO]}
  \BibitemShut {NoStop}%
\bibitem [{\citenamefont {Saga}\ \emph
  {et~al.}(2018{\natexlab{b}})\citenamefont {Saga}, \citenamefont {Tashiro},\
  and\ \citenamefont {Yokoyama}}]{Saga:2018ont}%
  \BibitemOpen
  \bibfield  {author} {\bibinfo {author} {\bibfnamefont {S.}~\bibnamefont
  {Saga}}, \bibinfo {author} {\bibfnamefont {H.}~\bibnamefont {Tashiro}},\ and\
  \bibinfo {author} {\bibfnamefont {S.}~\bibnamefont {Yokoyama}},\ }\href
  {https://doi.org/10.1103/PhysRevD.98.083518} {\bibfield  {journal} {\bibinfo
  {journal} {Phys.\ Rev.\ D}\ }\textbf {\bibinfo {volume} {98}},\ \bibinfo
  {pages} {083518} (\bibinfo {year} {2018}{\natexlab{b}})},\ \Eprint
  {https://arxiv.org/abs/1807.00561} {arXiv:1807.00561 [astro-ph.CO]}
  \BibitemShut {NoStop}%
\bibitem [{\citenamefont {Roper~Pol}\ \emph {et~al.}(2019)\citenamefont
  {Roper~Pol}, \citenamefont {Mandal}, \citenamefont {Brandenburg},
  \citenamefont {Kahniashvili},\ and\ \citenamefont {Kosowsky}}]{Pol:2019yex}%
  \BibitemOpen
  \bibfield  {author} {\bibinfo {author} {\bibfnamefont {A.}~\bibnamefont
  {Roper~Pol}}, \bibinfo {author} {\bibfnamefont {S.}~\bibnamefont {Mandal}},
  \bibinfo {author} {\bibfnamefont {A.}~\bibnamefont {Brandenburg}}, \bibinfo
  {author} {\bibfnamefont {T.}~\bibnamefont {Kahniashvili}},\ and\ \bibinfo
  {author} {\bibfnamefont {A.}~\bibnamefont {Kosowsky}},\ }\href@noop {} {\
  (\bibinfo {year} {2019})},\ \Eprint {https://arxiv.org/abs/1903.08585}
  {arXiv:1903.08585 [astro-ph.CO]} \BibitemShut {NoStop}%
\bibitem [{\citenamefont {Vitenti}(2020)}]{sandro_vacuum}%
  \BibitemOpen
  \bibfield  {author} {\bibinfo {author} {\bibfnamefont {S.~D.~P.}\
  \bibnamefont {Vitenti}},\ }\href@noop {} {\bibfield  {journal} {\bibinfo
  {journal} {in preparation}\ } (\bibinfo {year} {2020})}\BibitemShut {NoStop}%
\end{thebibliography}%

\end{document}